\documentclass[12pt,a4paper,reqno]{amsart}
\usepackage{amsmath, amsfonts,
   amssymb, 
   amsthm, euscript}

\usepackage{diagrams}
 \diagramstyle[small,scriptlabels,
 midshaft,nohug]
 \newarrow{TeXto}----{->}
 \newcommand{\rto}{\rTeXto}
 \newcommand{\lto}{\lTeXto}
 \newcommand{\uto}{\uTeXto}
 \newcommand{\dto}{\dTeXto}
 
 \newcommand{\rdto}{\rdTeXto}
 \newcommand{\luto}{\luTeXto}

\newtheorem{theor}{Theorem}
\theoremstyle{definition}
\newtheorem*{theorNo}{Theorem}
\newtheorem{state}[theor]{Proposition}
\newtheorem{lemma}[theor]{Lemma}
\newtheorem{cor}[theor]{Corollary}
\newtheorem{conjecture}[theor]{Conjecture}
\newtheorem{define}{Definition}
\newtheorem*{defNo}{Definition}
\newtheorem{property}{Property}
\newtheorem{example}{Example}
\newtheorem{counterexample}[example]{Counterexample}
\theoremstyle{remark}
\newtheorem{rem}{Remark}
\newtheorem*{Finalrem}{Final remark}

\DeclareFontFamily{OML}{cyr}{}
\DeclareFontShape{OML}{cyr}{m}{n}{
   <5> <6> <7> <8> <9> gen * wncyr
   <10> <10.95> <12> <14.4> <17.28> <20.74> <24.88> wncyr10
  }{}
\DeclareSymbolFont{rusletters}{OML}{cyr}{m}{n}
\DeclareSymbolFontAlphabet{\rusmath}{rusletters}
\DeclareMathSymbol\re{\rusmath}{rusletters}{"03}
\newcommand{\cEv}{\EuScript{E}}

\newcommand{\pinner}{\mathbin{\mathchoice
   {\hbox{\vrule width0.6em depth0pt height0.4pt
   \vrule width0.4pt depth0pt height0.8ex}}
   {\hbox{\vrule width0.6em depth0pt height0.4pt
   \vrule width0.4pt depth0pt height0.8ex}}
   {\hbox{\kern0.14em
   \vrule width0.48em depth0pt height0.4pt
   \vrule width0.4pt depth0pt height0.6ex\kern0.14em}}
   {\hbox{\kern0.1em
   \vrule width0.39em depth0pt height0.4pt
   \vrule width0.4pt depth0pt height0.5ex\kern0.1em}}}}
\newcommand{\inner}{\pinner\,}

\newcommand{\BBR}{\mathbb{R}}\newcommand{\BBC}{\mathbb{C}}
\newcommand{\BBN}{\mathbb{N}}
\newcommand{\BBZ}{\mathbb{Z}}
\newcommand{\EuA}{{{\EuScript A}}}
\newcommand{\cA}{\mathcal{A}}
\newcommand{\cB}{\mathcal{B}}
\newcommand{\cC}{\mathcal{C}}
\newcommand{\cE}{\mathcal{E}}
\newcommand{\cEL}{\mathcal{E}_{\IL}}
\newcommand{\cEEL}{{\cE}_{\text{\textup{EL}}}}
\newcommand{\cELiou}{{\cE}_{\text{\textup{Liou}}}}
\newcommand{\cEToda}{{\cE}_{\text{\textup{Toda}}}}
\newcommand{\cF}{\mathcal{F}}
\newcommand{\cH}{\mathcal{H}}

\newcommand{\cL}{\mathcal{L}}
\newcommand{\cP}{\mathcal{P}}
\newcommand{\cU}{\mathcal{U}}
\newcommand{\bc}{{\mathbf{c}}}
\newcommand{\bE}{\mathbf{E}}
\newcommand{\bR}{\mathbf{r}}
\newcommand{\bU}{\mathbf{U}}
\newcommand{\bun}{\mathbf{1}}
\newcommand{\ga}{\mathfrak{a}}
\newcommand{\gf}{\mathfrak{f}}
\newcommand{\hgf}{\smash{\hat{\mathfrak{f}}}}
\newcommand{\gh}{\mathfrak{h}}
\newcommand{\gm}{\mathfrak{m}}
\newcommand{\gothg}{\mathfrak{g}}
\newcommand{\gA}{\mathfrak{A}}
\newcommand{\gB}{\mathfrak{B}}
\newcommand{\veps}{\varepsilon}
\newcommand{\vph}{\varphi}
\newcommand{\dd}{\partial}
\newcommand{\Id}{{\mathrm d}}
\newcommand{\ID}{{\mathrm D}}
\newcommand{\IL}{{\mathrm L}}

\newcommand{\rme}{{\mathrm{e}}}
\newcommand{\rmN}{{\mathrm{N}}}
\newcommand{\uu}{{\underline{u}}}
\newcommand{\uv}{{\underline{v}}}
\newcommand{\uw}{{\underline{w}}}

\DeclareMathOperator{\img}{im}
\DeclareMathOperator{\dom}{dom}
\DeclareMathOperator{\id}{id}
\DeclareMathOperator{\sym}{sym}
\DeclareMathOperator{\cosym}{cosym}

\DeclareMathOperator{\arcsinh}{arcsinh}

\DeclareMathOperator{\Hom}{Hom}
\DeclareMathOperator{\End}{End}
\DeclareMathOperator{\Der}{Der}
\DeclareMathOperator{\CDiff}{\mathcal{C}Diff}
\DeclareMathOperator{\Diff}{Diff}
\DeclareMathOperator{\ord}{ord}
\DeclareMathOperator{\volume}{vol}
\DeclareMathOperator{\ad}{ad}

\newcommand{\lshad}{[\![}
\newcommand{\rshad}{]\!]}

\newcommand{\by}[1]{\textit{{#1}}}
\newcommand{\jour}[1]{\textit{{#1}}}
\newcommand{\vol}[1]{\textbf{{#1}}}
\newcommand{\book}[1]{\textrm{{#1}}}

\newcommand{\ib}[3]{ \{\!\{ {#1},{#2} \}\!\}_{{#3}} }
\newcommand{\ob}[2]{ \mathbf{[} {#1},{#2} \mathbf{]} }

\title[Involutive distributions of operator\/-\/valued evolutionary
derivations]%
{Involutive distributions of operator\/-\/valued evolutionary vector
fields}

\date{December 3, 2007; revised April 28, 2008;
in final form August 4, 2008.} 

\author[A. V. Kiselev]{Arthemy V. Kiselev} 
\thanks{
        \textit{Address}:
Mathematical Institute, University of Utrecht, P.O.Box~80.010, 3508~TA Utrecht, The Netherlands.
\textit{E-mails}: [\texttt{A.V.Kiselev},
\texttt{J.W.vandeLeur}]\texttt{\symbol{"40}uu.nl}%
}

\author[J. W. van de Leur]{Johan W. van de Leur}
\thanks{Partial financial support from
Twente 2006 Conference on Lie groups, 
7th International Confenence `Symmetry in Nonlinear
Mathematical Physics' (Kiev, 
2007), and 5th International Workshop `Nonlinear Physics: Theory and
Experiment' (Gallipoli, 
2008) is gratefully acknowledged.}



\begin{document}

\begin{abstract}
Converting a pioneering idea of V.~V.~Sokolov 
\textit{et al.}~\cite{SokolovUMN} to a geometric object,
we introduce a well\/-\/defined notion of linear matrix operators in total
derivatives, whose images in the Lie algebras $\gothg$ of evolutionary
vector fields on jet spaces are closed with respect to the commutation,
\begin{equation}\tag{*}\label{ADefFrob}
[\img A,\img A]\subseteq\img A.
\end{equation}
The images generate involutive distributions
on infinite jet bundles $J^\infty(\pi)$ over fibre bundles $\pi$,
and the operators induce Lie algebra structures $[\,,\,]_A$
on $\Omega=\dom A/\ker A$.
We postulate that
the operators are classified by the vector and covector transformation laws for
their domains $\smash{\dom A}$, which are related to sections
of~$\pi$ by Miura substitutions.\ The gauge of $\smash{\dom A}$ may be
independent from transformations of the images $\img A$.\ This is
a generalization of the classical theory with
$\smash{\dom A}\simeq\img A\subseteq\sym\cE$ for recursion operators for
integrable systems~$\cE$,
and with $\smash{\dom A}\simeq\smash{\widehat{\img A}}\supseteq\cosym\cE$
for Hamiltonian operators (here the Miura substitutions are the identity
mappings in both cases, and the gauge transformations are 
uniquely correlated).
In particular, recursion operators $A\in\End_\Bbbk\sym\cE$ that
satisfy~\eqref{ADefFrob} are solutions of the classical Yang\/--\/Baxter
equation $[A\vph_1,A\vph_2]=A\bigl([\vph_1,\vph_2]_A\bigr)$
for the Lie algebra~$\sym\cE\ni\vph_1,\vph_2$.


If, for $r$~linear differential operators $A_i\colon\Omega\to\gothg$
with a common domain $\Omega$, pairwise commutators of their images hit the
sum of images again, we endow the spaces
$\EuA=\bigoplus_{i=1}^r\Bbbk\cdot A_i$ of the operators with a bilinear bracket
\[
\ob{A_i}{A_j}=\sum\nolimits_{k=1}^r A_k\circ \bc_{ij}^k,\qquad
\bc_{ij}^k\in\Diff(\Omega\times\Omega\to\Omega),\quad 1\leq i,j\leq r
\]
that satisfies the Jacobi identity.
A class of such algebras 
is given by operators that generate Noether symmetries
of hyperbolic Euler\/--\/Lagrange systems of Liouville type; we calculate
explicitly the operators $A_i$ and the structural constants $\bc_{ij}^k$.
Thus 
we give an exhaustive description of higher symmetries
for all 2D~Toda chains associated with semi\/-\/simple 
Lie algebras of rank~$r$, completing the 
results of~\cite{LeznovSmirnovShabat,Meshkov198x,Sakovich,Shabat,SokolovUMN}.

The bracket $\ob{\,}{\,}$ of operators
is anti\/-\/symmetric for the 
domains $\Omega$ 
composed by vectors and is
symmetric for $\Omega$ with covector transformations. 
In the latter case, we reveal a flat non\/-\/Cartan affine connection
in the triples $\Omega\xrightarrow{\EuA}\gothg$
such that symmetric bi\/-\/differential Christoffel symbols~$\Gamma_{ij}^k$
are encoded by the structural constants $\bc_{ij}^k$, and such that completely
integrable commutative hierarchies $\gA\subset\img\EuA$ are the geodesics.

We demonstrate that the notion of Lie algebroids over
infinite jet bundles~$J^\infty(\pi)$
does not repeat the construction over finite\/-\/dimensional
base manifolds, when condition~\eqref{ADefFrob} is fulfilled by the anchors.
To correlate them 
with Lie algebra homomorphisms $A\colon\Omega\to\gothg$, we introduce formal
differential complexes over Lie algebras $\bigl(\Omega,[\,,\,]_A\bigr)$.
\end{abstract}
\enlargethispage{1.2\baselineskip}\maketitle\thispagestyle{empty}




\newpage\thispagestyle{plain}\noindent%
SIGMA (2008), submitted.\hfill Preprint $\smash{\text{IH\'ES}}$-M/07/38\\
\centerline{\rule{\textwidth}{0.7pt}}

\subsection*{Introduction}
Let $M^n$~be a smooth $n$-\/dimensional real manifold
and $\pi\colon E^{n+m}\to M^n$ be a fibre bundle.
The space of linear differential operators over $\BBR$ on the infinite jet
bundle $J^\infty(\pi)\xrightarrow{\pi_\infty}M^n$
contains operators~$A$ in total derivatives
whose images in the $C^\infty(J^\infty(\pi))$-\/module
$\Gamma(\pi_\infty^*(\pi))=\Gamma(\pi)\mathbin{{\otimes}_{C^\infty(M^n)}}
C^\infty(J^\infty(\pi))$ of evolutionary derivations are closed
with respect to the commutation,
\begin{equation}\label{IDefFrob}
[\img A,\img A]\subseteq\img A.
\end{equation}
We say that such operators~$A$ are
\emph{Frobenius} in view of the fact that their images constitute
involutive distributions on~$J^\infty(\pi)$. Each Frobenius operator transfers
the Lie algebra structure of evolutionary vector fields
to the bracket $[\,,\,]_A$ on the quotient of its domain by the
kernel and determines an $\BBR$-\/homomorphism of Lie algebras.
The Hamiltonian operators that correspond to the Poisson bi\/-\/vectors
with vanishing Schouten bracket are an example, among many others;
e.g., the 2D~Toda systems $\cEToda=\{u^i_{xy}=\exp(K^i_{\,j}u^j)\}$
associated with the Cartan matrices $K$ determine a class of
Frobenius operators that take values in the spaces
$
\sym\cEToda$ of their (Noether) higher
infinitesimal symmetries.

The main problem we address is the coordinate\/-\/independent definition
of operators that satisfy~\eqref{IDefFrob}, and we perform the analysis of their algebraic and geometric properties. We conclude that, up to a special case,
there are two principal classes of Frobenius operators, which generalize the recursions
$\sym\cE\to\sym\cE$ and the Hamiltonian structures $\cosym\cE\to\sym\cE$ for
integrable systems~$\cE$, respectively. The essential distinction of generic
Frobenius operators~$A$ from the recursion or Hamiltonian operators is that the
coordinate transformations in the domains
of~$A$ can be not correlated with the
transformations in their images. In this sense, the two arising classes of
Frobenius operators correspond to vector and covector\/-\/type transformation laws
for their domains~$\Omega^1(\xi_\pi)$, which are defined by $\pi$ and
another fibre bundle~$\xi$ over~$M^n$.

For example, if $\xi=\pi$ and hence
$\Omega^1(\pi_\pi)=\Gamma\bigl(\pi_\infty^*(\pi)\bigr)$, then Frobenius
endomorphisms $R\in\End_\BBR\Gamma(\pi_\infty^*(\pi))$ of generating
sections~$\vph$ of evolutionary vector fields are not only recursions for
the Lie algebra $(\gothg(\pi),[\,,\,])$ of derivations. Indeed, they induce
nontrivial deformations $[\,,\,]_{R}$ of the standard structure~$[\,,\,]$
on $\Gamma\bigl(\pi_\infty^*(\pi)\bigr)$ 
via 
the classical Yang\/--\/Baxter equation
\begin{equation}\label{IFrobRec}
[R\vph_1,R\vph_2]=R\bigl([\vph_1,\vph_2]_R\bigr),\qquad
\vph_1,\vph_2\in\Gamma(\pi_\infty^*(\pi)).
\end{equation}
Frobenius operators of second kind, which generalize the Hamiltonian
structures, are assinged in this paper to hyperbolic Euler\/--\/Lagrange
systems of Liouville type. To do that, we give an exhaustive description of
higher symmetries for these integrable systems. In particular,
our explicit formula~\eqref{Square} for the operators is valid for all
2D~Toda chains associated with root systems of complex semi\/-\/simple Lie
algebras of rank~$r$. These operators determine commutative hierarchies of
KdV\/-\/type evolution equations which are Noether symmetries of the
Euler\/--\/Lagrange systems. The operators specify
factorizations of higher Hamiltonian structures for the hierarchies.

We introduce a Lie\/-\/type algebra structure on spaces of linear differential
operators in total derivatives. We consider the linear spaces
$\EuA=\bigoplus_{i=1}^r\Bbbk\cdot A_i$ of operators
$A_i\colon\Omega^1(\xi_\pi)\to\Gamma\bigl(\pi_\infty^*(\pi)\bigr)$
with a common domain $\Omega^1(\xi_\pi)$ such that the sums of images of~$A_i$ are
closed under the commutation of vector fields,
\begin{equation}\label{ICommutClosure}
\Bigl[\sum\nolimits_i\img A_i,\sum\nolimits_j\img A_j\Bigr]\subseteq
\sum\nolimits_k\img A_k.
\end{equation}
By setting $\ob{A_i}{A_j}(p,q)\mathrel{{:}{=}}\bigl[A_i(p),A_j(q)\bigr]$
for $p,q\in\Omega^1(\xi_\pi)$ and $1\leq i,j\leq r$,
we endow the linear spaces~$\EuA$ with a Lie\/-\/type bracket
\begin{equation}\label{ICommuteOperators}
\ob{A_i}{A_j}=\sum\nolimits_{k=1}^r A_k\circ \bc_{ij}^k
\end{equation}
that satisfies the Jacobi identity. The bi\/-\/differential structural
constants $\bc_{ij}^k\colon\Omega^1(\xi_\pi)\times\Omega^1(\xi_\pi)
\to\Omega^1(\xi_\pi)$ are skew\/-\/symmetric for vector
pre\/-\/images of evolutionary fields and are symmetric for covectors
that constitute the domain~$\Omega^1(\xi_\pi)$ of~$A_i$. In the latter case,
we calculate the constants $\bc_{ij}^k$ explicitly.
This algebraic problem and our approach are different from Sato's
formalism~\cite{VdLeurKac} of pseudodifferential operators because, e.g.,
the compositions of Frobenius operators
$
\Omega^1(\xi_\pi)\to\Gamma(\pi_\infty^*(\pi))$ are not defined even in the
Hamiltonian case, 
and hence they do not constitute multiplicative associative algebras with unit.
Therefore the Lie\/-\/type algebra structure
relies heavily on
the intrinsic geometry of the jets~$J^\infty(\pi)$
as the base manifolds for all fibre bundles.
The bracket we propose, in particular, for Frobenius recursions~$R$ is
different from the Richardson\/--\/Nijenhuis
bracket~$\lshad\,,\,\rshad$ that determines their Lie
super\/-\/algebra structure~\cite[\S5.3]{Opava} by using the composition of
operators, although we use similar geometric techniques.

In this paper, we pass 
from the (wide) category of fibre bundles
over~$M^n$ to a more narrow category of bundles and
horizontal modules over the infinite jets~$J^\infty(\pi)$.
Then we investigate which objects survive and which there appear anew.
We conclude that the operators that obey~\eqref{IDefFrob} are closely related
to Lie algebroids over smooth finite\/-\/dimensional manifolds~$M^n$.
We show that Lie algebroids over $J^\infty(\pi)$, with the anchors
given by Frobenius operators, can not be defined as straightforward
generalizations of the finite\/-\/dimensional case, but this is achieved with a much more refine construction. To this end, for each Frobenius operator~$A$,
we generate a formal differential complex over Lie algebra
$\bigl(\Omega^1(\xi_\pi),[\,,\,]_A\bigr)$ 
and suggest to determine the Lie algebroids using representations of the
differential in this complex through homological vector fields.

On the other hand, we discover that Frobenius operators determine a generalization of the affine geometry such that bi\/-\/differential Christoffel
symbols~$\Gamma_{ij}^k$ are encoded by the structural constants $\bc_{ij}^k$ of
the algebras~$\EuA$. We prove that Frobenius operators~$A$ determine flat
connections in the triples
$\bigl(\Omega^1(\xi_\pi),\Gamma(\pi_\infty^*(\pi)),A\bigr)$
consisting of two Lie algebras and a morphism, and we interpret commutative
hierarchies in the images of Frobenius operators as the geodesics. This is
different from the classical realization of integrable systems as geodesic
flows on infinite\/-\/dimensional Lie groups~\cite{ArnoldKhesin}.

Our main result is the following. We formulate a well\/-\/defined notion of linear
differential operators whose images are closed w.r.t.\ the commutation,
see~\eqref{IDefFrob}, and we associate infinitely many such operators to
Euler\/--\/Lagrange hyperbolic systems of Liouville type. Here, in particular,
we solve an old problem~\cite{Leznov,LeznovSmirnovShabat,Shabat} in geometry of
Leznov\/--\/Saveliev's 2D~Toda chains related to semi\/-\/simple complex Lie algebras: We describe their infinitesimal symmetries by an explicit formula and
calculate all commutation relations in these Lie algebras.

\tableofcontents

The paper is organized as follows.
First in section~\ref{SecBasic} we outline our basic concept.
Then
in section~\ref{SecHam} we summarize important properties of the
Poisson structures for completely integrable evolutionary systems.
We introduce the notion of nondegenerate operators in total derivatives
and formulate a conjecture that simplifies the search for Poisson structures of~PDE.

In section~\ref{SecDef} we define the Frobenius operators,
establish the chain rule 
for them,
describe the differential Frobenius complex, 
and construct flat 
connections in the triples $\bigl(\Omega^1(\xi_\pi),\gothg(\pi),A\bigr)$
of two Lie algebras and a morphism.
Also, 
we describe an inductive method for the reconstruction of the Sokolov brackets
on the domains of nondegenerate Frobenius operators.
We illustrate it using the dispersionless $3$\/-\/component Boussinesq
system~\eqref{d-B}, see~\cite{Nutku,JK3Bous}, that admits a family of
Frobenius operators~\eqref{A12dBous} and yields the Frobenius
recursion~\eqref{FRec}.

In section~\ref{SecCompat} we propose the definitions of
the linear compatibility and
the strong compatibility of Frobenius operators.
We endow the linear spaces of both linear and strong compatible
Frobenius operators with the Lie\/-\/type algebra structure. 
We show that the Magri schemes 
provide commutative examples of such algebras. 
The structural constants encode the bi\/-\/differential Christoffel symbols,
which are transformed under reparametrizations by a direct analogue of the
rules for the classical connection $1$-\/forms.

In section~\ref{SecLiou} we assign the 
algebras of Frobenius operators~$A$
to Liouville\/-\/type Euler\/--\/Lagrange systems
(in particular, to the 2D~Toda chains associated with the root systems of
semi\/-\/simple Lie algebras~\cite{DSViniti84,Leznov,Shabat}).
To this end, we describe the generators $\vph=A(\cdot)$ of their 
higher symmetry algebras and calculate all commutation rules. 
As a by\/-\/product, we find 
the Hamiltonian structures for KdV\/-\/type systems related to these 
hyperbolic Darboux\/-\/integrable equations.

Finally, we discuss open problems in the theory of Frobenius structures.\\%
\centerline{\rule{1in}{0.7pt}}

\smallskip Let us fix some notation; the language of jet bundles is contained,
e.g., in~\cite{ClassSym,Opava,Manin1979,Olver}.
In the sequel, everything is real and $C^\infty$-\/smooth.
By $J^\infty(\pi)$ we denote the infinite jet space over a fibre
bundle $\pi\colon E^{n+m}\to M^n$,
we set $\pi_\infty\colon J^\infty(\pi)\to M^n$, and denote by $[u]$
the differential dependence on $u$ and its derivatives.
Put $\cF(\pi)=C^\infty(J^\infty(\pi))$, which is understood as the
inductive limit of filtered algebras~\cite{ClassSym} and hence each
function from $\cF(\pi)$ depends on finitely many coordinates
on~$J^\infty(\pi)$.  
The $\pi$-\/vertical evolutionary derivations
$\cEv_\vph=\sum_\sigma D_\sigma(\vph)\cdot\dd/\dd u_\sigma$ 
are described by the sections~$\vph\in\Gamma(\pi_\infty^*(\pi))
=\Gamma(\pi)\mathbin{{\otimes}_{C^\infty(M^n)}}
C^\infty(J^\infty(\pi))$ of the
induced fibre bundle~$\pi_\infty^*(\pi)$.
The shorthand notation for this $\cF(\pi)$-\/module is
$\varkappa(\pi)\equiv\Gamma(\pi_\infty^*(\pi))$.
For all $\psi$ such that $\cEv_\vph(\psi)$ makes sense,
the linearizations $\ell_\psi^{(u)}$ are defined by
$\ell_\psi^{(u)}(\vph)=\cEv_\vph(\psi)$, where $\vph\in\varkappa(\pi)$.
Denote by $\bar{\Lambda}^n(\pi)$ the highest $\pi$-\/horizontal forms on~$J^\infty(\pi)$.
For any 
$\cF(\pi)$-\/module $\gh=\Gamma(\pi_\infty^*(\xi))
=\Gamma(\xi)\mathbin{{\otimes}_{C^\infty(M^n)}}C^\infty(J^\infty(\pi))$
of sections of an induced bundle over~$M^n$, 
we use the notation $\xi_\pi\equiv\pi_\infty^*(\xi)$ and denote by
$\smash{\hat{\gh}}=\Hom_{\cF(\pi)}\bigl(\gf,\bar{\Lambda}^n(\pi)\bigr)$ the dual module. Examples of~$\xi$ will be given later.

Frobenius operators map $\gf\to\varkappa(\pi)$, where the module $\gf$ is one of
the following restrictions of~$\gh\subseteq\Gamma(\xi_\pi)$ onto the image of a
Miura differential substitution $w=w[u]\colon J^\infty(\pi)\to\Gamma(\xi)$:
We have that either
\[
\gf=\varkappa(\xi){\bigr|}_{w\colon J^\infty(\pi)\to\Gamma(\xi)}\quad
\text{or}\quad
\gf=\hat{\varkappa}(\xi){\bigr|}_{w\colon J^\infty(\pi)\to\Gamma(\xi)}.
\]
In particular, $\xi=\pi$ for recursion operators $R\in\End_\BBR\varkappa(\pi)$,
and $\gf=\hat{\varkappa}(\pi)$ 
for Hamiltonian operators; 
here we set $w=\id\colon\Gamma(\pi)\to\Gamma(\xi)$ in both cases.

Denote by
$\gothg(\pi)=\bigl(\varkappa(\pi),[\,,\,]\bigr)$ 
the Lie algebra of evolutionary vector fields~$\cEv_\vph$ 
with the standard bracket of sections.
Let $\bigl(\Omega^1(\xi_\pi),[\,,\,]_A\bigr)$ denote the Lie algebra that
is isomorphic to $\gf/\ker A$ as a vector space and is
endowed with the Lie bracket $[\,,\,]_A$  by a Frobenius operator
$A\colon\gf\to\gothg(\pi)$. Definitions are discussed in detail in
sections~\ref{SecHam} and~\ref{SecDef}.

\section{Basic concept}\label{SecBasic}
The problem of construction and classification of
operators in total derivatives that satisfy~\eqref{IDefFrob}
was suggested first in~\cite{SokolovUMN}. The operators $A$ were
regarded there as non\/-\/skew\/-\/adjoint generalizations of the Hamiltonian
operators, whose images are closed with respect to the standard Lie
algebra structure on the modules of evolutionary fields. 
The operators appeared in~\cite{SokolovUMN} in local coordinates
in the context of the scalar Liouville\/-\/type equations
$\cEL=\bigl\{u_{xy}^i=F(u,u_x,u_y;x,y),1\leq i\leq m\bigr\}$,
whose infinite groups of conservation laws
$\bigl[f(x,[w])\,\Id x\bigr]+\bigl[\bar{f}(y,[\bar{w}])\,\Id y\bigr]$
are differentially generated by finitely many densities
\begin{equation}\label{IInt}
w_1,\ldots,w_r\in\ker D_y\bigr|_{\cEL},\qquad
\bar{w}_1,\ldots,\bar{w}_{\bar{r}}\in\ker D_x\bigr|_{\cEL}.
\end{equation}
Let us recall the main motivating example that begins the theory of Frobenius operators.

\begin{example}[The Liouville equation]\label{ExLiou}
Consider the scalar Liouville equation
\begin{equation}\label{ELiou}
\cELiou=\{\cU_{xy}=\exp(2\cU)\}.
\end{equation}
The differential generators $w$, $\bar{w}$ of its conservation laws
$[\eta]=\bigl[f(x,[w])\,\Id x\bigr]+\bigl[\bar{f}(y,[\bar{w}])\,\Id y\bigr]$
are
\begin{equation}\label{KdVSubst}
w=\cU_x^2-\cU_{xx} \quad\text{and}\quad \bar{w}=\cU_y^2-\cU_{yy}
\end{equation}
such that $D_y(w)\doteq0$ and $D_x(\bar{w})\doteq0$ by virtue ($\doteq$)
of~$\cELiou$ and its differential consequences. The operators
\begin{equation}\label{SquareSok}
\square=\cU_x+\tfrac{1}{2}D_x\quad\text{and}\quad \bar{\square}=\cU_y+\tfrac{1}{2}D_y
\end{equation}
factor higher and Noether's symmetries
\[
\vph=\square\bigl(\phi(x,[w])\bigr),\quad
\vph_\cL=\square\Bigl(\frac{\delta\cH(x,[w])}{\delta w}\Bigr); \qquad
\bar{\vph}=\bar{\square}\bigl(\bar{\phi}(y,[\bar{w}])\bigr),\quad
\bar{\vph}_\cL=\bar{\square}\Bigl(\frac{\delta\bar{\cH}(y,[\bar{w}])}{\delta\bar{w}}\Bigr)
\]
of the Euler\/--\/Lagrange equation~\eqref{ELiou}
for any smooth $\phi,\bar{\phi}$ and $\cH,\bar{\cH}$.
Note that the operator $\square=\tfrac{1}{2}D_x^{-1}\circ
\bigl(\ell_w^{(\cU)}\bigr)^*$ is obtained using the adjoint linearization
of~$w$, and similarly for~$\bar{\square}$.

Each of the images of~\eqref{SquareSok} is closed w.r.t.\ the commutation such that
\[
\bigl[\square(p),\square(q)\bigr]=\square\bigl(
  \cEv_{\square(p)}(q)-\cEv_{\square(q)}(p)+\ib{p}{q}{\square}\bigr),\qquad
\text{here }\ib{p}{q}{\square}=D_x(p)\cdot q-p\cdot D_x(q),
\]
and same for $\bar{\square}$;
the evolutionary derivations $\cEv_{(\cdot)}$ are given
in~\eqref{EvCoord} on p.~\pageref{EvCoord} in local coordinates.
The symmetry algebra $\sym\cELiou\simeq\img\square+\img\bar{\square}$ is the sum of images of~\eqref{SquareSok},
and the two summands commute between each other,
$  
[\img\square,\img\bar{\square}]\doteq0
$ on~$\cELiou$. 
Therefore,
\begin{equation}\label{IStrongSquare}
[\img\square+\img\bar{\square},\img\square+\img\bar{\square}]\subseteq
\img\square+\img\bar{\square}.
\end{equation}

The Frobenius operator~$\square$ factors higher symmetries
of the potential modified KdV equation
\begin{equation}\label{IpmKdV}
\cE_{\text{pmKdV}}=\{\cU_t=-\tfrac{1}{2}\cU_{xxx}+\cU_x^3=\square(w)\},
\end{equation}
whose commutative hierarchy is composed by
Noether's symmetries $\vph_\cL\in\img(\square\circ\delta/\delta w)$ of the Liouville equation~\eqref{ELiou}.
The operator~$\square$ factors the second Hamiltonian structure $B_2=\square\circ A_1\circ\square^*$ for~$\cE_{\text{pmKdV}}$, here $A_1=D_x^{-1}$.

\enlargethispage{\baselineskip}
The generator~$w$ of conservation laws for~$\cELiou$ provides the Miura substitution~\eqref{KdVSubst}
from $\cE_{\text{pmKdV}}$ to the Korteweg\/--\/de Vries equation
\begin{equation}\label{IKdV}
\cE_{\text{KdV}}=\{w_t=-\tfrac{1}{2}w_{xxx}+3ww_x\}.
\end{equation}
The second Hamiltonian structure for~$\cE_{\text{KdV}}$ is factored to
the product $\hat{A}_2=\square^*\circ\hat{B}_1\circ\square$, where
$\hat{B}_1=D_x$ is the first Hamiltonian structure for the modified~KdV
(see diagram~\eqref{Diag} on p.~\pageref{Diag}).
The domain of the Frobenius operator~$\square$ contains sections of the
cotangent bundle $\cosym\cE_{\text{KdV}}$ to the KdV equation. 
%
The bracket $\ib{\,}{\,}{\square}$ on the domain of~$\square$ is equal to the
bracket $\ib{\,}{\,}{\hat{A}_2}$ induced on the domain of the
operator~$\smash{\hat{A}_2}$ (which is Hamiltonian and hence its image is
closed under commutation) for~$\cE_{\text{KdV}}$.
\end{example}


Example~\ref{ExLiou} is reproduced for all $m$-\/component 2D~Toda chains
\begin{equation}\label{IEToda}
\cEToda=\Bigl\{u^i_{xy}=\exp\bigl(\sum_{j=1}^m K^i_{\,j}u^j\bigr),
1\leq i\leq m\Bigr\}
\end{equation}
associated with semi\/-\/simple complex Lie algebras~\cite{Leznov,
LeznovSmirnovShabat}. 
For example, all the 2D~Toda chains~\eqref{IEToda} with the matrix~$K$
symmetrizable by a vector~$\vec{a}$ (that is, 
$\kappa_{ij}\mathrel{{:}{=}} a_i k^i_{\,j}=\kappa_{ji}$, no summation over~$i$)
admit the conserved density 
$w^1=\langle\kappa u_x,u_x\rangle/2-\langle\vec{a},u_{xx}\rangle$.
At the same time, the chains~\eqref{IEToda} may admit other conserved
densities. 
In the fundamental paper~\cite{Shabat}, A.~B.~Shabat
\textit{et al.} proved the existence of maximal ($r=\bar{r}=m$)
sets~\eqref{IInt} of conserved densities for~$\cEToda$ if and only if
the matrix~$K$ 
is the Cartan matrix of a root system
for a semi\/-\/simple Lie algebra of rank~$r$.
Further, Ref.~\cite{Shabat95} contains an explicit procedure that 
yields special systems of jet coordinates for~\eqref{IEToda};
with respect to them, all coefficients of the characteristic equation
$D_y(w)\doteq0$ on~$\cEToda$ become linear, whence its first integrals
$w^1$,\ $\ldots$,\ $w^r$ are obtained.
The differential orders of~$w,\bar{w}$
grow as $r$~grows, and the formulas are big already for
the Lie algebra~$G_2$, see~\cite{LeznovSmirnovShabat,Protaras}.
We claim that the orders of $w^i$ with respect to Shabat's momenta~$\gm_j$
(see below) coincide with the gradations for the principal realizations of the basic (\textit{i.e.}, simplest nontrivial highest weight) representations of
the corresponding affine Lie algebras.

The generators $\vph=\square(\vec{\phi}\bigl(x,[w]\bigr)$ of higher symmetry
algebras for Liouville\/-\/type equations are factored by matrix operators
$\square$ in total derivatives~\cite{Demskoi2004,TodaLawsActa,Meshkov198x}.
For the Euler\/--\/Lagrange Liouville\/-\/type systems, the operators provide
Noether symmetries $\vph_{\cL}$ with $\vec{\phi}=\delta\cH(x,[w])/\delta w$,
see also~\cite{Sakovich, VestnikNoether, TMPhGallipoli},
whence the description of all symmetries follows. For this class of
hyperbolic equations, which incorporates~\eqref{IEToda},
commutative Lie subalgebras of higher Noether symmetry
algebras yield completely integrable KdV\/-\/type
hierarchies~\cite{DSViniti84,JMathSci2004,TMPhGallipoli}.
The generators $w,\bar{w}$ of conservation laws
for~$\cEL$ induce Miura's transformations $w=w[u]$ between the KdV\/-{} and
modified KdV\/-\/type hierarchies upon $w$ and~$u$, respectively.
The operators~$\square$ factor higher Poisson structures
for the evolutionary systems and, moreover,
prescribe the nonlocalities that arise in these structures.
For instance, the first integral $w^1$ for systems~\eqref{IEToda} with a
symmetrizable matrix~$K$ yields the class $\vph=\bigl(u_x+\vec{\Delta}\cdot D_x
\bigr)\bigl(\vec{\phi}(x,[w^1])\bigr)$ of symmetries for~$\cEToda$,
here $\Delta^i=\sum_{j=1}^m k^i_j$; this always yields the second Poisson
structure for KdV in the upper\/-\/left corner $\bigl(\hat{A}_k\bigr)_{11}$ of
the Hamiltonian operators determined by the entire operator~$\square$ 
(see~\cite{TMPhGallipoli} for details).
Several formulas for the operators~$\square$ were known
from~\cite{LeznovSmirnovShabat,Shabat} for 2D~Toda chains~\eqref{IEToda}
associated with semi\/-\/simple Lie algebras of low ranks,
but not in the general case.

Under assumption that the densities~\eqref{IInt} are known, we obtain the
explicit formula~\eqref{Square} for these operators~$\square$.
We derive all the
commutation relations for~$\sym\cEL$; they are encoded by bi\/-\/differential
operators $\bc_{ij}^k$ that act on $\vec{\phi}$'s. We prove that
the $r$-\/tuples $\vec{\phi}$ obey the variational covector transformation laws
$\vec{\phi}\mapsto\tilde{\phi}=
\bigl[\bigl(\ell_{\tilde{w}}^{(w)}\bigr)^*\bigr]^{-1}(\vec{\phi})$ under reparametrizations~$\tilde{w}=\tilde{w}[w]$ of~\eqref{IInt}. Therefore
the operators~$\square$ assigned to Liouville\/-\/type systems~$\cEL$ are
natural examples of well\/-\/defined Frobenius structures.

\smallskip
The second motivation to study linear differential operators subject
to~\eqref{IDefFrob} is much more abstract, c.f.\ \cite{Opava}.
Let us compare geometry of ODE and of PDE from the following viewpoint.
For a linear $\Bbbk$-\/space~$V$, one can
study representations $\gothg\to\End_\Bbbk(V)$ of Lie algebras~$\gothg$ on~$V$
and thus endow the linear spaces of endomorphisms with Lie algebra structures.
On the other hand, let $\cF$ be a commutative associative algebra with unit,
e.g., a $\Bbbk$-\/algebra $\cF_{-\infty}=C^\infty(M^n)$ or an
$\cF_{-\infty}$-\/algebra $\cF(\pi)=C^\infty(J^\infty(\pi))$, and let
$\dd\colon\cF\to\cF$~be a derivation. Consider two left $\cF$-\/modules $P,Q$
and the space of linear differential operators $\Diff\bigl(P\to Q\bigr)$.
A question: Are there any natural algebraic structures on this linear space?
If $P=Q$ (e.g., $P=Q=\sym\cE$ of a differential equation~$\cE$, and we
deal with recursion operators, see~\cite{SokDemskojKN,JK3Bous}),
then one has the associative composition $A\circ B$ and the formal commutation
$A\circ B-B\circ A$ for $A,B\in P$, but what else?
And what if $P\neq Q$? In this paper, we give an affirmative answer on the above
question whenever~$\cF=\cF(\pi)$.

In geometry of PDE~\cite{ClassSym,Opava,Manin1979}, classical constructions such as
the differentials of Hamiltonians, bi\/-\/vectors $\tfrac{\dd}{\dd p}\wedge
\tfrac{\dd}{\dd q}$, etc., appear as zero\/-\/order terms%
\footnote{To facilitate the exposition, in appendix~\ref{BiTable} we summarize the notation for Hamiltonian evolutionary PDE
and its analogy with the structures for 
ordinary differential equations. However, it is a very delicate matter to restore the full parallel between finite\/-\/dimensional manifolds~$M$
and horizontal bundles over $J^\infty(\pi)$ with its own Cartan connection.
Essentially, theorems either are converted to definitions or become false.
This is the case of Lie algebroids, which we discuss in sections~\ref{SecLieAlgd} and~\ref{SecAlg}.} in Taylor expansions
of the \emph{variational} derivatives, variational bi\/-\/vectors, or
variational Poisson\/--\/Nijenhuis structures~\cite{JKGolovko2008,Lstar,Dirac}. Likewise, we discover that the
Lie\/-\/type brackets, which we introduce on linear spaces $\EuA\subset
\CDiff\bigl(\gf\to \varkappa(\pi)\bigr)$ of Frobenius operators, encode
bi\/-\/differential Christoffel symbols of flat connections.
At the same time, we show by counterexample~\ref{StuckAt}
on p.~\pageref{StuckAt} that Frobenius operators are not
the anchors of Lie algebroids over the jet spaces~$J^\infty(\pi)$.

We begin with the classical Hamiltonian formalism for~ODE
on finite\/-\/dimensional manifolds~$M$. Bearing it in mind,
we pass to the PDE setting and formulate the assertions.

Let $\cP\in\Gamma\bigl(\bigwedge^2(TM)\bigr)$ be a bi\/-\/vector
with vanishing Schouten bracket $\lshad\cP,\cP\rshad=0$.
Using the coupling $\langle\,,\,\rangle$ on $TM\times T^*M$ and
a nondegenerate Poisson bi\/-\/vector~$\cP$, one
transfers the Lie algebra structure~$[\,,\,]$ on~$\Gamma(TM )$
to $[\,,\,]_{\cP}$ on $\Gamma(T^*M)\ni\psi_1,\psi_2$ and obtains the
Koszul\/--\/Dorfman\/--\/Daletsky\/--\/Karas\"ev bracket~\cite{Dorfman}
\begin{equation}\label{IKDBr}
[\psi_1,\psi_2]_\cP=\IL_{\cP\psi_1}(\psi_2)-\IL_{\cP\psi_2}(\psi_1)+\Id\bigl(\cP(\psi_1,\psi_2)\bigr),
\end{equation}
here $\IL$~is the Lie derivative. By~\cite{YKSMagri},
the bracket~\eqref{IKDBr} is uniquely defined if, first,

$\bullet$\quad $[\,,\,]_{\cP}$ is a derivation of $C^\infty(M)$ with respect to
the $C^\infty(M)$-\/module structure of $\Gamma(T^*M)$, that is,
if we have that
\begin{equation}\label{ILeibnitz}
[\psi_1,f\cdot\psi_2]_{\cP}=\IL_{\cP\psi_1}(f)\cdot\psi_2+f\cdot
[\psi_1,\psi_2]_{\cP}
\end{equation}
for any $f\in C^\infty(M)$ and $\psi_1,\psi_2\in\Gamma(T^*M)$, and second, if

$\bullet$\quad $[\Id h_1,\Id h_2]=-\Id\,\{h_1,h_2\}_{\cP}$ holds for any
$h_1,h_2\in C^\infty(M)$. 
Then $\Bigl(\bigl(T^*M,[\,,\,]_{\cP}\bigr)\xrightarrow{\cP}
\bigl(TM,[\,,\,]\bigr)\Bigr)$ is a Lie algebroid~\cite{Vaintrob}
with the morphism~$\cP$ over the smooth manifold~$M$.

The de Rham differential $\Id_{\text{dR}}$ on
$\bigwedge^\bullet(T^*M)$ is defined in the complex over the Lie
algebra $\bigl(\Gamma(TM),[\,,\,]\bigr)$ by using Cartan's formula. If
the Poisson bi\/-\/vector~$\cP$
has the inverse symplectic two\/-\/form $\cP^{-1}$ such that
$\cP^ {-1}[x,y]=[\cP^{-1}x,\cP^{-1}y]_{\cP}$, then the
differential~$\Id_{\text{dR}}$ is correlated with the
Koszul\/--\/Schouten\/--\/Gerstenhaber bracket~$\lshad\,,\,\rshad_{\cP}$
on $\bigwedge^\bullet(T^*M)$ by
$\Id_{\text{dR}}=\lshad\cP^{-1},\cdot\rshad_{\cP}$. The differential
$\Id_{\text{dR}}$ on $\bigwedge^\bullet(T^*M)$ is
intertwined~\cite{Koszul23,JK1988} with the Poisson differential
$\partial_{\cP}=\lshad\cP,\cdot\rshad$ on $\bigwedge^\bullet(TM)$ by
\[ 
\bigl(\bigwedge\nolimits^{k+1}\cP\bigr)\bigl(\lshad\cP^{-1},\Psi\rshad_{\cP}\bigr) +
\lshad\cP,\bigl(\bigwedge\nolimits^k\cP\bigr)(\Psi)\rshad=0, \qquad
\forall\,\Psi\in\bigwedge\nolimits^k(T^*M).
\] 

The trivial infinitesimal deformations
\begin{equation}\label{BrNijenhuis}
[x,y]_{\rmN}\mathrel{{:}{=}}[\rmN x,y]+[x,\rmN y]-\rmN\bigl([x,y]\bigr)=\frac{\Id}{\Id\lambda}\Bigr|_{\lambda=0}\rme^{-\lambda\rmN}\,\bigl[\rme^{\lambda\rmN}(x),\rme^{\lambda\rmN}(y)\bigr]
\end{equation}
of the standard Lie algebra structure~$[\,,\,]$ on the tangent bundle
$TM\ni x,y$ over smooth manifolds~$M$ were described in~\cite{YKSMagri}
using the recursion operators~$\rmN\colon\Gamma(TM)\to\Gamma(TM)$. If the
Nijenhuis torsion
\begin{equation}\label{NijenhuisRec}
\lshad\rmN,\rmN\rshad^{\text{fn}}(x,y)=[\rmN x,\rmN y]-\rmN\bigl([x,y]_{\rmN}\bigr)
\end{equation}
for an endomorphism~$\rmN$ vanishes, then the Lie brackets $[\,,\,]_{\rmN^k}$
obtained 
by iterations of the Nijenhuis recursion~$\rmN$ are pairwise compatible.

The Nijenhuis and Poisson structures~$(\rmN,\cP)$, whenever satisfying
two compatibility conditions~\cite{JKGolovko2008,YKSMagri}, generate
infinite hierarchies of pairwise compatible Poisson structures
$\rmN^k\circ\cP$, $k\geq0$, which means that linear
combinations $\lambda_1\rmN^{k_1}\circ\cP+\lambda_2\rmN^{k_2}\circ\cP$
remain Poisson for any~$\lambda_1:\lambda_2$.
This assertion~\cite{YKSMagri}, which is valid for
finite\/-\/dimensional manifolds~$M$, admits a straightforward
generalization to the infinite\/-\/dimensional case when the base
manifold is the jet space~$J^\infty(\pi)$ over a fibre bundle
$\pi\colon E^{m+n}\to M^n$, see~\cite{IgoninVV,Lstar}, such that the
concept of Poisson\/--\/Nijenhuis' structures is applicable verbatim
to evolutionary PDE~\cite{JKGolovko2008}.
Also, this construction yields the Lie 
algebroids
$\Bigl(\bigl(\bar{H}^n(\pi),\{\,,\,\}_{\cP}\bigr)
  \xrightarrow{\cEv_{P\circ\cE}}
\bigl(\varkappa(\pi),[\,,\,]\bigr)\Bigr)$.

Let $P$~be a Hamiltonian operator for a system of differential equations
$\cE=\{w^1=0$,\ $\ldots$,\ $w^r=0\}$. The left\/-\/hand sides $w^i$ belong to
some $\cF(\pi)$-\/module $\gf$ of sections of an $r$-\/dimensional fibre bundle
over~$J^\infty(\pi)$, and the operator $P\colon\hgf\to\varkappa(\pi)$ takes
sections from the dual of~$\gf$ to evolutionary fields. In practice, the equations
$u^i_t=f_i[u]$ in determined evolutionary systems~$\cE$ are labelled by the
dependent variables~$u^i$ in the bundle~$\pi$, which establishes the isomorphism
$\gf\simeq\varkappa(\pi)$. 
However, the system~$\cE$ as a geometric object in~$J^\infty(\pi)$ admits arbitrary
reparametrizations $\tilde{w}=\tilde{w}[w]$ for its components. They are not anyhow
correlated with admissible changes $\tilde{u}=\tilde{u}[u]$ of the dependent
variables. Summarizing, we see that the Hamiltonian operators for~PDE are
well\/-\/defined under unrelated transformations of their domains~$\hgf$ and
images~$\varkappa(\pi)$. Let us formalize this property for a wider class of
operators.

Frobenius linear differential operators in total derivatives,
with images closed under the commutation, are well defined
as follows. Let $\pi$~and $\xi$~be fibre bundles over~$M^n$, let $w$~be
a fibre coordinate in~$\xi$, and construct the infinite jet bundle
$\xi_\infty\colon J^\infty(\xi)\to M^n$. Consider the $\cF(\xi)$-\/module of
sections $\varkappa(\xi)=\Gamma\bigl(\xi_\infty^*(\xi)\bigr)$ of the induced
fibre bundle~$\xi_\infty^*(\xi)$, and denote its $\bar{\Lambda}^n$-\/dual
by~$\hat{\varkappa}(\xi)$. Suppose further that there is a Miura substitution
$J^\infty(\pi)\to\Gamma(\xi)$ which embeds both
$\cF(\xi)$-\/modules into $\Gamma\bigl(\pi_\infty^*(\xi)\bigr)$.
(For instance, the substitution determines a system of differential equations
$\cE=\{w^1[u]=0,\ldots,w^r[u]=0\}$.) We denote
the substitution by the same letter~$w$, because from now on we take the
restrictions of $\varkappa(\xi)$ and $\hat{\varkappa}(\xi)$ onto its image.

$\bullet$\quad Frobenius operators of first kind are
$A\colon\varkappa(\xi)\bigr|_{w}\to\varkappa(\pi)$.
Under any diffeomorphisms $\tilde{u}=\tilde{u}[u]\colon
J^\infty(\pi)\to\Gamma(\pi)$ and $\tilde{w}=\tilde{w}[w]\colon
J^\infty(\xi)\to\Gamma(\xi)$, 
the operators~$A$ of first kind are transformed according to
\begin{equation}\label{IFrobKK}
A\mapsto\tilde{A}=\ell_{\tilde{u}}^{(u)}\circ A\circ\ell_w^{(\tilde{w})}
  \Bigr|_{\substack{w=w[u]\\u=u[\tilde{u}]}}.
\end{equation} 

\enlargethispage{0.7\baselineskip}
$\bullet$\quad Frobenius operators of second kind are linear mappings
$A\colon\hat{\varkappa}(\xi)\bigr|_{w}\to\varkappa(\pi)$.
For any differential changes of coordinates $\tilde{u}=\tilde{u}[u]$ and
$\tilde{w}=\tilde{w}[w]$ in $\pi$ and $\xi$, respectively, the operators obey
\begin{equation}\label{IFrobAK}
A\mapsto\tilde{A}=\ell_{\tilde{u}}^{(u)}\circ A\circ
 \bigl(\ell_{\tilde{w}}^{(w)}\bigr)^*
  \Bigr|_{\substack{w=w[u]\\u=u[\tilde{u}]}}.
\end{equation} 

$\bullet$\quad Finally (the degenerate case), if
no gauge of~$\Gamma(\xi)$ is allowed in a given setting,
then Frobenius operators $A\colon\Gamma\bigl(\pi_\infty^*(\xi)\bigr)\to
\varkappa(\pi)$ are transformed by $A\mapsto\tilde{A}=\ell_{\tilde{u}}^{(u)}
\circ A\bigr|_{u=u[\tilde{u}]}$ under $\tilde{u}=\tilde{u}[u]$.

Frobenius operators of first and second kind generalize linear recursion
operators $R\in\End_\BBR\varkappa(\pi)$ and Hamiltonian operators
$P\colon\hat{\varkappa}(\pi)\to\varkappa(\pi)$, respectively.
The domains of~$A$ 
may not be composed by sections of the tangent bundle
$\pi_\infty^*(\pi)$ to $J^\infty(\pi)$ or, respectively, of its dual,
which was constructed in~\cite{KuperCotangent}
(see also~\cite{IgoninVV,Lstar}). Thence Frobenius operators are described
neither by the Poisson bi\/-\/vectors nor by the recursion $(1,1)$-\/tensors.

For the same reason, the coupling $\langle\,,\,\rangle$ on
$\gothg\times\Gamma(\xi_\pi)$ is missing. However, Frobenius operators
$A\colon\gf\subseteq\Gamma(\xi_\pi)\to\varkappa(\pi)$
transfer the Lie algebra structure~$[\,,\,]$ for evolutionary vector fields
to $[\,,\,]_A$ on the quotients
$\Omega^1(\xi_\pi)=\Gamma(\xi_\pi)\bigr|_{w}/\ker A$ of the respective
$\cF(\pi)$-\/submodules~$\gf$ of~$\Gamma(\xi_\pi)$.
The Koszul bracket $[\,,\,]_A$ is defined by $A\bigl([\psi_1,\psi_2]_A\bigr)=
[A\psi_1,A\psi_2]$ for any $\psi_1,\psi_2\in\Omega^1(\xi_\pi)$.
Thus the operators determine the Lie algebra $\BBR$-\/homomorphisms
\begin{equation}\label{ILieAlgHom}
A\colon\bigl(\Omega^1(\xi_\pi),[\,,\,]_A\bigr)\to\bigl(\gothg(\pi),[\,,\,]\bigr).
\end{equation}
For any $\pi$\/-horizontal $\gothg$-\/module~$K$ there is a
flat connection
\begin{subequations}\label{IConnect}
\begin{gather}
\nabla^A\colon\Der_{\text{Int}}\bigl(\Omega^1(\xi_\pi),\gothg(\pi)\bigr)\to
\Der
\bigl(\gothg(\pi),K\bigr)\\
\intertext{that lifts inner derivations of~$\Omega^1(\xi_\pi)$ to $K$-\/valued
derivations of~$\gothg(\pi)$ by the formula}
\nabla^A_{[\psi,\cdot]_A}=\bigl[A(\psi),\cdot\bigr],\qquad
  \psi\in\Omega^1(\xi_\pi).
\end{gather}
\end{subequations}
The connection is $\Omega^1(\xi_\pi)$-\/linear,
$\nabla^A_{\phi\times[\psi,\cdot]_A}=A(\phi)\times\nabla^A_{[\psi,\cdot]_A}$,
w.r.t.\ the Lie multiplications
$\phi\times{\cdot}=[\phi,\cdot]_A$ by any $\phi\in\Omega^1(\xi_\pi)$, here
$A(\phi)\times{\cdot}=[A(\phi),\cdot]$. The connection~$\nabla^A$ is flat due
to the Jacobi identity. Thence the commutative hierarchies whose flows belong
to the image of~$A$ are the geodesics with respect to~$\nabla^A$.

The `variational' analogue $[\,,\,]_A$ of the Koszul\/--\/Dorfman
bracket~\eqref{IKDBr} equals
\begin{equation}\label{IKoszul}
[\psi_1,\psi_2]_A=\cEv_{A\psi_1}(\psi_2)-\cEv_{A\psi_2}(\psi_1)+\ib{\psi_1}{\psi_2}{A},\qquad \psi_1,\psi_2\in\Omega^1(\xi_\pi),
\end{equation}
where the evolutionary derivations are~\eqref{EvCoord}. Generally, the bi\/-\/differential term $\ib{\,}{\,}{A}$ is neither a Lie algebra structure nor a cocycle.
The chain rule~\eqref{ChainRuleFormula} for the brackets $\ib{\,}{\,}{A}$ and
$\ib{\,}{\,}{A\circ\Delta}$ determined by two Frobenius operators $A$
and~$A\circ\Delta$ follows from the evolutionary summands in~\eqref{IKoszul}.

An explicit formula for the Zhiber\/--\/Sokolov
bracket\footnote{The bracket~$\ib{\,}{\,}{P}$ is not the
Adler\/--\/Kirillov\/--\/Kostant bracket applied in~\cite{DSViniti84} to
construction of the Drinfel'd\/--\/Sokolov systems.}
$\ib{\,}{\,}{P}$ in the Hamiltonian case
$P\colon\hat{\varkappa}(\pi)\to\varkappa(\pi)$ is given by~\eqref{EqDogma}.
We reconstruct $\ib{\,}{\,}{\square}$ for a class of Frobenius
operators\footnote{We shall denote by~$\square$ the Frobenius operators of second
kind that are associated with Liouville\/-\/type systems of~PDE.}
$\square\colon\cosym\cE_{\text{KdV}}\to\sym\cE_{\text{pmKdV}}\subset
\sym\cE_{\text{Toda}}$ that factor higher symmetries of the 2D~Toda
chains~\eqref{IEToda} and
the Hamiltonian structures for the 
associated KdV\/-\/type evolution equations~\cite{DSViniti84}.
Thus Zhiber\/--\/Sokolov's pioneering idea in~\cite{SokolovUMN} to study
operators with property~\eqref{IDefFrob} as generalizations of the
Hamiltonian operators
$P\colon\hat{\varkappa}(\pi)\to\gothg(\pi)$ was motivated by
the examples~$\square\colon\gf\to\varkappa(\pi)$
of a different geometric origin with $\gf\ncong\hat{\varkappa}(\pi)$.
The bracket $\ib{\,}{\,}{R}$ for Frobenius recursions
$R\in\End_\BBR\varkappa(\pi)$ is unknown.
For Frobenius operators~$A$ that are nondegenerate in the sense
of~\eqref{NoKernelsIntersect}, we formulate an inductive procedure for
reconstruction of~$\ib{\,}{\,}{A}$.

The Leibnitz rule~\eqref{ILeibnitz}, which is
the first axiom for the Lie algebroids, is lost for
the Koszul brackets~\eqref{IKoszul} even in the Hamiltonian case
$\gf=\hat{\varkappa}(\pi)$. Therefore the understanding of
Frobenius structures as the Lie algebroids over the infinite jet spaces,
which is one of our goals, is essentially nontrivial and requires
the construction of a formal differential complex over
$\bigl(\Omega^1(\xi_\pi),[\,,\,]_A\bigr)$.
Namely, to each operator~\eqref{ILieAlgHom} we assign
the Frobenius complex\footnote{The presence of the two evolutionary summands in~\eqref{IKoszul} is the reason why we stay in the $\Hom$\/-\/category and do not pass to the more narrow $\CDiff$\/-\/category of the operators in total derivatives.}
\begin{multline}
\gothg(\pi)\xrightarrow{\text{const}}
\Hom_\BBR\bigl(\Omega^1(\xi_\pi),\gothg(\pi)\bigr)\xrightarrow{A\circ[\,,\,]_A}
\Hom_\BBR\bigl(\bigwedge\nolimits^2\Omega^1(\xi_\pi),\gothg(\pi)\bigr)\\
{}\xrightarrow{\text{J}}
\Hom_\BBR\bigl(\bigwedge\nolimits^3\Omega^1(\xi_\pi),\gothg(\pi)\bigr)
  \to\cdots.\label{IFrobComplex}
\end{multline}
The first inclusion in~\eqref{IFrobComplex} consists of the commutation
$\nabla^A_{(\cdot)}\vph_0$ with fixed elements $\vph_0\in\gothg(\pi)$,
the second arrow is the composition of the Koszul
bracket~$[\,,\,]_A$ with~$A$, and the third arrow calculates the
right\/-\/hand side of the Jacobi identity.
The homological vector fields $Q^2=0$ that encode the Frobenius complex~\eqref{IFrobComplex} assign the Lie algebroid structures over the jet spaces~$J^\infty(\pi)$ to Frobenius operators using the approach of~\cite{Vaintrob}.


We say that the Frobenius operators are \emph{linear compatible} if their
arbitrary linear combinations retain the same property~\eqref{IDefFrob}.
Families of $N$~linear compatible operators $A_1$,\ $\ldots$,\ $A_N$
induce 
deformations of~\eqref{ILieAlgHom} such that the Sokolov brackets obey
\[
\ib{\,}{\,}{\sum\limits_{i=1}^N \lambda_iA_i}=\sum_{i=1}^N
   \lambda_i\cdot\ib{\,}{\,}{A_i}.
\]
The linear compatibility 
extends the frames $\bigsqcup_{i}\BBR\cdot A_i$ to the linear
spaces~$\EuA=\bigoplus_{i}\BBR\cdot A_i$.

\enlargethispage{\baselineskip}
The operators $A_1$, $\ldots$, $A_N$ are called \emph{strong compatible} if,
by~\eqref{ICommutClosure}, the
commutators of evolutionary fields $\cEv_{A_k(\cdot)}$
in their images belong to the sum of these images. That is, for any $i,j\in1,\ldots,N$ and $p,q\in\Omega^1\bigl(\xi_\pi\bigr)=\gf/\bigcap_i\ker A_i$ we have
\begin{equation}\label{IntrExpansion}
[A_i(p),A_j(q)]= 
A_j\bigl(\cEv_{A_i(p)}(q)\bigr)-A_i\bigl(\cEv_{A_j(q)}(p)\bigr)
+\sum_{k=1}^N A_k\bigl(\Gamma^k_{ij}(p,q)\bigr)
\in\sum_{\ell=1}^N\img A_\ell.
\end{equation}
In this way, the Koszul brackets~\eqref{IKoszul} in the domains of
each~$A_\ell$ merge to the collective decomposition~\eqref{ICommutClosure}
of the commutators. For example, the commutation closure~\eqref{IStrongSquare}
shows that the operators~\eqref{SquareSok} are strong
compatible 
and that
\begin{align}
\Gamma_{\square\square}^{\square}&=D_x\otimes\bun-\bun\otimes D_x,&
\Gamma_{\overline{\square}\,\overline{\square}}^{\overline{\square}}&=
D_y\otimes\bun-\bun\otimes D_y,\notag\\
\Gamma_{{\square}\overline{\square}}^{{\square}}&=D_y\otimes\bun,\qquad
\Gamma_{{\square}\overline{\square}}^{\overline{\square}}=-\bun\otimes D_x,&
\Gamma_{\overline{\square}{\square}}^{{\square}}&=-\bun\otimes D_y,\qquad
\Gamma_{\overline{\square}{\square}}^{\overline{\square}}=D_x\otimes\bun,
\label{GammaSquareSok}
\end{align}
where the notation is obvious.
Note that $\Gamma_{{\square}\overline{\square}}^{{\square}}(p,q)\doteq
\Gamma_{{\square}\overline{\square}}^{\overline{\square}}(p,q)\doteq
\Gamma_{\overline{\square}{\square}}^{{\square}}(q,p)\doteq
\Gamma_{\overline{\square}{\square}}^{\overline{\square}}(q,p)\doteq0$
on~$\cELiou$ for any $p(x,[w])$ and~$q(y,[\bar{w}])$.
The bi\/-\/differential symbols~$\Gamma_{ij}^k$ 
satisfy~$\Gamma_{ij}^k(p,q)=-\Gamma_{ji}^k(q,p)$ and other normalizations.
They are transformed by a suitable variant of the classical formula
$\tilde{\Gamma}=g\Gamma g^{-1}+\Id g\,g^{-1}$ for connection $1$-\/forms $\Gamma$
and reparametrizations~$g$.

If the closure condition~\eqref{ICommutClosure} is fulfilled for $N$~Frobenius
operators $A_1$,\ $\ldots$,\ $A_N$, then we take $\ob{A_i}{A_j}(p,q)
\mathrel{{:}{=}}\bigl[A_i(p),A_j(q)\bigr]$ as the definition of a Lie\/-\/type
bracket~\eqref{ICommuteOperators} between them. 
Each operator~$A_\ell$ spans the one\/-\/dimensional 
algebra that is described by 
$\Gamma_{\ell\ell}^{\ell}=\ib{\,}{\,}{A_\ell}$, see~\eqref{IKoszul}.
We extend the bracket of operators onto the linear span
$\bigoplus_{\ell=1}^N \BBR\cdot A_\ell$ if, in addition,
the Frobenius operators are linear compatible.
The constants~$\Gamma_{ij}^k$ are bi\/-\/differential
Christoffel's symbols for the flat affine connection $\{\nabla^{A_\ell},1\leq\ell\leq N\}$ defined in~\eqref{IConnect}.
The Magri schemes for completely integrable systems~\cite{Dorfman,YKSMagri}
yield examples of commutative algebras of Frobenius operators.

Relaxing the normalization $\Gamma^k_{\ell\ell}=\ib{\,}{\,}{A_\ell}\cdot\delta^k_{\,\ell}$ but retaining~\eqref{ICommutClosure},
we arrive at a wider class of linear differential
operators~$A_i$ and Lie\/-\/type brackets~\eqref{ICommuteOperators} on the
spaces~$\EuA=\bigoplus_{i=1}^N\BBR\cdot A_i$. Namely, the image of a particular
operator may hit the images of other operators under the commutation.  
Now we formulate the assertion. 

\enlargethispage{\baselineskip}
\begin{theorNo}
Let $\kappa$ be a real constant nondegenerate symmetric $(m\times m)$-\/matrix.
Consider a hyperbolic Euler\/--\/Lagrange system $\cEL=\{\delta\cL/\delta
u=0\}$ which, in a suitable system of coordinates, is determined by a
Lagrangian $\cL=[L\,\Id x\Id y]$ with the density $L=-\tfrac{1}{2}\langle\kappa
u_x,u_y\rangle-\mathrm{H}_{\IL}(u;x,y)$. Set $\gm=\dd L/\dd u_y$.
Suppose further that the system~$\cEL$ is Liouville\/-\/type and admits
$r$~conserved densities $w[\gm]=(w^1,\ldots,w^r)\in\ker D_y\bigr|_{\cEL}$.
Let them be minimal such that $f\in\ker D_y\bigr|_{\cEL}$ implies $f=f(x,[w])$.
Introduce the operator
\begin{equation}\label{ISquare}
\square=\bigl(\ell_w^{(\gm)}\bigr)^*.
\end{equation}
Then we claim the following:
{\renewcommand{\theenumi}{\roman{enumi}}\begin{enumerate}
\item
All (up to $x\leftrightarrow y$) Noether symmetries $\vph_\cL$ of the
Lagrangial~$\cL$ for~$\cEL$ are
\[
\vph_\cL=\square\Bigl(\frac{\delta\cH(x,[w])}{\delta w}\Bigr)\qquad
\text{for any $\cH$.}
\]
\item
All (up to $x\leftrightarrow y$) symmetries $\vph$ of the system~$\cEL$ are
\[
\vph=\square\bigl(\phi(x,[w])\bigr)\qquad\text{for any $\phi=(\phi^1,\ldots,
  \phi^r)\in\gf=\hat{\varkappa}(\xi)\bigr|_{w=w[\gm]}$.}
\]
\item
In the chosen system of coordinates, the image of the operator~$\square$ is
closed with respect to the commutation in the Lie algebra~$\sym\cEL$.
\item
Under a diffeomorphism $\tilde{w}=\tilde{w}[w]$, the $r$-\/tuples~$\phi$ are
transformed by
\[
\phi\mapsto\tilde{\phi}=
   \bigl[\bigl(\ell_{\tilde{w}}^{(w)}\bigr)^*\bigr]^{-1}(\phi).
\]
Therefore, under any reparametrization $\tilde{u}=\tilde{u}[u]$ of
the dependent variables $\vec{u}={}^t(u^1,\ldots,u^m)$ in equation~$\cEL$,
and under a simultaneous change $\tilde{w}=\tilde{w}[w]$, the
operator~$\square$ obeys~\eqref{IFrobAK}. Consequently, the operator
$\square$~satisfies~\eqref{IDefFrob} in any system of coordinates, and hence
$\square\colon\hat{\varkappa}(\xi)\bigr|_{w=w[\gm]}\to\sym\cEL$ is a Frobenius operator of second kind. 
\item
The operator
\[
\cP=\square^*\circ\bigl(\ell_\gm^{(u)}\bigr)^*\circ\square\colon
\hat{\varkappa}(\xi)\bigl|_{w=w[\gm[u]]}\to
     \varkappa(\xi)\bigl|_{w=w[\gm[u]]}
\]
is Hamiltonian. 
\item
The bracket $\ib{\,}{\,}{\square}$ on the domain~$\gf$ of
the operator~$\square$ satisfies the equality
\[
\ib{\,}{\,}{\square}=\ib{\,}{\,}{\cP}.
\]
Its right\/-\/hand side is calculated explicitly by using
formula~\eqref{EqDogma} that is valid for Hamiltonian operators~$\cP$.
This yields the commutation relations in the Lie algebra~$\sym\cEL$.
\item
All coefficients of the operator~$\cP$ and of the
bracket~$\ib{\,}{\,}{\square}$ are differential functions
of the minimal conserved densities~$w$ for~$\cEL$.
\end{enumerate}}
\end{theorNo}

\section{Hamiltonian operators for PDE}\label{SecHam}
\noindent%
In this section we recall necessary definitions and introduce standard
notation, which follows~\cite{ClassSym,Dorfman,Dubr,Lstar,Opava,Olver}.

\subsection{Symmetries and conservation laws}
Let $\pi\colon E^{n+m}\to M^n$ be a vector bundle over an
$n$\/-\/dimensional manifold~$M$ and
$J^\infty(\pi)$ be the infinite jets over this bundle.
By definition, set $\pi_\infty\colon J^\infty(\pi)\to M^n$.
Put~$f[u]=f(u,u_x,\ldots,u_\sigma)$, here $u$ denotes the
components of a vector ${}^t(u^1$, $\ldots$, $u^m)$ and
$|\sigma|<\infty$.

On $J^\infty(\pi)$, there is the Cartan distribution~$\cC$
of $n$-\/dimensional planes that project without degeneration onto $TM^n$
under~$\pi_{\infty,*}$. The distribution is spanned by the total derivatives~$D_{x^i}$, $1\leq i\leq n$, whose action on $\cF(\pi)=C^\infty(J^\infty(\pi))$
is defined by restriction onto the jets $j_\infty(s)$ of sections $s\in\Gamma(\pi)$,
\begin{equation}\label{DefDxByJs}
j_\infty(s)\bigl(D_{x^i}(f)\bigr)=\frac{\dd}{\dd x^i}\Bigl(j_\infty(s)(f)\Bigr),
\qquad \forall s\in\Gamma(\pi),\quad \forall f\in\cF(\pi).
\end{equation}
In coordinates, we have
\begin{equation}\label{CartanConnection}
D_{x^i}=\frac{\dd}{\dd x^i}+\sum\limits_{j,\sigma}u^j_{\sigma+1_i}\cdot
   \frac{\dd}{\dd u^j_\sigma}.
\end{equation}
The Cartan connection $\widehat{\smash{\phantom{D_{x}}}}\colon
\frac{\dd}{\dd x^i}\mapsto D_{x^i}$ is flat.

Denote by $\varkappa(\pi)$ the $\cF(\pi)$-\/module
$\Gamma(\pi_\infty^*(\pi))=\Gamma(\pi)\mathbin{{\otimes}_{C^\infty(M^n)}}
C^\infty(J^\infty(\pi))$. The sections
$\vph=(\vph^1,\ldots,\vph^m)\in\Gamma(\pi_\infty^*(\pi))$
of the induced fibre bundle ``look like''
sections $s(x)={}^t(s^1(x),\ldots,s^m(x))\in\Gamma(\pi)$, but their
components $\vph^i(x,[u])\in\cF(\pi)$ are functions on the jet space~$J^\infty(\pi)$, see also~\eqref{LookFromOutside}. In coordinates, the evolutionary
derivations with sections $\vph\in\varkappa(\pi)$ are\footnote{The
notation~$\cEv_\vph$ makes no confusion with
$\cE=\bigl\{u_t=\vph(x,[u])\bigr\}$, because almost always we identify
such evolutionary systems with $\pi$-\/vertical
derivations~$\cEv_\vph\in\ID^v(J^\infty(\pi))$. A synonimic notation
$\re_\vph$ or $\dd_\vph$~is used for~\eqref{EvCoord} in the literature.}
\begin{equation}\label{EvCoord}
\cEv_\vph=\sum_{i,\sigma} D_\sigma(\vph^i)\cdot\frac{\dd}{\dd
u^i_\sigma}.
\end{equation}
Their invariant definition is $\bigl[\cEv_\vph,D_{x^i}\bigr]=0$ for
$\cEv_\vph\in\ID^v(J^\infty(\pi))$.
We shall use the permutability of evolutionary fields~\eqref{EvCoord}
with total derivatives~\eqref{CartanConnection} many times.

The commutators of evolutionary fields~\eqref{EvCoord} endow $\varkappa(\pi)$ with a Lie algebra structure,
\begin{equation}\label{EvBracket}
\bigl[\cEv_{\vph_1},\cEv_{\vph_2}\bigr]=\cEv_{[\vph_1,\vph_2]},\quad
\text{where }[\vph_1,\vph_2]=\cEv_{\vph_1}(\vph_2)-\cEv_{\vph_2}(\vph_1)
\end{equation}
which is defined by the componentwise action.
In what follows, we identify evolutionary vector fields~$\cEv_\vph$ with their generating sections~$\vph$. We denote by $\gothg(\pi)$ the Lie algebra $\bigl(\varkappa(\pi), [\,,\,]\bigr)$ with the bracket~\eqref{EvBracket}, and we
call it the \emph{standard} Lie algebra structure on~$\varkappa(\pi)$.
By definition, a recursion~$R$ for~$\gothg(\pi)$ is a linear differential
operator $\varkappa(\pi)\xrightarrow{R}\varkappa(\pi)$ in total derivatives.

The linearizations (the Fre\-ch\'et derivatives),
\begin{equation}\label{Linearization}
\ell_\psi=\sum_\sigma\frac{\dd\psi^i}{\dd u^j_\sigma}\cdot D_\sigma,
\qquad \psi\in\Gamma(\pi_\infty^*(\xi))
\text{ for some bundle~$\xi$ over~$M^n$,}
\end{equation}
are correlated with~\eqref{EvCoord} by $\cEv_\vph(\psi)=\ell_\psi(\vph)$.

Under a change $\tilde{u}=\tilde{u}[u]$ of fibre coordinates in~$\pi$, the
generating sections~$\vph$ of~\eqref{EvCoord} are transformed by
$\vph\mapsto\tilde{\vph}=\ell_{\tilde{u}}^{(u)}(\vph)$. Hence the recursion
operators obey
\begin{equation}\label{ChangeCoordRec}
R\mapsto\tilde{R}=\ell_{\tilde{u}}^{(u)}\circ R\circ\ell_u^{(\tilde{u})}
 \Bigr|_{u=u[\tilde{u}]}.
\end{equation}

Let $\Id_h=\widehat{\Id}_{\text{dR}(M^n)}$ be the Cartan lifting of
the de Rham differential on~$M^n$,
\begin{equation}\label{dh}
\Id_h=\sum_{i=1}^n\Id x^i\otimes D_{x^i}.
\end{equation}
Let $\bar{\Lambda}^i(\pi)$ be the $\cF(\pi)$-\/module of $i$-th horizontal forms on~$J^\infty(\pi)$.
Take the cohomology $\bar{\Lambda}^i(\pi)/\{\mathop{\mathrm{im}}\Id_h\}$
w.r.t. the horizontal differential~$\Id_h$ and denote it by~$\bar{H}^i(\pi)$.
So, elements $\bigl[H(x,[u])\,\Id\volume(M^n)\bigr]\in\bar{H}^n(\pi)$
are equivalent
if they differ by exact terms~$\Id_h\eta$,   
where $\eta\in\bar{H}^{n-1}(\pi)$.
The equivalence classes $\cH\in\bar{H}^{n}(\pi)$ of the highest horizontal cohomology will be called the Hamiltonians or the Lagrangians, whenever it is appropriate.

The Cartan differential on $J^\infty(\pi)$ is $\Id_\cC=\Id_{\text{dR}(J^\infty(\pi))}-\Id_h$. The pair of differentials $(\Id_h,\Id_\cC)$ generates the variational bi\/-\/complex and its $\cC$-\/spectral sequence.
Passing to the horizontal cohomology at each term of the variational
bi\/-\/complex, construct the first term $E_1(\pi)$ of the $\cC$-\/spectral
sequence.
Choose a restriction of $\Id_\cC$ onto $\bar{H}^n(\pi)$, which is a corner of the bi\/-\/complex, to be the Euler variational derivative.
(That is, apply the Cartan differential~$\Id_\cC$
and then perform multiple integration by parts, which does not alter the equivalence class modulo~$\Id_h$.)
Denote by $\bE$ the Euler variational derivative $(\delta/\delta u^1,\ldots,
\delta/\delta u^m)$ with respect to~$u={}^t(u^1,\ldots,u^m)$.
Recall that the operator of linearization for the image of the variational
derivative is self\/-\/adjoint,
\begin{equation}\label{Helmholz}
\ell_{\mathstrut\bE(\cH)}=\ell^*_{\bE(\cH)},\qquad \cH\in\bar{H}^n(\pi).
\end{equation}
For any $\cH=[H\,\Id\,\text{vol}(M^n)]\in\bar{H}^n(\pi)$ we have
$\bE(\cH=\ell_H^*(1)$. Consequently, under any change of variables
$\tilde{u}=\tilde{u}[u]\colon J^\infty(\pi)\to\Gamma(\pi)$,
a section $\psi=\bE(\cH)$ in the image of the variational derivative is
transformed by
$\psi\mapsto\tilde{\psi}=\bigl[\bigl(\ell_{\tilde{u}}^{(u)}\bigr)^*\bigr]^{-1}
(\psi)\bigr|_{u=u[\tilde{u}]}$.

For any fibre bundle $\xi$ over $M^n$, the $\cF(\pi)$-\/modules
$\Gamma(\pi_\infty^*(\xi))$ will be called horizontal.
For any horizontal module~$\gf$, denote by $\hgf$ the dual $\cF(\pi)$\/-module
$\hgf=\Hom_{\cF(\pi)}\bigl(\gf,\bar{\Lambda}^n(\pi)\bigr)$.

We study differential equations in jet bundles, considering them together with all their differential\/-\/algebraic consequences, which we assume to exist.
Let $\cE\subset J^\infty(\pi)$ be an equation, and
let the differential ideal $I(\cE)$ of the equation be generated in
coordinates by $\cE=\{F^1=0,\ldots,F^r=0\}$, here $F^i\in\cF(\pi)$.
In other words, the $r$-\/tuple ${}^t(F^1,\ldots,F^r)$ is a section of
a horizontal bundle $\gf=\Gamma(\pi_\infty^*(\xi))$ for $\xi$
with $r$-\/dimensional fibres over~$M^n$.
Note that $\ell_F\colon\varkappa(\pi)\to\gf$.

\begin{rem}\label{Unfortunate}
There are two particular examples of~$\xi$ that, unfortunately, often
lead to confusion and tempt us to ignore the general setting:
\begin{itemize}
\item
Determined autonomous evolutionary systems
$\cE=\bigl\{u^i_t=\vph^i\bigl(x^1,\ldots,x^n,[u]\bigr)$,
$1\leq i\leq m\bigr\}$.
Then $\gf\simeq\varkappa(\pi)$.
\item
Euler\/--\/Lagrange systems
$\cEEL=\bigl\{\bE(\cL)=0$, $\cL\in\bar{H}^n(\pi)\bigr\}$.
Then $\gf\simeq\hat{\varkappa}(\pi)=\Hom_{\cF(\pi)}\bigl(\varkappa(\pi),
\bar{\Lambda}^n(\pi)\bigr)$.
\end{itemize}
In both cases, the number of equations~$r$ equals the number of
unknowns~$m$, which can be used misleadingly to index the equations. We
emphasize that, in principle, the reparametrizations of $F\in\gf$ that
determine the system $\cE=\{F=0\}$ and of the dependent variables $u$
in $J^\infty(\pi)$ are not correlated at all.
\end{rem}

In this paper, we consider
\begin{itemize}
\item
Hamiltonian operators $P\colon\hat{\varkappa}(\cE)\to\varkappa(\cE)$ for determined evolutionary systems~$\cE$ (see section~\ref{SecPoisson}),
\item
Noether non\/-\/skew\/-\/adjoint operators
$A\colon\hat{\varkappa}(\cE)\to\varkappa(\cE)$ for the same class of equations
(see section~\ref{SecTwoSpace}) and inverse Noether operators
$\omega\colon\varkappa(\cE)\to\hat{\varkappa}(\cE)$ (see
Example~\ref{ExFrobRec}),
\item
Frobenius operators $\square\colon\hat{\varkappa}(\cE_1)\to\sym\cE_2$ that
factor symmetry flows $\cE_2$ on Euler\/--\/Lagrange equations~$\cEEL$
(see section~\ref{SecLiou}),
\item
Frobenius recursion operators $R\in\End_\BBR\varkappa(\pi)$
(see Example~\ref{ExFrobRec}), and
\item
Frobenius operators $A\colon\gf\subseteq\Gamma(\pi_\infty^*(\xi))\to\varkappa(\pi)
$ of general nature (see section~\ref{SecAlg}).
\end{itemize}


Let $\cE=\{F=0\}$ be a system of differential equations given by a section~$F\in\gf$.
Let a class $[\eta]\in{\bar{H}^{n-1}\bigr|}_\cE$ be a conservation law.
In other words, the divergence
\begin{equation}\label{Divergence}
\Id_h\eta=\tilde{\nabla}(F)
\end{equation}
vanishes
on the ideal $I(\cE)$ of algebraic\/-\/differential consequences of~$\cE$,
and this is realized by $r$ operators $\tilde{\nabla}=(\nabla^1,\ldots,\nabla^r)$
in total derivatives. The coefficients of~$\tilde{\nabla}$ belong to
$C^\infty(\cE)$ and the operators take values in the module $\bar{\Lambda}^n(\pi)$ of highest horizontal forms.

Represent the $n$-th horizontal form $\Id_h\eta$ as $\tilde{\nabla}(F)=\bigl\langle 1,\nabla(F)\bigr\rangle$,
where $\nabla$ coincides with $\tilde{\nabla}$ in any local coordinates but takes values in $\gf$
instead of $\bar{\Lambda}^n(\pi)$. Here $1$ stands for the $r$-\/tuple $(1,\ldots,1)$
and $\langle\,,\,\rangle\colon\hgf\times\gf\to\bar{\Lambda}^n(\pi)$ is the coupling.
Integrating by parts and introducing the adjoint operator $\nabla^*$, we obtain
$\langle 1,\nabla(F)\rangle=\langle\nabla^*(1),F\rangle\mod\text{im}\,\Id_h$.
The section $\nabla^*(1)\in\hgf$ is called the generating section of a conservation law $[\eta]$
for the equation~$\cE=\{F=0\}$.
By construction, $\psi=\nabla^*(1)\in\hgf$ has as many components
as there are equations in the system.
Under a reparametrization $F=\Delta(\tilde{F})$ of equations that constitute the
system~$\cE$, where $\Delta$~is a linear\footnote{If the nondegenerate
differential substitution~$\Delta$ is nonlinear, e.g.,
$F=\tilde{F}+\tilde{F}^2$, then $\psi$ becomes a nonlinear
operator on the module~$\gf$. This is a difficulty of the theory, see
footnote~${}^{\text{\ref{PotentialGeneral}})}$ on p.~\pageref{PotentialGeneral}.
Hence only linear coordinate transformations are declared admissible in the module
\emph{of equations}; in the sequel, nonlinear changes of fibre coordinares will be
allowed for horizontal bundles of other geometric nature.}
differential operator in total derivatives, the section $\psi\in\hgf$ is
transformed by $\psi\mapsto\tilde{\psi}=\Delta^*(\psi)$.
The generating sections $\psi$ of conservation laws are solutions of
the equation $\ell_F^*(\psi)\doteq0$ on~$\cE$; this follows from the formulas
$0\equiv\bE(\Id_h\eta)=\ell_{\langle\psi,F\rangle}^*(1)=\ell_F^*(\psi)
 +\ell_\psi^*(F)\doteq0$, where the second summand is set to zero by restriction onto the consequences of $\cE=\{F=0\}$.

We conclude that $\psi\in\hat{\varkappa}(\cE)$ for determined evolutionary
systems (this is why the notation $\psi\in\cosym\cE$ is used),
and $\psi\in\varkappa(\cEEL)$ for Euler\/--\/Lagrange systems~$\cEEL$.


\begin{lemma}[\cite{Opava}]\label{LGradient}
Let $\cE=\{F\equiv u_t-f(x,[u])=0\}$ be an evolutionary system.
For any conservation law $[\eta]=[\rho\,\Id x+\cdots]$ such that
$\Id_h(\eta)=\tilde{\nabla}(F)=\langle\nabla^*(1),F\rangle$,
the generating section $\psi=\nabla^*(1)$ of $[\eta]$
is the `gradient' $\psi=\bE(\rho)$ of the conserved density~$\rho$.
\end{lemma}

For the Euler\/--\/Lagrange equations $\cEEL=\{\bE(\cL)=0\}$,
the generating sections $\psi\in\hgf(\cEEL)$ of conservation laws
have a geometric nature of symmetries. This is indeed so.

\begin{theor}[Noether]\label{ThNoether} 
Let $\cEEL=\{\bE(\cL)=0\}$ be the Euler\/--\/Lagrange equation for a
Lagrangian $\cL\in\bar{H}^n(\pi)$. Then the
evolutionary derivation $\cEv_\vph$ is a Noether symmetry of the
Lagrangian, 
$\cEv_\vph(\cL)=0$, if and only if
$\vph$ is the generating section of a conservation law
$[\eta]$ such that $\Id_h\eta=0$ on~$\cEEL$.
\end{theor}


\begin{lemma}[\cite{Opava}, see a proof
   in~\cite{JMathSci2004}]\label{LCommuteEuEv}
The relation
\[
\bE(\cEv_\vph(\rho))=\cEv_\vph(\bE(\rho))+\ell_\vph^*(\bE(\rho))
\]
holds for any $\vph\in\varkappa(\pi)$ and $\rho\in\bar{\Lambda}^n(\pi)$.
\end{lemma}


Consequently, any Noether symmetry $\vph_\cL\in\sym_\cL\cEEL$ of a Lagrangian
is a symmetry of the Euler\/--\/Lagrange equation $\cEEL=\{F=0\}$
(and Lemma~\ref{LCommuteEuEv} shows also why the converse is not true).
For this reason, mappings of generating sections of conservation laws to
symmetries of differential equations will be called \emph{Noether} in what
follows. We have demonstrated that the identity transformation is Noether
for Euler\/--\/Lagrange equations~$\cEEL=\{\bE(\cL)=0\}$.

In the sequel, we apply Lemma~\ref{LCommuteEuEv} for other purposes as well;
namely, it will allow us to induce a bracket~\eqref{TwoGrads} on~$\hgf$
starting with a bracket~\eqref{TwoDens} of conservation laws for evolutionary systems~$\cE$.

\subsection{Poisson structures}\label{SecPoisson}
In the preceding section we have recalled that the image of Euler derivative
is dual to the module $\varkappa(\pi)$ of evolutionary fields.
The standard concept of Hamiltonian dynamics for PDE
(see~\cite{Dorfman,Lstar,Opava}) is based on a class of Hamiltonian operators
that map $\hat{\varkappa}(\pi)\to\varkappa(\pi)$
and which induce a Poisson structure on~$\bar{H}^n(\pi)$
due to the existence of a coupling $\langle\,,\,\rangle\colon
\hat{\varkappa}(\pi)\times\varkappa(\pi)\to\bar{\Lambda}^n(\pi)$.

In this section we consider Hamiltonian operators.
First we regard them as mappings of $\cF(\pi)$-\/modules,
and then as Lie algebra homomorphisms. Next, we restrict these homomorphisms
onto evolutionary systems and, finally, to hierarchies of systems.

Now we recall necessary algebraic constructions.

For any vector space $V$ over~$\BBR$,
let $\Delta\in\Hom_\BBR(\bigwedge^k V,V)$ and
$\nabla\in\Hom_\BBR(\bigwedge^l V,V)$. Denote by
$\Delta[\nabla]\in\Hom_\BBR(\bigwedge^{k+l-1}V,V)$ the
\emph{action} $\Delta[\cdot]\colon\Hom_\BBR(\bigwedge^N V,V)\to
\Hom_\BBR(\bigwedge^{N+k-1}V$, $V)$ of $\Delta$ on $\nabla$, which is
given by the formula
\[
\Delta[\nabla](a_1,\ldots,a_{k+l-1})=
\sum_{\sigma\in S_{k+l-1}^l} (-1)^\sigma
\Delta(\nabla(a_{\sigma(1)},\ldots,a_{\sigma(l)}),a_{\sigma(l+1)},
\ldots,a_{\sigma(k+l-1)}),
\]
where $a_i\in V$ and $S_m^k\subset S_m$
denotes the \emph{unshuffles}. The unshuffles are 
permutations such that $\sigma(1)<\sigma(2)<\ldots<\sigma(k)$ and
$\sigma(k+1)<\ldots<\sigma(m)$ for all $\sigma\in S_m^k$; note that
$\sigma(i)$~is the index of the object placed onto the $i$-th
position under the permutation, unlike in~\cite{Kassel}.
The \emph{Schouten} (\emph{Richardson\/--\/Nijenhuis})
\emph{bracket}\footnote{
The definition of Schouten bracket is valid over any
field $\Bbbk$ such that $\text{char}\,\Bbbk\neq2$.}
$\lshad\Delta,\nabla\rshad\in\Hom_\BBR(\bigwedge^{k+l-1}V,V)$ of
$\Delta$ and $\nabla$ is~\cite{Dorfman,ForKac}
\begin{equation}\label{SchoutenBr}
\lshad\Delta,\nabla\rshad=\Delta[\nabla]-(-1)^{(k-1)(l-1)}\nabla[\Delta].
\end{equation}
Thus the commutator $\lshad X,Y\rshad=[X,Y]$ of two evolutionary vector
fields $X,Y\in\gothg(\pi)$ is skew\/-\/symmetric. The bracket
$\lshad A_1,A_2\rshad\in\bigwedge^3 V$
of two bi\/-\/vectors is symmetric w.r.t.\ $A_1$ and~$A_2$,
but it is skew\/-\/symmetric w.r.t.\ its
arguments $\psi_1$,\ $\psi_2$,\ $\psi_3\in V^*$.

\begin{define}
A linear skew\/-\/adjoint $(m\times m)$-\/matrix operator
$A\colon\hat{\varkappa}(\pi)\to\varkappa(\pi)$ in total derivatives
is called \emph{Hamiltonian} if, for any $\cH_1,\cH_2\in\bar{H}^n(\pi)$,
the bi\/-\/linear skew\/-\/symmetric bracket\footnote{We respect
Dirac's notation. Our choice of the signs in~\eqref{PoissonEquiv}
is such that we multiply by bra-{} covectors $\langle\psi{\mid}$ from the left
and have the ket-\/vectors ${\mid}\vph\rangle$ standing on the right.
However, sometimes we use the reversed coupling
$\langle\,,\,\rangle\colon\gf\times\hgf\to
\bar{\Lambda}^n(\pi)$ for convenience.}
\[
\{\,,\,\}_A\colon\bar{H}^n(\pi)\times\bar{H}^n(\pi)\to\bar{H}^n(\pi),\qquad
\{\mathcal{H}_1,\mathcal{H}_2\}_A \mathrel{{:}{=}}
  \bigl\langle\bE({\mathcal{H}}_1),A(\bE({\mathcal{H}}_2))\bigr\rangle
\]
satisfies the Jacobi identity
\begin{equation}\label{Jacobi3sums}
\sum_{\circlearrowright}\{\{\cH_1,\cH_2\}_A,\cH_3\}_A=0.
\end{equation}
By construction of the Poisson bracket $\{\,,\,\}_A$,
its equivalent definitions are
\begin{equation}\label{PoissonEquiv}
\{\cH_1,\cH_2\}_A=   
\langle \psi_1,A(\psi_2)\rangle
\eqsim \cEv_{A(\psi_2)}(\cH_1) \eqsim -\cEv_{A(\psi_1)}(\cH_2)
\end{equation}
modulo ($\eqsim$) exact terms, here $\psi_i=\bE(\cH_i)$.
The bracket $\{\,,\,\}_A$ endows $\bar{H}^n(\pi)$ with
a Lie algebra structure over~$\BBR$.
Two Hamiltonian operators $A_1$ and $A_2$ are called \emph{compatible} if
their linear combinations $\lambda_1A_1+\lambda_2A_2$ are Hamiltonian as well.
\end{define}

We postpone examples of Hamiltonian operators to section~\ref{SecTwoSpace},
where Noether operators will be considered. In example~\ref{ProjSolut} on
p.~\pageref{ProjSolut} we investigate a wider class of Hamiltonian operators
defined on $\ga$-\/modules over commutative $\Bbbk$-\/algebras~$\ga$.

\enlargethispage{\baselineskip}
\begin{lemma}
Suppose that evolution equations $\cE=\{F=u_t-f[u]=0$, $F\in\gf\}$
are enumerated by
the dependent variables~$u$, which means that the isomorphism $\gf\simeq\varkappa
(\pi)$ is being used. Then, under a fibre coordinate change $\tilde{u}=
\tilde{u}[u]$ in~$\pi$ that preserves the evolutionary form of the equations~$\cE$, a Hamiltonian operator~$A$ is transformed by
\begin{equation}\label{ChangeCoordHamOp}
A\mapsto\tilde{A}=\ell_{\tilde{u}}^{(u)}\circ A\circ
  \bigl(\ell_{\tilde{u}}^{(u)}\bigr)^*\Bigr|_{u=u[\tilde{u}]}.
\end{equation}
Formula~\eqref{ChangeCoordHamOp} is also valid for any Noether
operator~$A\colon\hat{\varkappa}(\pi)\to\varkappa(\pi)$.
\end{lemma}

Hamiltonian operators $A$ can be regarded~\cite{Dorfman,Lstar}
as the variational Poisson bi\/-\/vectors with vanishing Schouten
brackets $\lshad A,A\rshad=0$ such that the Poisson bracket
$\{\cH_1,\cH_2\}_A=\lshad\lshad A,\cH_2\rshad,\cH_1\rshad$ is a derived
bracket in the sense of~\cite{YKS}.
The compatibility condition for the Poisson bi\/-\/vectors
is $\lshad A_1,A_2\rshad=0$.

\enlargethispage{\baselineskip}
By definition, put
\begin{equation}\label{ActOnCoeff}
\ell_{A,\psi}(\vph)\mathrel{{:}{=}}\bigl(\cEv_{\vph}(A)\bigr)(\psi)
\end{equation}
for any $\vph\in\varkappa(\pi)$,
$\psi\in\gf=\Gamma(\pi_\infty^*(\xi))$ for some~$\xi$, and
an operator $A\in\CDiff(\gf,\varkappa(\pi))$ in total derivatives.
Note that $\ell_{A,\psi}$ is an operator in total derivatives w.r.t.\ its
argument~$\vph$ and w.r.t.\ $\psi$ (but not w.r.t.\ the coefficients of~$A$),
and hence the adjoint $\ell_{A,\psi}^*$ is well defined.
We emphasize that the notation $\ell_{A,\psi}$ is not
the same as the linearization~$\ell_{A(\psi)}$,
which was introduced in~\eqref{Linearization}.

\begin{lemma}[\cite{Opava}]\label{AuxLemma}
Let $A$ 
be a matrix operator in total derivatives as in~\eqref{ActOnCoeff}.
Then one has
$$
\ell_{A,\psi_1}^*(\psi_2)=\ell_{A^*,\psi_2}^*(\psi_1)
$$
for any sections $\psi_1$, $\psi_2$ of a horizontal module
over~$J^\infty(\pi)$.
\end{lemma}

\begin{state}[A criterion of $\lshad A,A\rshad=0$, 
\cite{Opava}]\label{Criterion}
A 
skew\/-\/adjoint ope\-ra\-tor
$A\in\CDiff(\hat{\varkappa}(\pi)$,\ $\varkappa(\pi))$
in total derivatives is Hamiltonian if and only if the relation
\begin{equation}\label{CriterionFormula}
\ell_{A,\psi_1}(A(\psi_2))-\ell_{A,\psi_2}(A(\psi_1)) =
  A\bigl(\ell_{A,\psi_2}^*(\psi_1)\bigr)
\end{equation}
holds for all $\psi_1$, $\psi_2\in\hat{\varkappa}(\pi)$.
The r.h.s.\ of formula~\eqref{CriterionFormula} is skew\/-\/symmetric
w.r.t.\ $\psi_1$,~$\psi_2$. 
\end{state}


The proof is informative in itself. It amounts to a straightforward
calculation of the value of variational Schouten bracket $\lshad A,A\rshad$
on three Hamiltonians $\cH_1$, $\cH_2$, $\cH_3\in\bar{H}^n(\pi)$.
Let $\psi_i=\bE(\cH_i)$ be the respective gradients.
The Jacobi identity $\lshad A,A\rshad(\cH_1$,\ $\cH_2$,\ $\cH_3)=0$
can be expressed
as $\langle\mathbf{b}(\psi_1,\psi_2),\psi_3\rangle=0$, where $\mathbf{b}$ is a differential operator w.r.t.\ each argument
and $\langle\,,\,\rangle\colon\gf\times\hgf\to\bar{\Lambda}^n(\pi)$.
Since $\psi_3\in\hgf$ is arbitrary, we have $\mathbf{b}(\psi_1,\psi_2)=0$
for all $\psi_1,\psi_2\in\hgf$. The calculation shows that
$\mathbf{b}(\psi_1,\psi_2)$~is equal to the left\/-\/hand side minus
the right\/-\/hand side of~\eqref{CriterionFormula}.

\begin{lemma}[{\cite[\S7.8]{Opava}}]\label{LBaseEuler}
Consider an operator in total derivatives
$J\in\CDiff_{(l)}(\hat{\varkappa}(\pi),P)$ which is
skew\/-\/sym\-met\-ric w.r.t.\ its $l$ arguments that belong
to~$\hat{\varkappa}(\pi)$
and which takes values in an $\cF(\pi)$-\/module~$P$.
If for all $\cH_1$, $\ldots$, $\cH_l\in\bar H^n(\pi)$ one has
$J(\bE(\cH_1),\ldots,\bE(\cH_l))=0$, then $J\equiv0$.
\end{lemma}

Lemma~\ref{LBaseEuler} implies that
the Jacobi identity~$J(\psi_1,\psi_2,\psi_3)=0$ can be
verified for elements $\psi_i\in\text{im}\,\bE$ only
(in particular, this is done in the proof of Proposition~\ref{Criterion}).

\begin{proof}[Proof of Proposition~\textup{\ref{Criterion}}]
Let $\cH_\alpha$, $\cH_\beta$, and $\cH_\gamma$ be Hamiltonians.
The Jacobi identity is
\begin{multline}
\{\{\cH_\alpha,\cH_\beta\}_A,\cH_\gamma\}_A
 +\{\{\cH_\beta,\cH_\gamma\}_A,\cH_\alpha\}_A
 +\{\{\cH_\gamma,\cH_\alpha\}_A,\cH_\beta\}_A=
-\sum_{{\circlearrowright}} \cEv_{A(\psi_\gamma)}
(\langle A(\psi_\alpha),\psi_\beta\rangle)\\
{}=-\sum_{\circlearrowright}\Bigl[
\langle \cEv_{A(\psi_\gamma)}(A)(\psi_\alpha),\psi_\beta\rangle
 +\langle A(\cEv_{A(\psi_\gamma)}(\psi_\alpha)),\psi_\beta\rangle
 +\langle A(\psi_\alpha),\cEv_{A(\psi_\gamma)}(\psi_\beta)\rangle\Bigr]=0.
\label{Jacobi3sumsNew}
\end{multline}
Consider the elements of the second sum,
\begin{multline*}
\langle A(\cEv_{A(\psi_\gamma)}(\psi_\alpha)),\psi_\beta\rangle
=\langle \psi_\beta, A(\cEv_{A(\psi_\gamma)}(\psi_\alpha))\rangle
=-\langle A(\psi_\beta), \cEv_{A(\psi_\gamma)}(\psi_\alpha)\rangle\\
=-\langle A(\psi_\beta), \ell_{\psi_\alpha} (A(\psi_\gamma))\rangle
=-\langle \ell_{\psi_\alpha}^* (A(\psi_\beta)),A(\psi_\gamma)\rangle=
(\text{by~\eqref{Helmholz}})\\
=-\langle \ell_{\psi_\alpha} (A(\psi_\beta)),A(\psi_\gamma)\rangle
=-\langle A(\psi_\gamma),\ell_{\psi_\alpha} A(\psi_\beta)\rangle.
\end{multline*}
Substituting this back in~\eqref{Jacobi3sumsNew}, we obtain
\begin{multline*}
0=-\sum_{\circlearrowright} \langle (\cEv_{A(\psi_\gamma)}(A))(\psi_\alpha),
\psi_\beta\rangle
+\Bigl[\sum_{\circlearrowright}\langle A(\psi_\gamma),\ell_{\psi_\alpha}
A(\psi_\beta)\rangle
-\sum_{\circlearrowright}\langle A(\psi_\alpha),\ell_{\psi_\beta} A(\psi_
\gamma)\rangle\Bigr]\\
{}=-\langle (\cEv_{A(\psi_\gamma)}(A))(\psi_\alpha),\psi_\beta\rangle
-\langle (\cEv_{A(\psi_\alpha)}(A))(\psi_\beta),\psi_\gamma\rangle
-\langle (\cEv_{A(\psi_\beta)}(A))(\psi_\gamma),\psi_\alpha\rangle.
\end{multline*}
Now set $\alpha=3$, $\beta=2$, $\gamma=1$; thence we have
\begin{equation}\label{ExpandOneSum}
0=-\langle (\cEv_{A(\psi_1)}(A))(\psi_3),\psi_2\rangle
-\langle (\cEv_{A(\psi_3)}(A))(\psi_2),\psi_1\rangle-
\langle (\cEv_{A(\psi_2)}(A))(\psi_1),\psi_3\rangle.  
\end{equation}
Consider the first summand,
\begin{subequations}\label{EachSummand}
\begin{align}
\langle (\cEv_{A(\psi_1)}(A))(\psi_3),\psi_2\rangle&=
\langle (\ell_{A,\psi_3}(A(\psi_1))),\psi_2\rangle=
\langle A(\psi_1),\ell_{A,\psi_3}^*(\psi_2)\rangle=
(\text{by Lemma~\ref{AuxLemma}})\notag\\
{}&=\langle A(\psi_1),\ell^*_{A^*,\psi_2}(\psi_3)\rangle
=\langle \ell_{A^*,\psi_2}(A(\psi_1)),\psi_3\rangle\notag\\
{}&=-\langle \ell_{A,\psi_2}(A(\psi_1)),\psi_3\rangle.\\
\intertext{Next, the second summand in~\eqref{ExpandOneSum} is equal to}
\langle (\cEv_{A(\psi_3)}(A))(\psi_2),\psi_1\rangle&=
\langle \psi_1,\ell_{A,\psi_2}(A(\psi_3))\rangle=
\langle \ell^*_{A,\psi_2}(\psi_1),A(\psi_3)\rangle=
-\langle A(\ell^*_{A,\psi_2}(\psi_1),\psi_3\rangle.\\
\intertext{Now consider the third term in the right\/-\/hand side
of~\eqref{ExpandOneSum},}
\langle (\cEv_{A(\psi_2)}(A))(\psi_1),\psi_3\rangle&=
\langle (\ell_{A,\psi_1}(A(\psi_2)),\psi_3\rangle.
\end{align}
\end{subequations}
Substituting~\eqref{EachSummand} in~\eqref{ExpandOneSum},
we finally obtain
$$
\langle \ell_{A,\psi_2}(A(\psi_1)),\psi_3\rangle
+\langle A(\ell^*_{A,\psi_2}(\psi_1),\psi_3\rangle
-\langle (\ell_{A,\psi_1}(A(\psi_2)),\psi_3\rangle=0,
$$
whence follows~\eqref{CriterionFormula}.
The proof is complete.
\end{proof}

Actually, Proposition~\ref{Criterion} states that images of Hamiltonian
operators are closed w.r.t.\ the commutation. This is readily seen
from~\eqref{DefKoszulHamWhy} below, which contains the left\/-\/hand side
of~\eqref{CriterionFormula}. Note that, by Lemma~\ref{LBaseEuler},
the assertion holds for
Hamiltonian evolutionary vector fields $\cEv_{A(\psi)}$ which may
not possess Hamiltonians $\cH$ such that $\psi=\bE(\cH)$.
However, the Lie algebra of Hamiltonian evolutionary vector fields
$\cEv_{A(\bE(\mathcal{H}))}\in\gothg(\pi)$
that do possess the Hamiltonians~$\mathcal{H}$ is correlated by
a morphism of Lie algebras  
with the Lie algebra $\bigl(\bar{H}^n(\pi),\{\,,\,\}_A\bigr)$ as follows.

\begin{state}[{\cite[\S27]{Dorfman}}]\label{ThDorfmanTriplette}
The Euler derivative $\bE$ and Hamiltonian operators~$A$ determine
the Lie algebra morphisms
\begin{equation}\label{SeqLieAlgHom}
\left(\bar{H}^n(\pi),\{\,,\,\}_A\right)\xrightarrow{\,\bE\,}
\left(\hat{\varkappa}(\pi),[\,,\,]_A\right)\xrightarrow{\,\cEv_{A(\cdot)}\,}
\left(\gothg(\pi),[\,,\,]\right),
\end{equation}
such that
\[
A\bigl([\psi_1,\psi_2]_A\bigr)=[A\psi_1,A\psi_2],\qquad 
\psi_1,\psi_2\in\hat{\varkappa}(\pi)
\]
and\footnote{Hence the image of the
Euler operator $\bE$ is closed with respect to the commutation.}
\[
\bigl[\bE(\cH_1),\bE(\cH_2)\bigr]_A=\bE\bigl(\{\cH_1,\cH_2\}_A\bigr),\qquad
\cH_1,\cH_2\in\bar{H}^n(\pi).
\]
The correlation between the Poisson bracket $\{\,,\,\}_A$ on~$\bar{H}^n(\pi)$,
the Koszul\/--\/Dorfman\/--\/Daletsky\/--\/Karas\"ev bracket
$[\,,\,]_A$ on the quotient
\[
\Omega^1(\pi)\mathrel{{:}{=}}\hat{\varkappa}(\pi)/\ker A,
\]
and the standard commutator
$[\,,\,]$ of evolutionary fields $\cEv_{A(\cdot)}$, see~\eqref{EvBracket},
is 
\begin{subequations}
\begin{gather}\label{DefKoszulHam}
[\psi_1,\psi_2]_A=
\cEv_{A(\psi_1)}(\psi_2)-\cEv_{A(\psi_2)}(\psi_1)+\ib{\psi_1}{\psi_2}{A}=
\bE\bigl(\langle\psi_1,A(\psi_2)\rangle\bigr),\\
\intertext{where}
A\bigl(\ib{\psi_1}{\psi_2}{A}\bigr)=
 \bigl(\cEv_{A\psi_1}(A)\bigr)(\psi_2)
 -\bigl(\cEv_{A\psi_2}(A)\bigr)(\psi_1).\label{DefKoszulHamWhy}
\end{gather}
\end{subequations}
For $A$ Hamiltonian, both $\ker A\subset\bigl(\hat{\varkappa}(\pi),
 [\,,\,]_A\bigr)$
and $\text{\textup{im}}\,A\subset\bigl(\gothg(\pi),[\,,\,]\bigr)$ are
ideals in the respective Lie algebras.
\end{state}

Formula~\eqref{CriterionFormula} 
provides the expression for the Sokolov bracket~$\ib{\,}{\,}{A}$,
\begin{equation}\label{EqDogma}
\ib{\psi_1}{\psi_2}{A}=\ell_{A,\psi_1}^*(\psi_2),\qquad
\psi_1,\psi_2\in\hat{\varkappa}(\pi),
\end{equation}
which is valid for Hamiltonian operators~$A$ if they are
nondegenerate\footnote{A coordinate\/-\/free definition of
the nondegenerate Hamiltonian operators means that
they have trivial kernels on the $\ell^*$\/-\/coverings
over the jet bundle~$J^\infty(\pi)$,
see~\cite{Lstar} for the construction of $\Delta$-\/coverings over~PDE.}
in the sense of~\eqref{NoKernelsIntersect}, see below.
In coordinates, we have that for a Hamiltonian operator $A=\|\sum_\tau
A_\tau^{\alpha\beta}D_\tau\|$ and $\psi_1,\psi_2\in\hat{\varkappa}(\pi)$,
\[
\ib{\psi_1}{\psi_2}{A}^i=\sum_{\sigma,\alpha} (-1)^\sigma
 \Bigl(D_\sigma\circ\Bigl[\sum_{\tau,\beta} D_\tau(\psi_1^\beta)\cdot
 \frac{\dd A_\tau^{\alpha\beta}}{\dd u^i_\sigma}\Bigr]\Bigr)
 \bigl(\psi_2^\alpha\bigr).
\]
Another way of calculating Dorfman's bracket $[\,,\,]_A$
for any Hamiltonian operator~$A$ is based on the
equivalence~\eqref{PoissonEquiv} and further use of Lemma~\ref{LCommuteEuEv},
see~\eqref{TwoGrads} in section~\ref{SecTwoSpace}.

The property $A\colon\hat{\varkappa}(\cE)\to\varkappa(\cE)$
of Hamiltonian operators 
remains valid for their restrictions onto the equations they determine.
We have it as follows.

\begin{lemma}[\cite{Opava}]\label{HamFromTo}
Let $A$ be a Hamiltonian operator and consider a
Hamiltonian evolutionary system
\begin{equation}\label{HamEvolEq}
u_t=A(\bE(\cH)),\qquad \cH\in\bar{H}^n(\pi).
\end{equation}
Then the operator~$A$ takes the generating sections
$\psi=\bE(\rho)\in\hat{\varkappa}(\cE)$
of conservation laws $[\eta]=[\rho\,\Id x+\cdots]$ for~\eqref{HamEvolEq}
to generating sections $\vph\in\varkappa(\cE)$ of
symmetries $\cEv_{\vph}\in\sym\cE$ for this system.
\end{lemma}


Next, recall that the Schouten bracket~\eqref{SchoutenBr}
of variational bi\/-\/vectors satisfies the 
Jacobi identity
\begin{equation}\label{Jacobi4Schouten}
\lshad\lshad A_1,A_2\rshad,A_3\rshad+
\lshad\lshad A_2,A_3\rshad,A_1\rshad+
\lshad\lshad A_3,A_1\rshad,A_2\rshad=0.
\end{equation}
Hence the original Jacobi identity
$\lshad A,A\rshad(\psi_1,\psi_2,\psi_3)=0$ for the
arguments of $A$ implies that $\dd_A=\lshad A,{\cdot}\,\rshad$ is a
differential, giving rise to the Poisson cohomology $H_A^k(\pi)$.
Obviously, the Casimirs $\mathbf{c}\in\bar H^n(\pi)$ such that
$\lshad A,\mathbf{c}\rshad=0$ for a Hamiltonian operator~$A$
constitute~$H_A^0(\pi)$.

\begin{theor}[The Magri scheme \cite{Dorfman,Magri}]\label{ThMagri}
Suppose $\lshad A_1,A_2\rshad=0$, $\cH_0\in H_{A_1}^0(\pi)$ is a Casimir
of~$A_1$, and $H_{A_1}^1(\pi)=0$. Then for any $k>0$ there is a
Hamiltonian $\cH_k\in\bar{H}^n(\pi)$ such that
\begin{equation}\label{MagriResolve}
\lshad A_2,\cH_{k-1}\rshad=\lshad A_1,\cH_k\rshad.
\end{equation}
Put $\vph_k\mathrel{{:}{=}}A_1\bigl(\bE(\cH_k)\bigr)$.
The Hamiltonians $\cH_i$, $i\geq0$, pairwise Poisson commute w.r.t.\ either
$A_1$ or $A_2$, the densities of~$\cH_i$ are conserved on any
equation $u_{t_k}=\vph_k$, and the evolutionary derivations
$\cEv_{\vph_k}$ pairwise commute for all~$k\geq0$.
\end{theor}

We emphasize that Theorem~\ref{ThMagri} holds for the Hamiltonians
which belong to the linear subspaces $S\subset\bar{H}^n(\pi)$ spanned by the
descendants of the Casimirs~$\cH_0\in H_{A_1}^0(\pi)$.
The commutativity of flows~$\vph_k$ and formulas~\eqref{PoissonEquiv}
imply that the phase volume $\iint_S\langle\psi_1,A(\psi_2)\rangle$
$\Id\psi_1\Id\psi_2=0$ is conserved on the subspaces~$S$.

The Magri scheme starts from any Hamiltonians
$\cH_{k-1},\cH_k\in\bar{H}^n(\pi)$ that satisfy~\eqref{MagriResolve},
but we always intend to operate with maximal subspaces of $\bar{H}^n(\pi)$,
and therefore we require $\lshad A_1,\cH_0\rshad=0$ such that the sequences of Hamiltonians can not be extended with~$k<0$.

\begin{proof}[Proof of Theorem~\textup{\ref{ThMagri}}]
The main homological equality~\eqref{MagriResolve} is established by
induction on~$k$. Starting with a Casimir~$\cH_0$, we obtain
\[
0=\lshad A_2,0\rshad=\lshad A_2,\lshad A_1,\cH_0\rshad\rshad =
 -\lshad A_1,\lshad A_2,\cH_0\rshad\rshad\mod\lshad A_1,A_2\rshad=0
\]
using the Jacobi identity~\eqref{Jacobi4Schouten}.
The first Poisson cohomology $H^1_{A_1}(\pi)=0$ is
trivial by assumption of the theorem, and hence the closed element
$\lshad A_2,\cH_0\rshad$ in the kernel of $\lshad A_1,\cdot\rshad$
is exact: $\lshad A_2,\cH_0\rshad=\lshad A_1,\cH_1\rshad$
for some~$\cH_1$.

For~$k\geq1$, we have
\[
\lshad A_1,\lshad A_2,\cH_k\rshad\rshad =
-\lshad A_2,\lshad A_1,\cH_k\rshad\rshad =
-\lshad A_2,\lshad A_2,\cH_{k-1}\rshad\rshad =0
\]
using the Jacobi identity~\eqref{Jacobi4Schouten}
for the Schouten bracket~\eqref{SchoutenBr}
and by $\lshad A_2,A_2\rshad=0$.
Hence $\lshad A_2,\cH_k\rshad=\lshad A_1,\cH_{k+1}\rshad$
by~$H^1_{A_1}(\pi)=0$, and thus we proceed infinitely.
\end{proof}

Correlated Magri's schemes for coupled hierarchies are
further considered in section~\ref{SecLiou}, see diagram~\eqref{Diag}.

\subsection{Lie algebroids}\label{SecLieAlgd}
Hamiltonian operators determine Lie algebra homomorphisms~\eqref{SeqLieAlgHom}
that map the Poisson bracket~\eqref{PoissonEquiv}
to the standard Lie structure~\eqref{EvBracket} on~$\varkappa(\pi)$.
In the finite\/-\/dimensional case, this yields the Lie algebra homomorphisms
$P\colon\Gamma(T^*M)\to\Gamma(TM)$ to sections of the tangent bundles for
smooth manifolds~$M$. Recall that a very important class of the Lie algebra
homomorphisms to $\Gamma(TM)$ is then provided by the Lie algebroids.

\begin{define}[\cite{Vaintrob}]\label{DefLieAlgd}
A \emph{Lie algebroid} over an $n$-\/dimensional
manifold~$M^n$ is a vector bundle $\Omega\to M^n$ whose
sections $\Gamma\Omega$ are equipped with a Lie algebra
structure $[\,,\,]_\ga$ together with a morphism
of vector bundles $\ga\colon\Omega\to TM$,
called the \emph{anchor}, such that the Leibnitz rule
\begin{equation}\label{LeibnitzInAlg}
[\psi_1,f\cdot\psi_2]_\ga=f\cdot[\psi_1,\psi_2]_\ga
  +\bigl(\ga(\psi_1)f\bigr)\cdot\psi_2
\end{equation}
holds for any
$\psi_1,\psi_2\in\Gamma\Omega$ and $f\in C^\infty(M^n)$.

Equivalently, a Lie algebroid structure on~$\Omega$ is
a homological vector field~$Q$ on $\Pi\Omega$
(take the fibres of $\Omega$, reverse their parities, and thus obtain
the new super\/-\/bundle $\Pi\Omega$ over~$M^n$).
The homological vector fields, which are differentials on
$C^\infty(\Pi\Omega)=\Gamma(\bigwedge^\bullet\Omega^*)$, equal
\begin{equation}\label{HomVF}
Q=\tilde{y}^i\ga_i^\alpha(x)\frac{\dd}{\dd x^\alpha}
 +\frac{1}{2}\tilde{y}^j\tilde{y}^i c^k_{ij}(x)\frac{\dd}{\dd\tilde{y}^k},
\qquad Q^2=0,
\end{equation}
where
\begin{itemize}
\item
$(x^\alpha)$ is a system of local coordinates near a point $x\in M^n$,
\item
$(y^i)$ are local coordinates along the fibres of~$\Omega$ and
$(\tilde{y}^i)$ are the respective coordinates on~$\Pi\Omega$, and
\item
$\ga(e_i)=\ga_i^\alpha(x)\cdot\dd/\dd x^\alpha$ is the image under
the anchor $\ga$ and $[e_i,e_j]_\ga=c^k_{ij}(x)e_k$ give
the structural constants for a local basis $(e_i)$ of sections~$\Gamma\Omega$,
respectively.
\end{itemize}
The proof of equivalence 
is straightforward,
see~\cite{Vaintrob,YKS} for details and other equivalent definitions.
\end{define}

\enlargethispage{\baselineskip}
\begin{example}
Consider the Cartan connection~\eqref{CartanConnection},
and set the anchor $\ga\colon\cC\to\Gamma(TM^n)$ to be
$\ga=\pi_{\infty,*}\circ j_\infty(s_0)_*$, which is
the restriction $j_\infty(s_0)_*(X)$ of horizontal fields~$X$ 
onto the jet $j_\infty(s_0)$ of a fixed section $s_0\in\Gamma(\pi)$
and further projection under $\pi_{\infty}\colon J^\infty(\pi)\to M^n$.
The corresponding homological vector field is the restriction
$j_\infty(s_0)^*(\Id_h)$ of the horizontal differential~\eqref{dh}
onto $j_\infty(s_0)$.
This is a generalization of a classical example $\Omega=TM^n$,
see~\cite{YKS}.
\end{example}

\begin{example}\label{PoissonAlgebroid}
Let $P\colon\hat{\varkappa}(\pi)\to\varkappa(\pi)$
be a Hamiltonian operator and
$\{\,,\,\}_P\colon\bar{H}^n(\pi)\times\bar{H}^n(\pi)\to\bar{H}^n(\pi)$
be the Poisson bracket. Consider the one\/-\/dimensional bundle
over $J^\infty(\pi)$ with the fibre
$\Id x^1\wedge\ldots\wedge\Id x^n=\Id\volume(M^n)$ at each point of~$M^n$.
The equivalence classes of sections constitute an $\cF(\pi)$\/-\/module
and are endowed with the Lie algebra structure~$\{\,,\,\}_P$.
Set the anchor $\ga\mathrel{{:}{=}}\cEv_{P\circ\bE}$.
By~\eqref{PoissonEquiv}, the Leibnitz rule~\eqref{LeibnitzInAlg} holds:
\begin{equation}\label{LeibnitzPoissonAlg}
\{\cH_1,f\cdot\cH_2\}_P=f\cdot\{\cH_1,\cH_2\}_P+\ga(\cH_1)(f)\cdot\cH_2,
\qquad \cH_1,\cH_2\in\bar{H}^n(\pi),\quad f\in\cF(\pi).
\end{equation}
The sections $\cH\in\bar{H}^n(\pi)$ are parameterized by the Hamiltonian
densities $h\in\cF(\pi)$. Let $\tilde{h}$ be the functional coordinates
in~$\Pi\bar{H}^n(\pi)$, whose fibre is parity\/-\/dual to~$h$. Then the
homological vector field~$Q$ that encodes the Lie algebroid
$\bigl(\bar{H}^n(\pi),\{\,,\,\}_P\bigr)\xrightarrow{\ga}
 \bigl(\varkappa(\pi),[\,,\,]\bigr)$ is equal to
\[
Q=\int_{\cF(\pi)}\Bigl[
 \tilde{h}\cdot\cEv_{P(\bE([h\,\Id\volume(M^n)]))}
   +\frac{1}{2}\tilde{h}_2\tilde{h}_1\,\bigl\{
     [h_1\,\Id\volume(M^n)],[h_2\,\Id\volume(M^n)]\bigr\}_P\cdot
   \frac{\dd}{\dd\tilde{h}} \Bigr]\,\Id\cF(\pi).
\]

Proposition~\ref{ThDorfmanTriplette} states that $\ga$,
which is defined by a Hamiltonian operator~$P$, is a morphism of
Lie algebras. This property was not included into the definition of
Lie algebroids because, even in a general situation not related to
Hamiltonian operators, the property is a consequence
of the Leibnitz rule~\eqref{LeibnitzInAlg} and of the Jacobi identity
for $[\,,\,]_\ga$.
\end{example}

\begin{state}[\cite{YKSMagri}]\label{PropMorphismFollows}
The anchors~$\ga$ map the brackets $[\,,\,]_\ga$ between sections
of bundles~$\Omega$ over finite\/-\/dimensional manifolds~$M^n$
to the Lie bracket $[\,,\,]$ on the tangent bundle to~$M^n$.
\end{state}

The converse is not true. Namely, there are morfisms of Lie algebra
structures in the modules of sections of fibre
bundles over the base manifold that do not respect the
Leibnitz rule~\eqref{LeibnitzInAlg}. 
Moreover, and this is our claim of crucial importance, the converse of
Proposition~\ref{PropMorphismFollows} is not valid for Hamiltonian operators~$P$ in total derivatives and the morphisms
$\ga=\cEv_{P(\cdot)}\colon\Omega^1(\pi)\to\gothg(\pi)$, where Dorfman's bracket
$[\,,\,]_P$ on~$\Omega^1(\pi)$ is~\eqref{DefKoszulHam}.
Indeed, there is no Leibnitz rule over~$\cF(\pi)$ for~$[\,,\,]_P$.
This is the obstruction to an extension of Dorfman's bracket
$[\,,\,]_{\cP}$ of $1$-\/forms $\psi_1,\psi_2\in\Gamma(T^*M^n)$,
\begin{equation}\label{DDKBr}
[\psi_1,\psi_2]_{\cP}=\IL_{\cP\psi_1}\psi_2-\IL_{\cP\psi_2}\psi_1+
  \Id_{\text{dR}(M^n)}\bigl(\cP(\psi_1,\psi_2)\bigr),\qquad
 \cP\in\Gamma\Bigl(\bigwedge\nolimits^2 TM^n\Bigr),
\end{equation}
to Lie algebroid structures over the base manifolds~$J^\infty(\pi)$.

\enlargethispage{1.7\baselineskip}\begin{counterexample}\label{StuckAt}
Consider the first Hamiltonian structure $P_1=D_x$ for the KdV
equation~\eqref{IKdV}, and consider two conserved densities $\rho_1=w$
and $\rho_2=\tfrac{1}{2}w^2$. Obviously,
$\{\rho_1\,\Id x,\rho_2\,\Id x\}_{P_1}=0$, and hence Dorfman's bracket
$[\psi_1,\psi_2]_{P_1}$ of the gradients $\psi_i=\bE_w(\rho_i)$ also vanishes.
Now multiply $\rho_1$ by any nonlinear $f(x)\in C^\infty(M)$ and,
applying Euler derivative~$\bE_w$, get~$f\cdot\psi_1$.
First let us commute $f\cdot\psi_1$ with $\psi_2$ by $[\,,\,]_{P_1}$ and obtain
the equivalence class in the quotient $\hat{\varkappa}(\pi)/\ker P_1$,
\begin{align}
[f(x)\cdot 1,w]_{D_x} &= f'(x)+\text{const}.\label{DxStraight}\\
\intertext{On the other hand, the Leibnitz rule~\eqref{LeibnitzInAlg}
prescribes that this is equal to}
{} &\phantom{{}={}}
f(x)\cdot[1,w]_{D_x}-\cEv_{D_x(w)}\bigl(f(x)\bigr)\cdot1=0-0=0.\notag
\end{align}
This zero value can be achieved at any point $x_0\in M$ by choosing
$\text{const}\mathrel{{:}{=}}-f'(x_0)$ in~\eqref{DxStraight},
but not on the entire~$M$ at~once.

Analogously, the Leibnitz rule does not hold for multiplication
by~$f(x,[u])\in\cF(\pi)$, and we do not repeat the reasonings for the sake of brevity~only.
\end{counterexample}

This counterexample manifests the fundamental difference between jet
bundles $J^\infty(\pi)$ over $M^n$ and smooth manifolds themselves,
which can be regarded as the jet bundles over the point~$\{x_0\}$, $n=0$.

We conclude that, depending on a problem, we have the choice between, 
first, assuming that the Lie algebra homomorphism
$A\colon\Omega^1(\xi_\pi)\to\gothg(\pi)$ is enough for further reasonings,
and, second, postulating the Leibnitz rule~\eqref{LeibnitzInAlg}.
The former case~\eqref{SeqLieAlgHom} is realized by Frobenius operators.
In this sense, they can be regarded as the anchors in the analogues of
Lie algebroids over infinite jet bundles~$J^\infty(\pi)$.

However, if the Leibnitz rule is still needed, then we propose to resolve
the difficulty as follows. First we assign formal differential
complexes~\eqref{FrobComplex} to Frobenius operators, and the representations
of the complexes through homological vector fields will determine the Lie
algebroids (see~\eqref{HomVF} in the second part of
definition~\ref{DefLieAlgd}). This scheme can be further applied in the
Batalin\/--\/Vilkovisky (BV) formalism to construction and analysis of the
quantum Poisson 
manifolds~\cite{KontsZaboronsky,BarKonts}.

\subsection{The bracket of conservation laws}\label{SecTwoSpace}
Let us convert Lemma~\ref{HamFromTo} to a definition.
Suppose $\cE=\{F=0\mid F\colon J^\infty(\pi)\to\gf\}$~is a determined
evolutionary system, hence $\hgf(\cE)\simeq\hat{\varkappa}(\cE)$
and the coupling $\langle\,,\,\rangle\colon\hgf\times\varkappa\to
\bar{\Lambda}^n(\pi)$ is well defined.
In this section we consider the class of \emph{Noether}
(or `pre\/-\/Hamiltonian' \cite{Lstar,JK3Bous}) operators $A\colon\cosym\cE\to\sym\cE$
that map generating section of conservation laws for~$\cE$ to symmetries but may not be skew\/-\/adjoint or, even if it is so, may not
define a bracket that satisfies the Jacobi identity~\eqref{Jacobi3sums}.
In all cases, if a Noether operator for an autonomous evolution equation
$\cE=\{F=u_t-f(x,[u])=0\}$ does not depend on the time~$t$,
then $A$ satisfies the operator equation $\ell_F\circ A\doteq A\circ\ell_F$
on~$\cE$.

Using Noether operators,
one can induce skew\/-\/symmetric brackets between
conservation laws $\eta\in\bar{H}^{n-1}(\cE)$ and, second, between
their generating sections~$\hgf(\cE)$.

Let $\eta=\rho\,\Id x+\cdots\in\bar{\Lambda}^{n-1}(\cE)$ be a conserved
current for an evolutionary system~$\cE$.
Then, by Lemma~\ref{LGradient}, its generating section
$\psi_\eta=\bE(\rho)\in\hgf(\cE)$ is the Euler derivative of the conserved density.
For any $\vph\in\sym\cE$, the current $\cEv_\vph(\eta)$ is obviously
conserved on~$\cE$ and its generating section $\bE(\cEv_\vph(\rho))$
equals $\cEv_\vph(\psi)+\ell_\vph^*(\psi)$ by Lemma~\ref{LCommuteEuEv}.

Given an operator $A\colon\cosym\cE\to\sym\cE$
and any two conserved currents $\eta_1,\eta_2\in\bar\Lambda^{n-1}(\cE)$,
set\footnote{The arising
algebra $(\bar{H}^{n-1}(\cE),\lshad\,,\,\rshad_A)$ of conservation laws
converts the Lie algebra of Hamiltonians with a Poisson bracket to an algebra with the bracket specified by a Noether operator $A\colon\cosym\cE\to\sym\cE$.
This algebra could be called the ``current algebra'' if the term were not
already in use.}
\begin{subequations}\label{TwoBoth}
\begin{align}
 {}&\langle\!\langle\,,\,\rangle\!\rangle_A \colon
  \bar{\Lambda}^{n-1}(\cE)\times\bar{\Lambda}^{n-1}(\cE)\to
  \bar{\Lambda}^{n-1}(\cE),\label{TwoDens}\\
\langle\!\langle\eta_1,\eta_2\rangle\!\rangle_A
\stackrel{\text{def}}{{}={}}
  \cEv_{A(\psi_1)}(\eta_2)&-\cEv_{A(\psi_2)}(\eta_1)
 +\tfrac{1}{2}\langle\psi_1,A(\psi_2)\rangle
 -\tfrac{1}{2}\langle A(\psi_1),\psi_2\rangle,\notag\\
\intertext{and (see~\eqref{DDKBr}, which is~\cite[Eq.~(3.2${}'$)]{YKSMagri})}
 {}&\lshad\,,\,\rshad_A \colon \hgf(\cE)\times\hgf(\cE)\to
  \hgf(\cE).\label{TwoGrads}\\
\lshad\psi_1,\psi_2\rshad_A \stackrel{\text{def}}{{}={}}
 \cEv_{A(\psi_1)}(\psi_2)&-\cEv_{A(\psi_2)}(\psi_1)+
  \ell_{A(\psi_1)}^*(\psi_2)-\ell_{A(\psi_2)}^*(\psi_1)
 +\tfrac{1}{2}\bE\bigl(\langle\psi_1,A(\psi_2)\rangle
    -\langle A(\psi_1),\psi_2\rangle\bigr),\notag
\end{align}
\end{subequations}
By~\eqref{PoissonEquiv}, we have $\{\cH_1,\cH_2\}_A=-\cEv_{A(\bE(\cH_1))}(\cH_2)$ for Hamiltonian operators~$A$, and therefore
$\langle\!\langle\cH_1,\cH_2\rangle\!\rangle_A=-\{\cH_1,\cH_2\}_A$.
Hence we conclude that
formula~\eqref{TwoGrads} gives an alternative way to calculate the
bracket $\ib{\,}{\,}{A}$, see~\eqref{DefKoszulHam} and~\eqref{EqDogma}.
Indeed, this is achieved by omitting the first two summands in~\eqref{TwoGrads}
and setting to zero the application of evolutionary derivatives to~$\psi_1$
and~$\psi_2$ in what remains.
The equivalence between \eqref{EqDogma} and~\eqref{TwoGrads} implies
nontrivial identities even for the simplest 
scalar Hamiltonian operator~$D_x$.
Formula~\eqref{EqDogma} is more elegant because it contains only the action
by total derivatives onto $\psi_1,\psi_2\in\hgf(\cE)$ (note that the
partial derivatives $\dd/\dd u^i_\sigma$ are applied in it only to the coefficients of~$A$). At the same time, the last four summands
in~\eqref{TwoGrads} incorporate an application of the partial derivatives to
$\psi_1$ and~$\psi_2$. However, the skew\/-\/symmetry
$\ib{\psi_1}{\psi_2}{A}=-\ib{\psi_2}{\psi_1}{A}$ of the expression~\eqref{EqDogma} is not obvious.
Again, we recall that the notation~$\ell_{A,\psi}$, see~\eqref{ActOnCoeff},
is not the linearization~$\ell_{A(\psi)}$ given in~\eqref{Linearization}.

\begin{rem}\label{BrLawsJacobi}
Only skew\/-\/adjoint operators $T^*M^n\to TM^n$, which can be represented by
bi\/-\/vectors $\Gamma\bigl(\bigwedge^2TM^n\bigr)$, were considered
in~\cite[\S3.2]{YKSMagri}. This was why the antisymmetrization was not performed in~\eqref{DDKBr}, which is the realization of~\eqref{TwoGrads}
in the case of finite\/-\/dimensional manifolds~$M^n$.

Due to the antisymmetrization, brackets~\eqref{TwoBoth}
are invariant under adding self\/-\/adjoint operators to~$A$.
This shows that the brackets $\langle\!\langle\,,\,\rangle\!\rangle_A$ and
$\lshad\,,\,\rshad_A$ satisfy the Jacobi identity if and only if $A=P+S$,
where $P$~is a Hamiltonian operator and~$S^*=S$.
\end{rem}

\begin{example}
\label{Exd3Bous}
Consider the dispersionless $3$-component Boussinesq\/-\/type system~\cite{Nutku}
\begin{equation}\label{d-B}
u_t=ww_x+v_x,\quad v_t=-uw_x-3u_xw,\quad w_t=u_x,
\end{equation}
which is equivalent to the Benney\/--\/Lax equation.
Note that~\eqref{d-B} is scaling\/-\/invariant; e.g., the homogeneity weights
are $|u|=3/2$, $|v|=2$, $|w|=1$, $|D_x|=1$, and~$|D_t|=3/2$.

\enlargethispage{1.3\baselineskip}
In~\cite{JK3Bous}, two compatible Hamiltonian operators $\hat{A}_0$
and $\hat{A}_2$, and a self\/-\/adjoint Noether operator
$A_1\colon\cosym\cE\to\sym\cE$ were found for~\eqref{d-B}
by a calculation on the $\ell^*$-\/covering over~$\cE$ and using
the scaling invariance of the system.
The first Hamiltonian structure for~\eqref{d-B} is given by the operator
\begin{subequations}\label{A12dBous}
\begin{align}
\hat{A}_0&=\begin{pmatrix} D_x & 0 & 0\\ 0 & -4wD_x-2w_x & D_x \\ 0 & D_x & 0
    \end{pmatrix}.\label{dBA0}\\
\intertext{Sokolov's bracket $\ib{\,}{\,}{\hat{A}_0}$ in the inverse image
of $\hat{A}_0$
is obtained using formula~\eqref{EqDogma}, and we have that}
{}&{}\ib{\,}{\,}{\hat{A}_0}^u=\ib{\,}{\,}{\hat{A}_0}^v=0,\qquad
\ib{\vec{p}}{\vec{q}}{A_0}^w=2\bigl(p^vq^v_x-p^v_xq^v\bigr).
\label{A0Bracket}\\
\intertext{The self\/-\/adjoint Noether operator $A_1$ is}
A_1&=\begin{pmatrix} ww_x+v_x & -3u_xw-uw_x & u_x\\
-3u_xw-uw_x & -3w^2w_x-4v_xw-uu_x & v_x\\
u_x & v_x & w_x \end{pmatrix}.\label{A1dBous}\\
\intertext{By Remark~\ref{BrLawsJacobi}, the brackets
$\langle\!\langle\,,\,\rangle\!\rangle_{A_1}$ on $\bar{H}^1(\cE)$
and $\lshad\,,\,\rshad_{A_1}$ on $\hat{\varkappa}(\cE)$
vanish identically.
The second Hamiltonian operator for~\eqref{d-B} is
the linear combination}
\hat{A}_2&=A_1+A_2,\label{A2HamdBous}\\
\intertext{where, following~\cite{JK3Bous}, the operator
$A_2\colon\cosym\cE\to\sym\cE$ is equal to} 
A_2&=\begin{pmatrix}
(2w^2+4v)\,D_x+ww_x+v_x & -11uw\,D_x-(2u_xw+8uw_x) & 3u\,D_x \\
-11uw\,D_x-3u_xw-uw_x & h_1\,D_x+h_0 & 4v\,D_x \\
3u\,D_x+u_x & 4v\,D_x+2v_x & 2w\,D_x
\end{pmatrix},\label{A2dBous}
\end{align}
\end{subequations}
here we put $h_0=-(2uu_x+8vw_x+4v_xw+6w^2w_x)$ and $h_1=-(3u^2+16vw+6w^3)$.
The right\/-\/hand side of system~\eqref{d-B}
belongs to the image of each of the three operators $\hat{A}_0$, $A_1$,
and~$\hat{A}_2$:
\[  
\begin{pmatrix}u_t\\ v_t\\ w_t\end{pmatrix}=
 \hat{A}_0\begin{pmatrix}\tfrac{1}{2}w^2+v\\ u\\ uw\end{pmatrix}=
 A_1\begin{pmatrix}1\\ 0\\ 0\end{pmatrix}=
 A_2\begin{pmatrix}1\\ 0\\ 0\end{pmatrix}.
\]

Again, by Remark~\ref{BrLawsJacobi}, the bracket $\lshad\,,\,\rshad_{A_2}$
coincides with Dorfman's bracket $[\,,\,]_{\hat{A}_2}$ for the second
Hamiltnian operator, and both of them satisfy the Jacobi identity.
Thus Sokolov's bracket $\ib{\,}{\,}{\hat{A}_2}$ can be caclulated
not by~\eqref{EqDogma}, but using~\eqref{TwoGrads} and omitting
all terms with evolutionary derivations.

In what follows, the Noether operators $\hat{A}_0$, $A_1$, and
$\hat{A}_2\colon\cosym\cE\to\sym\cE$
for~\eqref{d-B} will be met again. We claim that the images of $A_1$
and $A_2$ are closed w.r.t.\ the commutation~\eqref{EvBracket}.
The images of arbitrary linear combinations
$\lambda_0\hat{A}_0+\lambda_1A_1+\lambda_2\hat{A}_2$
are also closed under commutation, and hence the three Frobenius operators
are linear compatible.
Moreover, the image of a pre\/-\/symplectic operator~$A_1^{-1}$
is closed w.r.t.\ either $[\,,\,]_{\hat{A}_0}$ or
$[\,,\,]_{\hat{A}_2}$ on~$\hat{\varkappa}(\cE)$, and therefore both
recursion operators $R_0=\hat{A}_0\circ A_1^{-1}$ and
$R_2=\hat{A}_2\circ A_1^{-1}\colon\sym\cE\to\sym\cE$ are Frobenius as well.
\end{example}

In the previous subsection we addressed the construction of Lie algebroids over
$J^\infty(\pi)$ and noticed a very important property of~\eqref{TwoBoth}:
For operators $A=P+S$ in total derivatives, the Leibnitz rule is valid
for the bracket~\eqref{TwoDens} but does not hold for~\eqref{TwoGrads}.

Not every skew\/-\/adjoint Noether operator is Hamiltonian since, in general,
the brackets~\eqref{PoissonEquiv} and~\eqref{TwoGrads} may not satisfy the Jacobi
identity.

\begin{lemma}[\cite{JKVerb3Vect}]\label{L3Vect}
Let $\cE=\{u_t=f(x,[u])\}$ be an evolutionary system of differential
order $\ord\ell_f>1$, and let its symbol be nondegenerate
on an open dense subset of~$\cE$. Then any skew\/-\/adjoint Noether
operator~$A\colon\cosym\cE\to\sym\cE$ is Hamiltonian.
\end{lemma}


Unfortunately, the non\/-\/vanishing of the determinant of the symbol (which
is the higher\/-\/order matrix $\sum_{|\sigma|=\text{max}}(\ell_f)_\sigma$)
depends on the system of local coordinates.

\begin{counterexample}
Consider the Kaup\/--\/Boussinesq system
\begin{equation}\label{KB}
u_t=uu_x+v_x,\qquad v_t=(uv)_x+\veps^2u_{xxx},
\end{equation}
whose symbol is degenerate and
which extends the assertion of Lemma~\ref{L3Vect}, see~\cite{JKVerb3Vect}.
System~\eqref{KB} is transformed by $w=v-\veps u_x$
to the second\/-\/order Kaup\/--\/Broer system
\[
u_t=uu_x+w_x+\veps u_{xx},\qquad w_t=(uw)_x-\veps w_{xx}.
\]
After the transformation,
the determinant of the symbol does not vanish anywhere
(of course, we assume $\varepsilon\neq0$).
At the same time, it is clear that the
third\/-\/order linearization of the right\/-\/hand side of the
Kaup\/--\/Boussinesq system~\eqref{KB} is nondegenerate
in the following sense. 
\end{counterexample}

\begin{define}
We say that a matrix operator~$A$ in total derivatives
is \emph{nondegenerate} if
\begin{equation}\label{NoKernelsIntersect}
\bigcap_\sigma \ker A_\sigma=\{0\},\qquad\text{where }
 A=\sum_\sigma A_\sigma\cdot D_\sigma.
\end{equation}
We shall consider nondegenerate Noether and recursion operators that are
transformed using~\eqref{ChangeCoordRec} and~\eqref{ChangeCoordHamOp},
respectively, under nondegenerate reparametrizations $\tilde{u}=\tilde{u}[u]$
of fibre coordinates. 
This makes the nondegeneracy~\eqref{NoKernelsIntersect} well defined.
\end{define}

\begin{example}\label{ExNondegenerateWithKernel} 
The Hamiltonian operator $\left(\begin{smallmatrix}0 & 0\\
0 & D_x\end{smallmatrix}\right)$ is degenerate.
The operator $\square=u_x+\tfrac{1}{2}D_x$, which was introduced
in~\eqref{SquareSok} and which will be made well defined by~\eqref{Frob2} below,
is nondegenerate, although its kernel
is spanned by~$\exp(-2u)$.
\end{example}

A conjecture that the assumptions of Lemma~\ref{L3Vect} can be weakened is
already present in~\cite{JKVerb3Vect}, but its proof~\cite{Gessler}, which is
quoted there, hinted no way how it can be generalized.

\begin{conjecture}
The assertion of Lemma~\ref{L3Vect} 
holds for skew\/-\/adjoint Noether operators $A\colon\cosym\cE\to\sym\cE$
on evolutionary systems~$\cE=\{u_t=f\}$ with $\ord\ell_f>1$,
if the linearizations~$\ell_f$ of the right\/-\/hand sides are
nondegenerate, see~\eqref{NoKernelsIntersect}, on open dense subsets of~$\cE$.
\end{conjecture}

\section{Frobenius operators}\label{SecDef}\label{SecAlg}
In this section we introduce the well\/-\/defined notion of Frobenius
operators, revealing the nature of their domains 
$\Omega^1(\xi_\pi)\subseteq\Gamma(\pi_\infty^*(\xi))$,
establish their properties,
and give examples of Frobenius operators. 
In particular, we recognize
Frobenius recursions $\varkappa(\pi)\to\varkappa(\pi)$
as higher differential order solutions of the
classical Yang\/--\/Baxter equation for the Lie algebra~$\gothg(\pi)$;
here we construct examples of the projector solutions which are related to
Liouville\/-\/type systems (e.g., to the $2$D~Toda chains).
We establish the chain rule for factorizations of Frobenius operators
and correlate the Sokolov brackets for Frobenius recursions 
that produce sequences of Lie algebra structures.
We show that Frobenius operators~$A$ induce flat connections in triples
$\bigl(\Omega^1(\xi_\pi),\gothg(\pi),A\bigr)$ of Lie algebras and
their morphisms. Finally, we assign formal differential complexes
over~$\Omega^1(\xi_\pi)$ to Frobenius operators. 

The Frobenius theorem~\cite{InvolutiveDistrib}, which deals with
involutive distributions over finite\/-\/di\-men\-si\-o\-nal manifolds,
gave the proper name to the class of operators defined here.

\subsection{Main definitions}
By Proposition~\ref{ThDorfmanTriplette},
the Poisson bracket~$\{\,,\,\}_A$ in the Lie algebra of
Hamiltonians~$\bar{H}^n(\pi)$
is correlated with the commutator~$[\,,\,]$
in the Lie algebra~$\gothg(\pi)$ of vector fields~$\cEv_{A(\cdot)}$.
The skew\/-\/adjoint operator $A$ in the Poisson bracket
$\{\cH_1,\cH_2\}_A=\langle\bE({\mathcal{H}}_1),
A(\bE({\mathcal{H}}_2))\rangle$
coincides with the operator $A$ whose image is a subalgebra of
evolutionary fields $\smash{{\cEv}_{A(\bE(\mathcal{H}))}}$.
Thus we have
\begin{equation}\label{KuperDefPoisson} 
\bigl[\cEv_{A(\bE(\cH_1))} , \cEv_{A(\bE(\cH_2))}\bigr] =
  \cEv_{A(\bE(\{\cH_1,\cH_2\}_A))}.
\end{equation}
This is readily seen from the Jacobi identity~\eqref{Jacobi4Schouten}
for the Schouten bracket~\eqref{SchoutenBr}. Indeed, the commutator of
two Hamiltonian vector fields equals
\[
\bigl[\lshad A,\cH_1\rshad,\lshad A,\cH_2\rshad\bigr]=
\lshad A,\lshad\cH_2,\lshad A,\cH_1\rshad\rshad\rshad +
\lshad\cH_2,\lshad\lshad A,\cH_1\rshad,A\rshad\rshad,
\]
whence the derived bracket $\lshad\cH_2,\lshad A,\cH_1\rshad\rshad=
\{\cH_1,\cH_2\}_A$ appears in the right\/-\/hand side of~\eqref{KuperDefPoisson}
and $\dd_A^2(\cH_1)=0$ is contained in the second summand.

In this section 
we reverse the status of the two Lie algebra
structures~$[\,,\,]\bigr|_{\mathrm{im}\,A}$ and~$\{\,,\,\}_A$,
giving the priority to the commutators of vector fields
and thus considering involutive distributions of evolutionary vector
fields $\cEv_\vph\in\gothg(\pi)$ whose generators
$\vph=A(\cdot)\in\varkappa(\pi)$ belong to images of matrix
operators~$A$ in total derivatives.

\begin{define}\label{DefFrob}\label{DefpHO}
The definition of a \emph{Frobenius operator} consists of two parts;
all notation is correlated with the previous material. 

\textbf{\S1 \textmd{(see~\cite{SokolovUMN})}.}\quad%
Prohibit any changes of coordinates in all modules of sections.
Consider a matrix linear differential operator~$A\colon\gf\to\varkappa(\pi)$
in total derivatives, where $\gf\subseteq\Gamma\bigl(\pi_\infty^*(\xi)\bigr)$
is a certain submodule that will be specified in~\S2 of this definition.
The operator~$A$ is \emph{Frobenius} 
if, first, its image 
in the $\cF(\pi)$-\/module $\varkappa(\pi)$ of generators
of evolutionary vector fields is closed w.r.t.\ the commutation:
\begin{equation}\label{EqDefFrob}
[\text{im}\,A, \text{im}\,A]\subseteq\text{im}\,A.
\end{equation}
The commutator $[\,,\,]\bigr|_{\mathrm{im}\,A}$ induces\footnote{%
Two sets of summands appear in the bracket of evolutionary vector
fields $A(\psi_1),A(\psi_2)$ that belong to the image of a Frobenius
operator~$A$:
\[\bigl[A(\psi_1),A(\psi_2)\bigr]=A\bigl(\cEv_{A(\psi_1)}(\psi_2)-
  \cEv_{A(\psi_2)}(\psi_1)\bigr)+\bigl(
  \cEv_{A(\psi_1)}(A)(\psi_2)-\cEv_{A(\psi_2)}(A)(\psi_1)\bigr).\]
In the first summand we have used the permutability of
evolutionary derivations and operators in total derivatives. The second
summand hits the image of~$A$ by construction.}
a skew\/-\/symmetric \emph{Koszul bracket} $[\,,\,]_A$
in the quotient $\Omega^1(\xi_\pi)=\gf/\ker A$ of the domains of~$A$:
\begin{subequations}\label{EqOplusB}
\begin{gather}
\bigl[A(\psi_1),A(\psi_2)\bigr]=A\bigl([\psi_1,\psi_2]_A),\qquad
\psi_1,\psi_2\in\Omega^1(\xi_\pi). 
   \label{EqOplusBBoth}\\
\intertext{The Koszul bracket, which is defined up to~$\ker A$, equals}
[\psi_1,\psi_2]_A=\cEv_{A(\psi_1)}(\psi_2)-\cEv_{A(\psi_2)}(\psi_1)+
  \{\!\{\psi_1,\psi_2\}\!\}_A.\label{EqOplusBKoszul}
\end{gather}
\end{subequations}
It contains two standard summands and
the \emph{Sokolov bracket} $\{\!\{\,,\,\}\!\}_A$.

\textbf{\S2.}\quad%
Let $w$~be a fibre coordinate in the bundle~$\xi$.
Allow coordinate reparametrizations, both in~$\xi$ and~$\pi$.
Suppose that there is a differential substitution $J^\infty(\pi)\to\Gamma(\xi)$
that yields the embedding $\colon\gf\hookrightarrow
\Gamma\bigl(\pi_\infty^*(\xi)\bigr)$ of an $\cF(\xi)$-\/module~$\gf$.
From now on, restrict~$\gf$ onto the image of the
substitution. Thence we denote the substitution by the same letter~$w[u]$
and retain the notation~$\gf$ for the restriction of the $\cF(\xi)$-\/module.
Put $\Omega^1(\xi_\pi)=\gf\bigr|_{w=w[u]}/\ker A$.

We postulate that, under diffeomorphisms
$\tilde{u}=\tilde{u}[u]\colon J^\infty(\pi)\to\Gamma(\pi)$
and $\tilde{w}=\tilde{w}[w]\colon J^\infty(\xi)\to\Gamma(\xi)$,
the transformation rules for the operator~$A$ are uniquely defined
by the fibre bundles~$\pi$ and~$\xi$.
This implies that $\gf$ is one of the following:\footnote{%
\label{PotentialGeneral}%
The axiom we accept means that, in this paper, we consider only
the tangent and cotangent bundles to (infinite\/-\/dimensional) smooth
manifolds. Therefore we do not attempt to define Frobenius operators on
$\cF(\xi)$-\/modules other than listed. Second, we do not consider
operators $\bigotimes_{i=1}^k\varkappa(\xi)\otimes\bigotimes_{j=1}^\ell
\hat{\varkappa}(\xi)\bigr|_{w=w[u]}\to\varkappa(\pi)$ whenever $k+\ell>1$.
This would generalize \S2 of this definition to the same extent
as the Schouten bracket~$\lshad\,,\,\rshad$ incorporates the commutator
$[\,,\,]$ and the Gerstenhaber bracket $\lshad\,,\,\rshad_{\cP}$ contains
the Koszul bracket~$[\,,\,]_{\cP}$. However, the full generality
must not be discarded (e.g., when the isomorphism
$\gf\simeq\varkappa(\pi)$ is \emph{not} used for Noether operators~$A$, see
Remark~\ref{Unfortunate}). Finally, note that, when allowing nonlinear
differential reperametrizations $\tilde{F}=\tilde{F}[F]\in\gf$ of
equations, we arrive at the theory of \emph{nonlinear} Frobenius
Noether operators.}
\begin{description}
\item[$\gf=\varkappa(\xi)\bigr|_{w=w[u]}$]
Frobenius operators of first kind are
\begin{equation}\label{Frob1}
A\colon\varkappa(\xi)\bigr|_{w}\to\varkappa(\pi).
\end{equation}
Under any diffeomorphisms $\tilde{u}=\tilde{u}[u]\colon
J^\infty(\pi)\to\Gamma(\pi)$ and $\tilde{w}=\tilde{w}[w]\colon
J^\infty(\xi)\to\Gamma(\xi)$, 
these operators are transformed according to
\begin{equation}\label{FrobKK}
A\mapsto\tilde{A}=\ell_{\tilde{u}}^{(u)}\circ A\circ\ell_w^{(\tilde{w})}
  \Bigr|_{\substack{w=w[u]\\u=u[\tilde{u}]}}.
\end{equation}
\item[$\gf=\hat{\varkappa}(\xi)\bigr|_{w=w[u]}$]
Frobenius operators of second kind are linear mappings
\begin{equation}\label{Frob2}
A\colon\hat{\varkappa}(\xi)\bigr|_{w}\to\varkappa(\pi).
\end{equation}
For any differential changes of coordinates $\tilde{u}=\tilde{u}[u]$ and
$\tilde{w}=\tilde{w}[w]$, these operators obey
\begin{equation}\label{FrobAK}
A\mapsto\tilde{A}=\ell_{\tilde{u}}^{(u)}\circ A\circ
 \bigl(\ell_{\tilde{w}}^{(w)}\bigr)^*
  \Bigr|_{\substack{w=w[u]\\u=u[\tilde{u}]}}.
\end{equation}
\end{description}
Sections $\psi$ that constitute the domains~$\Omega^1(\xi_\pi)$ of Frobenius
operators~\eqref{Frob1} of first kind are transformed as vectors~$\cEv_\psi=
\psi\cdot\dd/\dd w+\cdots$; in the case~\eqref{Frob2}, the sections~$\psi$
satisfy the rules valid for variational covectors~$\psi=\delta(\cdot)/\delta w$.

If Frobenius operator~$A$ is a recursion $\varkappa(\pi)\to\varkappa(\pi)$,
then $w=\text{id}$, and the transformations of the domain and image
of~$A$ are uniquely correlated. If $A$~is a Noether operator
$\hat{\varkappa}(\pi)\to\varkappa(\pi)$, then $w$~determines the system
$\cE=\{w[u]=0\}$ of differential equations, and, by default, we use the
misleading isomorphism $\gf\simeq\varkappa(\pi)$ for evolution equations
(see Remark~\ref{Unfortunate} on p.~\pageref{Unfortunate}).
Otherwise, there may be no constraint between transformations
of the domains and images.
In particular, the gauge group of~$\gf$ can be trivial, meaning that
transformations of $w$ in $\xi$ are still prohibited, as in~\S1:
\begin{description}
\item[$\gf=\Gamma\bigl(\pi_\infty^*(\xi)\bigr)$] These Frobenius operators
$A\colon\Gamma\bigl(\pi_\infty^*(\xi)\bigr)\to\varkappa(\pi)$ are transformed
by $A\mapsto\tilde{A}=\ell_{\tilde{u}}^{(u)}\circ A\bigr|_{u=u[\tilde{u}]}$
under $\tilde{u}=\tilde{u}[u]$.
\end{description}
Note that no substitution $w[u]$ is needed in the third (degenerate) case.
\end{define}

\begin{rem}[On the ``Frobenius theorem'']\label{RemIntegralManifolds}
Frobenius operators~$A\colon\gf\to\varkappa(\pi)$ specify involutive
distributions of vertical symmetries
$\cEv_{A(\cdot)}\in\ID^v(J^\infty(\pi))$ of the 
Cartan distribution~$\cC=\langle\widehat{\dd/\dd x^i} 
\rangle\subset\ID(J^\infty(\pi))$, see~\eqref{CartanConnection},
which is itself Frobenius.

Assume that for such an involutive distribution
$\langle\cEv_{A(\cdot)}\rangle$ there is an integral manifold (typically,
it would be infinite\/-\/dimensional), and suppose further that it is a
differential equation~$\cE$. (Note that by an equation we mean the
infinite prolongation, which does not always exist.) The definition of Frobenius
operators implies that $\cE$ admits infinitely many symmetries $\vph=A(\phi)$
which contain free functional parameters~$\phi\in\gf$. This property is close
but not equivalent to Definition~\ref{DefLiouType} of Liouville\/-\/type
differential equations (see p.~\pageref{DefLiouType} and
Remark~\ref{RemLiouVsSym} that follows).
\end{rem}

\enlargethispage{1.3\baselineskip}\begin{lemma}
The bracket $[\,,\,]_A$ is $\BBR$-\/bilinear,
skew\/-\/symmetric, and transfers the Jacobi identity\footnote{In what follows,
the nondegeneracy condition~\eqref{NoKernelsIntersect} specifies the case
when the Jacobiator $J(\psi_1,\psi_2,\psi_3)$ for $[\,,\,]_A$
vanishes identically, not being a nonzero element of $\ker A$
as in~\cite{MikhAssoc}.}
from the Lie algebra $\gothg(\pi)$ of evolutionary vector fields
to~$\Omega^1(\xi_\pi)$, see~\eqref{JacobiKoszul}.
The kernel of~$A$ is an ideal in~$\gf$. The image
of a Frobenius operator may not be an ideal in the space
of evolutionary derivations although it is a Lie subalgebra by definition.
\end{lemma}

\begin{rem}\label{RemNotCocycle}
The bi\/-\/differential skew\/-\/symmetric bracket $\{\!\{\,,\,\}\!\}_A\in
\CDiff 
\bigl(\bigwedge^2\Omega^1(\xi_\pi),\Omega^1(\xi_\pi)\bigr)$
does not generally satisfy the Jacobi identity. Indeed, for the Koszul
bracket~$[\,,\,]_A$ we have
\begin{align}
0&=\sum_\circlearrowright \bigl[ [\psi_1,\psi_2]_A,\psi_3\bigr]_A =
 \sum_\circlearrowright
  \bigl[ \cEv_{A(\psi_1)}(\psi_2)-\cEv_{A(\psi_2)}(\psi_1)
  +\ib{\psi_1}{\psi_2}{A}, \psi_3\bigr]_A \notag\\
{}&=\sum_\circlearrowright\Bigl\{
\cEv_{A\bigl(\cEv_{A(\psi_1)}(\psi_2)-\cEv_{A(\psi_2)}(\psi_1)\bigr)}(\psi_3)
-\cEv_{\psi_3}\bigl(\cEv_{A(\psi_1)}(\psi_2)-\cEv_{A(\psi_2)}(\psi_1)\bigr)
  \notag \\
{}&\qquad{}
+\ib{\cEv_{A(\psi_1)}(\psi_2)-\cEv_{A(\psi_2)}(\psi_1)}{\psi_3}{A}\notag\\
{}&\qquad{}
 +\cEv_{A\bigl(\ib{\psi_1}{\psi_2}{A}\bigr)}(\psi_3)
 -\underline{\cEv_{A(\psi_3)}\bigl(\ib{\psi_1}{\psi_2}{A}\bigr)}
 +\ib{\ib{\psi_1}{\psi_2}{A}}{\psi_3}{A}\Bigr\}. \label{JacobiKoszul}
\end{align}
The underlined summand with a derivation of the coefficients of
$\ib{\psi_1}{\psi_2}{A}$, which belong to $\cF(\pi)$, may not
vanish on the $\cF(\xi)$-\/module $\gf$, 
when the action of $\cEv_{A(\cdot)}$ onto the basis of sections
$\psi_i$ is set to zero
(see Remark~\ref{RemLstar} on p.~\pageref{RemLstar}). Note that the Jacobi
identity for $[\,,\,]_A$ then amounts to the last line
of~\eqref{JacobiKoszul}, which equals $\sum_\circlearrowright
\bigl[\ib{\psi_1}{\psi_2}{A},\psi_3\bigr]_A=0$
and which is a half of the equation for infinitesimal deformations
of~$[\,,\,]_A$.
\end{rem}

\begin{rem}
In this paper, we study only those operators whose images in~$\varkappa(\pi)$
are closed w.r.t.\ the standard bracket~\eqref{EvBracket}.
Suppose, however, that an endomorphism $R\in\End_\BBR\varkappa(\pi)$ is Frobenius and hence induces a new Lie algebra structure $[\,,\,]_R$ on
the quotient $\varkappa(\pi)/\ker R$ of its domain.
Contrary to Proposition~\ref{Criterion}, the images of
standard Hamiltonian operators are generally not closed w.r.t.\ the
Koszul brackets~$[\,,\,]_R$ induced by Frobenius recursions~$R$.

Although one could repeat the whole construction for operators
$\Omega^1(\xi_\pi)\to\varkappa(\pi)$ whose images are closed
w.r.t.~$[\,,\,]_R$, this does not produce a new formalism.
Indeed, the operators~$A$, whose images in~$\varkappa(\pi)$ are closed
w.r.t.~$[\,,\,]_R$, generate standard Frobenius operators~$R\circ A$,
and \emph{vice versa}.
At the same time, the new formalism appears for operators~$A$ whose
images in~$\varkappa(\pi)$ are closed w.r.t.\ the $\mathbf{r}$-\/brackets
\eqref{rBracket} 
given by $\mathbf{r}$-\/matrices for the Lie algebra~$\gothg(\pi)$.
\end{rem}\enlargethispage{\baselineskip}

\begin{rem}
Specifying the Koszul bracket~$[\,,\,]_A$ on~$\Omega^1(\xi_\pi)$ by
a Frobenius operator~$A$, we define Frobenius generalizations
$\omega\colon\varkappa(\pi)\to\Omega^1(\xi_\pi)$ of
\emph{symplectic} structures $\varkappa(\pi)\to
\hat{\varkappa}(\pi)$. To this end, we require that the
images of $\omega$ are closed w.r.t.\ $[\,,\,]_A$.
Since all modules are already known, the adaptation of \S2 of
Definition~\ref{DefFrob} for~$\omega$ is obvious.
\end{rem}

\subsubsection{Flat connection}
Let $A\colon\gf\to\varkappa(\pi)$
be a Frobenius operator. It provides the homomorphism of Lie algebras
\begin{equation}\label{LieAlgHom}
A\colon\bigl(\Omega^1(\xi_\pi),[\,,\,]_A\bigr)\to
\bigl(\gothg(\pi),[\,,\,]\bigr).
\end{equation}
Let $K$ be a $\gothg(\pi)$-\/module; typically, $K=\cF(\pi)$ or
any other horizontal $\cF(\pi)$-\/module (e.g., $\gothg(\pi)$ itself).
By the homomorphism~$A$,
the $\gothg(\pi)$-\/module $K$~is an $\Omega^1(\xi_\pi)$-\/module as well.
The Jacobi identity implies that the adjoint action by an element of Lie algebra $\Omega^1(\xi_\pi)$ is a derivation.
We further impose the condition of semi\/-\/simplicity:
we claim that
\[ 
[\psi_1,\cdot]_A=[\psi_2,\cdot]_A\quad\Longrightarrow\quad\psi_1=\psi_2,
\]
for equivalence classes $\psi_1,\psi_2\in\Omega^1(\xi_\pi)=\gf/\ker A$.

Now we define a connection $\nabla^A$ in the triple~\eqref{LieAlgHom},
\[
\nabla^A\colon\Der_{\text{Int}}\bigl(\Omega^1(\xi_\pi),\gothg(\pi)\bigr)\to
   \Der\bigl(\gothg(\pi),K\bigr).
\]
The construction is analogous to the definition of connections in triples
$(\cA,\cB,\imath)$, where $\cA$~is a $\Bbbk$-\/algebra, $\cB$~is
an $\cA$-\/algebra, and $\imath\colon\cA\to\cB$ is an algebra
homomorphism~\cite{JKConnection}.
In its turn, this generalizes the connections in fibre bundles
$E^{n+m}\xrightarrow{\pi}M^n$, when $\cA=C^\infty(M^n)$,
$\cB=C^\infty(E^{n+m})$, and $\imath\colon\cA\hookrightarrow\cB$. We set
\begin{equation}\label{Connection}
\nabla^A\colon A\circ[\psi,\cdot]_A\mapsto[A(\psi),\cdot].
\end{equation}
The above definition is $\Omega^1(\xi_\pi)$-\/linear. Indeed, for a derivation
$\Delta=[\psi,\cdot]_A$ we have that
\begin{equation}\label{IsConnect}
\nabla^A_{f\times\Delta}=A(f)\times\nabla^A_\Delta,\qquad
f\in\Omega^1(\xi_\pi),\quad
\Delta\in\Der_{\text{Int}}\bigl(\Omega^1(\xi_\pi)\bigr),
\end{equation}
where the multiplication ${\times}$ by~$f$ and by its image under~$A$ is
the adjoint action. Note that the right\/-\/hand side of the analogue
of~\eqref{IsConnect} in a classical definition of
$C^\infty(M^n)$-\/linearity of connections in fibre bundles~$\pi$ does
contain the image~$\imath(f)$ of the identical embedding
$\cA\hookrightarrow\cB$ and not $f\in\cA$~itself.

\begin{state}
Connection~\eqref{Connection} is flat:
\begin{equation}\label{FlatnessCondition}
\Bigl(\nabla^{A}_{p}\circ\nabla^{A}_{q}
 -\nabla^{A}_{q}\circ\nabla^{A}_{p}
 -\nabla^A_{[p,q]_A}\Bigr)\bigl(\vph\bigr)=0,
\quad\forall p,q\in\Omega^1(\xi_\pi),\quad\vph\in\gothg(\pi).
\end{equation}
\end{state}

\begin{proof}
The Jacobi identity for the
bracket~\eqref{EvBracket} of generating sections of evolutionary vector fields,
\[
\bigl[A(p),\bigl[A(q),\vph\bigr]\bigr]+
\bigl[A(q),\bigl[\vph,A(p)\bigr]\bigr]+
\bigl[\vph,\bigl[A(p),A(q)\bigr]\bigr]=0
\]
is the flatness condition~\eqref{FlatnessCondition}.
\end{proof}

\begin{cor}
If the flows of a commutative hierarchy~$\gA$ belong to
the image of a Frobenius operator~$A$, then the hierarchy is a geodesic
w.r.t.\ connection~\eqref{Connection}.
\end{cor}
\begin{proof}
Indeed, for any curve $\psi(\tau)\colon\BBR\to\Omega^1(\gA)$ located in the
inverse image of~$\gA$ under~$A$,
the covariant derivative $\nabla^A_{\psi(\tau)}A(\psi'(\tau))$
of the velocity $\psi'(\tau)$ vanishes along the curve.
\end{proof}

\begin{rem}
The Frobenius operators~\eqref{LieAlgHom} induce a morphism
$\Omega^1(\xi_\pi)\to\bigwedge^\bullet\gothg(\pi)$ to the Schouten algebra
of evolutionary polyvector fields, which is endowed with the
Schouten bracket~\eqref{SchoutenBr}. Flat connections~\eqref{Connection}
in triples~\eqref{LieAlgHom} are naturally extended to connections
in $\bigl(\Omega^1(\xi_\pi),\bigwedge^{\bullet}\gothg(\pi),A\bigr)$,
which are flat in the graded sense.
\end{rem}

\begin{rem}
The notion of connections for the triples (which generalize
$\cA=C^\infty(M^n)$, $\cB=C^\infty(E^{n+m})$, and
$\imath\colon\cA\hookrightarrow\cB$)
was proposed by Kra\-sil'\-shchik in~\cite{JKConnection}. It leads to
the Cartan cohomology, see~\cite[Ch.~5]{Opava}, which allows
to interpret symmetries and recursion operators as equivalence classes.
Therefore we expect to encounter some cohomology for the
connection~\eqref{Connection} in the triples~\eqref{LieAlgHom}.

\enlargethispage{1.3\baselineskip}
Of course, connection~\eqref{Connection} in the triples~\eqref{LieAlgHom}
is not the Cartan connection~\eqref{CartanConnection} on~$J^\infty(\pi)$.
This is readily seen from the fact that everything at hand
is $\pi$-\/vertical and is projected to 
zero vector fields on~$TM^n$ under $\pi_{\infty,*}$. (Everything would be
projected to \emph{the} point~$x_0$
if the jet bundle were a finite\/-\/dimensional manifold $N^m$ for
$\pi\colon N^m\to\{x_0\}$, see~\eqref{DxStraight}.)
Therefore, instead of a dual description of the $n$-\/dimensional distribution
$\cC\subset\ID(J^\infty(\pi))$ by Cartan $1$-\/forms, we face the problem
of dual representation of the involutive 
distributions~$\langle\cEv_{A(\cdot)}\rangle$.
\end{rem}

\subsubsection{Frobenius complex}
Now, using Cartan's formula, we construct a differential complex on
the Chevalley cohomology
$\Hom_\BBR\bigl(\bigwedge^k\Omega^1(\xi_\pi),\gothg(\pi)\bigr)$ with values
in an $\Omega^1(\xi_\pi)$-\/module~$\gothg(\pi)$. This is the Frobenius
complex assigned to an operator~$A\colon\Omega^1(\xi_\pi)\to\gothg(\pi)$.

For any~$k\geq0$ and $\omega_k\in\Hom_\BBR\bigl(\bigwedge^k\Omega^1(\xi_\pi),
\gothg(\pi)\bigr)$, define the differential
$\mathbf{d}\colon\omega_k\mapsto\mathbf{d}\omega_k$ by setting
\begin{multline*}
\mathbf{d}\omega_k\bigl(\psi_0,\ldots,\psi_k\bigr)=
\sum_i(-1)^i\nabla^A_{\psi_i}\omega\bigl(\psi_0,\ldots,\widehat{\psi_i},\ldots,\psi_k\bigr)\\
 {}+\sum_{i<j}(-1)^{i+j-1}\omega_k\bigl([\psi_i,\psi_j]_A,\psi_0,\ldots,
    \widehat{\psi_i},\ldots,\widehat{\psi_j},\ldots,\psi_k\bigr).
\end{multline*}
Hence we obtain the complex
\begin{multline}
\gothg(\pi)\xrightarrow{\text{const}}
\Hom_\BBR\bigl(\Omega^1(\xi_\pi),\gothg(\pi)\bigr)\xrightarrow{A\circ[\,,\,]_A}
\Hom_\BBR\bigl(\bigwedge\nolimits^2\Omega^1(\xi_\pi),\gothg(\pi)\bigr)\\
{}\xrightarrow{\text{J}}
\Hom_\BBR\bigl(\bigwedge\nolimits^3\Omega^1(\xi_\pi),\gothg(\pi)\bigr)
  \to\cdots.\label{FrobComplex}
\end{multline}
The first inclusion in~\eqref{FrobComplex} consists of the commutations
$\nabla^A_{(\cdot)}\vph_0$ with fixed elements $\vph_0\in\gothg(\pi)$,
the second arrow is the composition of the Koszul
bracket~$[\,,\,]_A$ with~$A$, and the third arrow calculates the
right\/-\/hand side of the Jacobi identity~\eqref{JacobiKoszul}.
The Poisson complex is a special case of~\eqref{FrobComplex}, extending it from the left with the $\cF(\pi)$-\/module of $n$-th horizontal
cohomology~$\bar{H}^n(\pi)$ for~$J^\infty(\pi)$.
Again, we assume the semi\/-\/simplicity: no elements
$\vph_0\in\gothg(\pi)$ commute with the entire Lie algebra~$\gothg(\pi)$.

The homological vector field representations (e.g., see~\eqref{dh})
for the differential~$\mathbf{d}$ allow to encode the Lie algebroids
$\Gamma(\xi_\pi)\to\varkappa(\pi)$ over infinite jet spaces~$J^\infty(\pi)$
in terms of~\eqref{HomVF}. 



\subsection{Examples of Frobenius operators}
Throughout this paper, we consider only the commutative case when
the base manifolds~$M^n$ are not $\BBZ_2$-\/graded and all
coordinates on the jet bundles $J^\infty(\pi)$ are permutable.
Also, we confine ourselves to the local setting and consider Frobenius matrix
differential operators which are polynomial in total derivatives.
However, the search for nonlocal Frobenius
(super-)\/operators can be performed using standard techniques,
see Remark~\ref{RemLstar} on p.~\pageref{RemLstar}.


Let us have some examples of Frobenius operators of second kind
and of Sokolov's brackets on their domains.
We start with Hamiltonian operators, which are the most well studied examples
of Frobenius structures.

\enlargethispage{\baselineskip}
\begin{example}\label{RemDogma} 
Every Hamiltonian operator is Frobenius.
Indeed, the criterion in Proposition~\ref{Criterion}
gives formula~\eqref{EqDogma} for the bracket $\ib{\,}{\,}{A}$
on the domains~$\hat{\varkappa}(\pi)$ of nondegenerate Hamiltonian
operators~$A$, which obey the law~\eqref{ChangeCoordHamOp} under a change of
coordinates.
This demonstrates that the formalism of Frobenius operators is a true
generalization of the Hamiltonian approach to nonlinear
evolutionary~PDE. The use of Frobenius operators can be a
helpful intermediate step in the search for and classification of the
Hamiltonian structures. An advantage of this approach is that it is
easier to solve first equation~\eqref{EqOplusB} w.r.t.\ operators~$A$
and filter out skew\/-\/adjoint solutions rather than to solve the
Jacobi identity in the form of~\eqref{EqDogma}. Thus, by
Proposition~\ref{Criterion}, a nondegenerate skew\/-\/adjoint Frobenius
operator~$A$ is Hamiltonian iff the Sokolov bracket
$\{\!\{\,,\,\}\!\}_A$ is equal to the r.h.s.\ of~\eqref{EqDogma}
up to~$\ker A$.
Also, we note that we do not require a restriction
of the operators $A$ onto differential equations~$\cE$
such that $A\colon\cosym\cE\to\sym\cE$, unlike in~\cite{Lstar}.
\end{example}

\begin{rem}\label{RemFalqui}
A recent version~\cite{Getzler,Falqui} of the Darboux theorem for
$(1+1)$D evolutionary systems implies that Hamiltonian operators~$A$ for
non\/-\/exceptional systems can be transformed to const${}\cdot D_x$.
The bracket $\{\!\{\,,\,\}\!\}_A$ is then zero, which is readily seen
from~\eqref{EqDogma}.
Hence the actual statement of the Darboux theorem for PDE is that the
bracket $\{\!\{\,,\,\}\!\}_A$ can be trivialized for such Hamiltonian operators.
\end{rem}

\begin{rem}\label{RemManyOps}
One bracket~$\ib{\,}{\,}{A}$ can correspond to several
operators~$A$ that satisfy~\eqref{EqDefFrob}. For example,
the second structure $A_2^{\text{KdV}}=-\tfrac{1}{2}D_x^3+2wD_x+w_x$
for the KdV equation~\eqref{IKdV}
determines the bracket $\ib{p}{q}{A_2^{\text{KdV}}}=pq_x-p_xq$, which
is also induced by the operators~\eqref{KdVWeightOps}, 
see Example~\ref{ExSohr} below,
and by $\smash{A_1^{(2)}}=D_x\circ(D_x+u)$, see Example~\ref{ExSanders}.
Actually, this Wronskian\/-\/based bracket (c.f.~\cite{ForKac})
is scattered through the text, see Example~\ref{ExHeav} on p.~\pageref{ExHeav}.
Hence there are fewer brackets $\ib{\,}{\,}{A}$ than there are Frobenius operators~$A$.
\end{rem}

Formula~\eqref{EqDogma} does not remain valid for arbitrary
Frobenius operators, which are generally 
non\/-\/skew\/-\/adjoint.
In appendix~\ref{AppLiouEBracket} we describe an inductive procedure that
reconstructs the bracket $\ib{\cdot}{\cdot}{A}$ for nondegenerate
operators~$A$, see~\eqref{NoKernelsIntersect}.
From now on, we consider only nondegenerate Frobenius operators.

\begin{example} 
Noether operators~\eqref{A12dBous} for the $3$-\/component dispersionless
Boussinesq\/-\/type system~\eqref{d-B} are Frobenius.
For the Frobenius operators $A_1,A_2\colon\cosym\cE\to\sym\cE$,
the transformation formulas~\eqref{ChangeCoordHamOp} classify them to the second
kind; the substitutions $J^\infty(\pi)\to\Gamma(\xi)$ are the identity
mappings. The same is obvious for Hamiltonian
operators $\hat{A}_0$ and $\hat{A}_2$ for this system.
However, unlike $\lshad\,,\,\rshad_{A_1}=0$ and
$\lshad\,,\,\rshad_{A_2}=\lshad\,,\,\rshad_{\hat{A}_2}$,
see~\eqref{TwoGrads} and Remark~\ref{BrLawsJacobi},
the components of Sokolov's brackets for $A_1$ and $A_2$ are obtained
using the inductive algorithm, and the result is~\eqref{SokBrABoth} on
p.~\pageref{SokBrABoth}. Bi\/-\/differential representations
for components of these brackets are formulated in appendix~\ref{AppBiDiffA12}.
\end{example}

In section~\ref{SecLiou} we shall consider a class of 
Frobenius operators $\square\colon\cosym\cE_1\to\sym\cE_2$ of second kind
that map (co-)\/tangent bundles for two hierachies of evolutionary systems
related by the substitution $w\colon\cE_2\to\cE_1$. 

Now we list several operators which are known in a fixed system of local
coordinates and whose images are then closed under commutation.
The transformation rules for their domains are unknown, although it may occur
that these reparametrizations are uniquely determined by a change of coordinates
in the images.

\begin{example}[KdV scaling weights]\label{ExSohr}
Let us fix the weights $|u|=2$, $|D_x|=1$ that originate from the scaling invariance of the KdV equation $u_t=-\tfrac{1}{2}u_{xxx}+3uu_x$,
see also~\eqref{IKdV}; we have that $|D_t|=3$.
Using the method of undetermined coefficients, we performed the search for scalar
Frobenius operators that are homogeneous w.r.t.\ the weights not
greater than~$7$. We obtained two compatible Hamiltonian operators
$A_1^{\text{KdV}}=D_x$ and $A_2^{\text{KdV}}=-\tfrac{1}{2}D_x^3+2u\,D_x+u_x$,
the generalizations $D_x^{2n+1}$ of~$D_x$, and the 
operator
\[ 
uu_{xxx}+3uu_{xx}\,D_x+3uu_x\,D_x^2+u^2\,D_x^3.
\]
Also, there are four non\/-\/skew\/-\/adjoint operators that
satisfy~\eqref{EqDefFrob},
\begin{align*}
A_4^{(6)}&=u^3-u_x^2,\qquad A_5^{(6)}=2u_x^2-uu_{xx}-2uu_x\,D_x+u^2\,D_x^2,\\
  \{\!\{p,q\}\!\}_{A_4^{(6)}}&=2u_x\cdot (p q_x-p_x q),\qquad
\{\!\{p,q\}\!\}_{A_5^{(6)}}=-2u_x\cdot (p q_x-p_x q)+u\cdot (p q_{xx}-p_{xx} q);
\end{align*}\begin{align*}
A_8^{(7)}&=u_x^2\,D_x-2uu_{xx}\,D_x-4uu_xD_x^2-4u^2\,D_x^3,\qquad
  \{\!\{p,q\}\!\}_{A_8^{(7)}}=u^2\cdot (p q_x-p_x q);\\
A_9^{(7)}&=-2u_xu_{xx}-u_x^2\,D_x,\qquad
\{\!\{p,q\}\!\}_{A_9^{(7)}}=8u_{xx}\cdot (p q_x-p_x q)+
2u_x\cdot (p q_{xx}-p_{xx} q).
\end{align*}
Finally, we have found the operators that contain arbitrahry functions:
$f(u)D_x^n$ and $f(u)u^2$ with vanishing brackets, and also we have got
\begin{align}
A_3&=f(u)u_x,&
   \{\!\{p,q\}\!\}_{A_3}&=f(u)\bigl(p_xq-pq_x\bigr);\notag\\
A_4&=f(u)\bigl(u\,D_x-u_x\bigr),&
   \{\!\{p,q\}\!\}_{A_4}&=f(u)\bigl(p q_x-p_x q\bigr).
\label{KdVWeightOps}
\end{align}
Formula~\eqref{EqDogma} is not valid for any of these
non\/-\/skew\/-\/adjoint operators.
\end{example}

\begin{example}
\label{Ex1st1st}
There is a three\/-\/parametric family of scalar first\/-\/order
Frobenius operators
$A=a(u,u_x)\,D_x+b(u,u_x)$ with brackets
$\ib{p}{q}{A}=c(u,u_x)\cdot(pq_x-p_xq)$. They are given by
\begin{equation}\label{pHHydro}
a=\gamma(u),\quad
b=\beta(u)\cdot u_x+\alpha\cdot\gamma(u),\quad
c=-\beta(u),
\end{equation}
where $\alpha$ is a constant and the functions $\beta$, $\gamma$ are
arbitrary. 
\end{example}

\subsection{Frobenius recursion operators}\label{SecFrobRec}
Within this section, we set $\xi\mathrel{{:}{=}}\pi$ and $w=u$. Thus we
consider Frobenius recursion operators $R\colon\varkappa(\pi)\to\varkappa(\pi)$
whose images are closed w.r.t.~\eqref{EvBracket}.
The standard Lie algebra structure $[\,,\,]$ is transferred to $[\,,\,]_R$ on
the quotients $\Omega^1(\pi_\pi)=\varkappa(\pi)/\ker R$
by the classical Yang\/--\/Baxter equation for~$\gothg(\pi)$,
\begin{equation}\label{FrobRec}
[R\vph_1,R\vph_2]=R\bigl([\vph_1,\vph_2]_R\bigr),\qquad
\vph_1,\vph_2\in\varkappa(\pi)/\ker R.
\end{equation}
For $R\neq\text{id}$, the new Lie bracket $[\,,\,]_R$ is
different from the original commutation 
of evolutionary vector fields in the image.

\begin{rem}\label{RemYB}
Frobenius recursion operators provide higher differential order
solutions of the classical Yang\/--\/Baxter equation for
the algebra~$\gothg(\pi)$. The parallel with the zero\/-\/order
theory of $\mathbf{r}$-\/matrices is as follows.
Let $\gothg$ be a Lie algebra with the Lie bracket $[\,,\,]$.
The classical $\bR$-\/matrix~\cite{Blaszak,Kassel,Reyman} is
a linear map $\bR\colon\gothg\to\gothg$ that endows $\gothg/\ker\bR$
with the second Lie product~$[\,,\,]_{\bR}$; for any $a,b\in\gothg$
set
\begin{equation}\label{rBracket}
[a,b]_{\bR} = [\bR a,b] + [a,\bR b].
\end{equation}
A sufficient condition for
an operator~$\bR$ of differential order zero to be an $\bR$-\/matrix is that
$\bR$~satisfies the Yang\/--\/Baxter equation YB$(\alpha)$:
\begin{equation}\tag{\ref{FrobRec}${}'$}\label{RKoszulBaxter}
[\bR a,\bR b] = \bR\bigl([a,b]_{\bR}\bigr) - \alpha[a,b],\qquad a,b\in\gothg,
\quad \alpha=0\text{ or }1.
\end{equation}
Now let a recursion differential operator $R\in\End_\BBR\varkappa(\pi)$
be Frobenius, then the Koszul
bracket~\eqref{EqOplusBKoszul} satisfies the Yang\/--\/Baxter
equation~\eqref{FrobRec} with~$\alpha=0$.

We conclude that, for 
Lie algebra~$\gothg(\pi)$ of evolutionary vector fields
on jet spaces~$J^\infty(\pi)$, the classical Yang\/--\/Baxter
equation~\eqref{FrobRec} admits solutions $\bigl(R,[\,,\,]_R\bigr)$ of
form~\eqref{EqOplusBKoszul} other than the standard
$\bR$-\/brackets~\eqref{rBracket}.
The two notions of $\bR$-\/matrices and Frobenius recursions $R\in\End_\BBR
\varkappa(\pi)$ as solutions of equation~\eqref{FrobRec} are very close.
However, the distinction between them is expressed by the respective brackets
$[\,,\,]_{\bR}$ and $[\,,\,]_R$ on their domain~$\varkappa(\pi)$:
the Koszul bracket~\eqref{EqOplusBKoszul} is calculated and the $\bR$-\/bracket
$[\,,\,]_{\bR}$ is postulated. This explains why \S1~of Definition~\ref{DefFrob}
incorporates the form of the Koszul bracket. 
\end{rem}

\begin{example}[Projector solutions]\label{ProjSolut}
Instead of the entire jet space~$J^\infty(\pi)$, consider
a Darboux\/-\/integrable Liouville\/-\/type hyperbolic
system~\cite{Shabat,SokolovUMN}
\[
\cE_\IL=\{u^1_{xy}=f^1(x,y;[u]),\ldots,u^m_{xy}=f^m(x,y;[u])\}
  \subset J^\infty(\pi),
\]
see Definition~\ref{DefLiouType} on p.~\pageref{DefLiouType} and a comment on
it in Remark~\ref{Vessiot}.
Essentially, $\cE_\IL\simeq\bigl\{J^\infty(\pi_x)\oplus J^\infty(\pi_y)\bigr/{\sim}$,
plus the rules for calculating $u_{x\ldots xy\ldots y}\bigr\}$,
where $\pi_x,\pi_y\colon\BBR^m\times\BBR\to\BBR$ are trivial
bundles over~$\BBR$ with the base coordinates~$x$ and~$y$, respectively,
and $\sim$~glues together their fibres with coordinates~$(u^1,\ldots,u^m)$.

In this notation, the symmetry algebra $\sym\cE_\IL$ is decomposed along the
$x$-{} and $y$-characteristics such that any $\vph\in\sym\cE_\IL$ is of
the form~\eqref{SymForLiou}, see p.~\pageref{SymForLiou},
\[
\vph=\square(\phi)+\bar{\square}(\bar{\phi}),
\]
where sections $\phi\in\hat{\varkappa}(\xi_x)\bigr|_{w}$ and
$\bar{\phi}\in\hat{\varkappa}(\xi_y)\bigr|_{\bar{w}}$ belong to certain
restrictions of the modules 
and the operators $\square$ and $\bar{\square}$ are Frobenius.
This is discussed in section~\ref{SecLiou} in detail.

Without loss of generality (we consider the $x$-\/components
of all structures, but one could replace $x\leftrightarrow y$
in what follows), set
\begin{equation}\label{ProjOnX}
R\colon\cEv_{\square(\phi)+\bar{\square}(\bar{\phi})}\mapsto
\cEv_{\square(\phi)}\in\sym\cE_\IL.
\end{equation}
The Liouville\/-\/type systems are so special that $R$~is
a recursion on~$\sym\cE_\IL$,
although $\ker R$ is very big under this projection. The image of~$R$ is closed
w.r.t.\ the restriction of bracket~\eqref{EvBracket} onto~$\cE_\IL$. Thus we obtain
an example of a Frobenius recursion operator.

Surprisingly, we can apply here the algebraic construction of Poisson
brackets, which was
introduced in~\cite{KuperCotangent}, to such projector recursions~$R$.
We recall that a homomorphism $\Pi\in\Hom_\ga(V,\End_\ga V)$ is
\emph{Hamiltonian} for a commutative $\Bbbk$-\/algebra~$\ga$ and an
$\ga$-\/module~$V$ if the \emph{Poisson bracket}
$\{v_1,v_2\}_\Pi\mathrel{{:}{=}}\bigl(\Pi(v_1)\bigr)(v_2)$
of $v_1,v_2\in V$
satisfies the commutation closure condition $\Pi\bigl(\{v_1,v_2\}_\Pi\bigr)
=\bigl[\Pi(v_1),\Pi(v_2)\bigr]$. The latter reduces to~\eqref{KuperDefPoisson}
if we choose $\Bbbk=\BBR$, $\ga=\cF(\pi)$, $V=\bar{H}^n(\pi)$, and
$\Pi=\cEv_{A(\bE(\cdot))}$.

In our case, we have $\ga=\BBR$ and~$V=\sym\cE_\IL$. Define
$\Pi\in\Hom_\BBR\bigr(\sym\cE_\IL,\End_\BBR\sym\cE_\IL\bigr)$ by
\[
\Pi(\vph)=\ad_{R(\vph)}\circ R,\qquad \vph\in\sym\cE_\IL,
\]
whence we obtain $\{\vph_1,\vph_2\}_\Pi=\bigl[R(\vph_1),R(\vph_2)\bigr]$ and
\[
\bigl[\bigl[R(\vph_1),R(\vph_2)\bigr],R(\vph_3)\bigr]=
\bigl[R(\vph_1),R\bigl(\bigl[R(\vph_2),R(\vph_3)\bigr]\bigr)\bigr]-
\bigl[R(\vph_2),R\bigl(\bigl[R(\vph_1),R(\vph_3)\bigr]\bigr)\bigr]
\]
for any $\vph_1,\vph_2,\vph_3\in\sym\cE_\IL$. Clearly, the above equality holds
by virtue of the Jacobi identity, since $[\text{im}\,R,\text{im}\,R]\subseteq
\text{im}\,R$ and~$R^2=R$.

This example describes a rare situation when a Poisson structure is
introduced on~$\sym\cE_\IL$ and not on~$\bar{H}^{n}(\pi)$.
The price we pay for this is that
the underlying algebra $\ga=\BBR$ and an $\BBR$-\/module structure of~$\sym\cE_\IL$
are very poor in comparison with $\cF(\pi)$ for~$\bar{H}^{n}(\pi)$, respectively.
\end{example}

Suppose a Frobenius operator is divisible by another Frobenius operator;
then their Koszul brackets are correlated as follows.

\begin{state}[The chain rule, c.f.~\cite{Sanders}]\label{ChainRule}
Suppose that the Frobenius operator~$A$ is nondegenerate,
see~\eqref{NoKernelsIntersect}, and the image of $A'= A\circ\Delta$ is
closed w.r.t.\ the commutation.
Then the brackets ${\{\!\{\,,\,\}\!\}}_{A}$
and ${\{\!\{\,,\,\}\!\}}_{A'}$ are related by the formula
\begin{equation}\label{ChainRuleFormula}
\Delta ({\{\!\{\xi_1,\xi_2\}\!\}}_{A\circ\Delta})=
\cEv_{A(\Delta (\xi_1))}(\Delta )(\xi_2)-\cEv_{A(\Delta (\xi_2))}(\Delta )(\xi_1)
+{\{\!\{\Delta (\xi_1),\Delta (\xi_2)\}\!\}}_{A}
\end{equation}
for any sections $\xi_1,\xi_2\in\gf\subseteq\Gamma(\pi_\infty^*(\xi))$ of
a horizontal fibre bundle in the domain of~$\Delta$.
\end{state}

This assertion provides a considerable simplification of the search for
new Frobenius operators using known ones and allows to reconstruct
Sokolov's brackets in the inverse images of Frobenius recursions
$R\in\End_\BBR\varkappa(\pi)$, factored by Frobenius operators
$A\colon\Omega^1(\xi_\pi)\to\varkappa(\pi)$.
A weaker version of Proposition~\ref{ChainRule} was formulated
in~\cite{Sanders} (where the operators were considered in a fixed system of
local coordinates).
This property is valid in the
Hamiltonian case although is not well known.

\begin{proof} 
Suppose $\xi_1,\xi_2\in\gf$ and $\psi_i=\Delta(\xi_i)$,
$\vph_i=A(\psi_1)$ for $i=1,2$. We have
\begin{subequations}
\begin{align}
[\varphi_1,\varphi_2]&=
(A\circ\Delta)\bigl(\cEv_{\varphi_1}(\xi_2)-\cEv_{\varphi_2}(\xi_1)
 +\ib{\xi_1}{\xi_2}{A\circ\Delta}\bigr).\label{Bottom}\\
\intertext{On the other hand, we recall that $\psi_i=\Delta(\xi_i)$
and deduce}
[\varphi_1,\varphi_2]&=
A\bigl(\cEv_{\varphi_1}(\psi_2)-\cEv_{\varphi_2}(\psi_1)
 +\ib{\psi_1}{\psi_2}{A}\bigr)\label{Medium}\\
{}&=(A\circ\Delta)
  \bigl(\cEv_{\varphi_1}(\xi_2)-\cEv_{\varphi_2}(\xi_1)\bigr)
 +A\bigl(\cEv_{\varphi_1}(\Delta)(\xi_2)
   -\cEv_{\varphi_2}(\Delta)(\xi_1)+\ib{\psi_1}{\psi_2}{A}\bigr).\notag
\end{align}
\end{subequations}
Now subtract~\eqref{Bottom} from~\eqref{Medium} and,
using the nondegeneracy~\eqref{NoKernelsIntersect} and hence omitting
the operator~$A$, we obtain the assertion.
\end{proof}

\begin{example}\label{ExSanders}
Historically, the class of scalar 
operators, first regarded in view of \S1 in Definition~\ref{DefpHO},
was studied in~\cite{SokolovUMN} using local coordinates.
In~\cite{Sanders}, Gelfand's symbolic method~\cite{Gelfand} was
applied in that setting, and it was argued that the class is infinite
and contains the (presumably for $\smash{A_1^{(2)}}$ etc.,
well\/-\/defined Frobenius) operators
\begin{align}
A^{(1)}_1&=D_x;\quad \lefteqn{A^{(2)}_1=D_x\circ(D_x+u);}&
  A^{(3)}_1&=A^{(2)}_1\circ(D_x+u);\notag\\
A^{(4)}_1&=A^{(3)}_1\circ(D_x+u),&
  A^{(4)}_2&=A^{(3)}_1\circ(D_x+2u);\notag\\
\intertext{here the superscripts denote the differential order
and the subscripts enumerate operators of equal order.
Further, there are two operators}
A^{(n)}_1&=A^{(n-1)}_1\circ(D_x+u),&
  A^{(n)}_2&=A^{(n-1)}_2\circ(D_x+(n-2)u)\notag\\
\intertext{for any odd $n\geq5$, and there are four operators}
A^{(n)}_1&=A_1^{(n-1)}\circ(D_x+u),&
  A^{(n)}_3&=A_1^{(n-1)}\circ(D_x+2u),\notag\\
A^{(n)}_4&=A_2^{(n-1)}\circ(D_x+(n-3)u),&
  A^{(n)}_2&=A_2^{(n-1)}\circ(D_x+(n-2)u) \label{SandersList}
\end{align}
for any even $n\geq6$.
These operators are homogeneous w.r.t.\ the weights
$|u|=|D_x|=1$, hence the weights coincide with the differential orders.

None of these operators is Hamiltonian. The brackets
$\ib{\,}{\,}{A_i^{(n)}}$ are reconstructed from the vanishing bracket
for $\smash{A_1^{(1)}}$ by using the chain rule (see
Proposition~\ref{ChainRule}); the first four of them are
\begin{align*}
\ib{p}{q}{A_1^{(2)}}&=p_xq-pq_x,\\
\ib{p}{q}{A_1^{(3)}}&=2u(p_xq-pq_x)+p_{xx}q-pq_{xx},\\
\ib{p}{q}{A_1^{(4)}}&=3(u^2+u_x)(p_xq-pq_x)+3u(p_{xx}q-pq_{xx})+
  p_{xxx}q-pq_{xxx},\\
\ib{p}{q}{A_2^{(4)}}&=6(u^2+2u_x)(p_xq-pq_x)+6u(p_{xx}q-pq_{xx})+
  2(p_{xxx}q-pq_{xxx})+p_xq_{xx}-p_{xx}q_x.
\end{align*}

It remains unknown whether operators~\eqref{SandersList}, which are
factored to products of primitive first order operators with integer
coefficients, exhaust all Frobenius operators with differential
polynomial coefficients and homogeneous w.r.t.\ the weights~$|u|=|D_x|=1$.
The transformation rules for these operators are not known and the
substitutions~$w$ for them, if they are not of the third (degenerate) kind,
have not been constructed.
\end{example}

Consider an operator $R=\hat{A}_2\circ\omega$ factored by a Hamiltonian
operator~$\hat{A}_2\colon\hat{\varkappa}(\pi)\to\varkappa(\pi)$ and
a pre\/-\/symplectic structure~$\omega\colon\varkappa(\pi)\to
\hat{\varkappa}(\pi)$ whose image is closed w.r.t.\ Dorfman's
bracket~$[\,,\,]_{\hat{A}_2}$. The recursion operator~$R$ is Frobenius, but
the bracket~$\ib{\,}{\,}{R}$, which can be reconstructued
using~\eqref{ChainRuleFormula}, is generally nonlocal.
The exception is given by operators~$\omega$ of zero
differential order if the determinants of their matrices do not vanish identically.

\begin{example}\label{ExFrobRec}
Again, let us consider Hamiltonian and Noether operators~\eqref{A12dBous}
for the dispersionless $3$-\/component Boussinesq\/-\/type system~\eqref{d-B},
which is a true source of illustrative examples for the
exposition.\footnote{The operator $A_2$ was labelled `first' and $A_1$ was
`second' with the lexicographical order $w\prec u\prec v$ of their
arguments $\vec{\psi}={}^t\bigl(\psi^w,\psi^u,\psi^v\bigr)
\in\hat{\varkappa}(\pi)$ and images $\vec{\vph}={}^t\bigl(\vph^w,\vph^u,\vph^v
\bigr)\in\varkappa(\pi)$ in~\cite{JK3Bous}.}
The image of the Hamiltonian operator~$\hat{A}_0\colon\cosym\cE$
$\to\sym\cE$ is
closed w.r.t.\ commutation~\eqref{EvBracket}. Note that the determinant of
the zero\/-\/order operator~\eqref{A1dBous} does not vanish identically,
hence $\omega=A_1^{-1}\colon\sym\cE\to\cosym\cE$ is an isomorphism of
modules for an open dense subset of solutions $s\in\Gamma(\pi)$ of~\eqref{d-B}.
Therefore $A_1^{-1}$ is a Frobenius structure w.r.t.\ $[\,,\,]_{\hat{A}_0}$.
By this argument, define
\begin{equation}\label{FRec}
R_0=\hat{A}_0\circ A_1^{-1}\colon\sym\cE\to\sym\cE.
\end{equation}
This recursion operator for~\eqref{d-B} is Frobenius.
Indeed, its image is contained in the
image of the Hamiltonian operator~$\hat{A}_0$,
whose Dorfman's bracket $[\,,\,]_{\hat{A}_0}$ is pushed forward
by the zero\/-\/order operator~$A_1$ 
to~$[\,,\,]_{R_0}$ on~$\sym\cE$. 

The bracket~$\ib{\,}{\,}{R_0}$ can be calculated explicitly, although the
resulting formulas are relatively long. For that, we use a \textsc{Maple}
program for the \textsc{Jets} environment~\cite{Jets}; the listing is
contained in appendix~\ref{AppRecJets}.
Each component of Sokolov's bracket~$\ib{\,}{\,}{R_0}$ for~\eqref{FRec}
is a fraction of differential polynomials and contains about $15,000$ summands
in the numerator. The denominators are the cubes of the determinant of
the matrix~$A_1$. Perhaps, these fractions are reducible.

\enlargethispage{\baselineskip}
In the same way, we construct the Frobenius
recursion~$R_2=\hat{A}_2\circ A_1^{-1}$ for~\eqref{d-B} using its second
Hamiltonian structure~\eqref{A2HamdBous} and the
bracket\footnote{In the next section we show why this compatibility occurs.}
$\ib{\,}{\,}{\hat{A}_2}=\ib{\,}{\,}{A_1}+\ib{\,}{\,}{A_2}$,
see~\eqref{SokBrABoth}. The bracket $\ib{\,}{\,}{R_2}$ is computed using a
slight modification of the program that is contained in
appendix~\ref{AppRecJets}. The modification amounts to a substitution
of~$\hat{A}_2$ and $\ib{\,}{\,}{\hat{A}_2}$ for
$\hat{A}_0$ and $\ib{\,}{\,}{\hat{A}_0}$, respectively.   
\end{example}

Finally, we specify a condition upon the brackets~$\ib{\,}{\,}{R^\ell}$
for sequences of Lie algebra structures~$[\,,\,]_{R^\ell}$ induced
on~$\varkappa(\pi)$ by Frobenius iterations~$R^\ell$ of
a recursion~$R\in\End_\BBR\varkappa(\pi)$. The condition~\eqref{RelSokPowers}
is a consequence of Proposition~\ref{ChainRule}.

\begin{state}
Let $R\colon\varkappa(\pi)\to\varkappa(\pi)$ be a Frobenius operator
and assume that its powers~$R^2,\ldots,R^k$ are Frobenius for some~$k>1$.
Then 
the brackets~$\ib{\,}{\,}{R^\ell}$ for each~$\ell\in[1,\ldots,k)$
satisfy the relation
\begin{multline}
R^{\ell-1}\bigl(\ib{\xi_1}{\xi_2}{R^\ell}\bigr)=
 \sum_{i=0}^{\ell-2} R^{\ell-i-2}\Bigl[
   \cEv_{R^\ell(\xi_1)}(R)\bigl(R^i\xi_2\bigr)-
   \cEv_{R^\ell(\xi_2)}(R)\bigl(R^i\xi_1\bigr)\Bigr]\\
 {}+\ib{R^{\ell-1}\xi_1}{R^{\ell-1}\xi_2}{R},\label{RelSokPowers}
\end{multline}
where $\xi_1,\xi_2$ belong to the domain of~$R^\ell$.
\end{state}

Under assumption that $R,\ldots,R^k$ are
Frobenius for $k>1$, is there any condition for $R^{k+1},\ldots,R^{k+\ell}$ to
be Frobenius as well? A well\/-\/known condition for Nijenhuis
operators~\cite{Dorfman,YKSMagri} originates from the standard cohomology
theory for Lie algebras~\cite{Dorfman,Fuchs,Gerstenhaber}. At the same time,
the nontrivial finite deformations $[\,,\,]_{R^k}$ of the standard Lie algebra
structure~$[\,,\,]$ on~$\gothg(\pi)$ are not the trivial infinitesimal
deformations~$[\,,\,]_{\mathrm{N}^\ell}$, see~\eqref{BrNijenhuis},
which are obtained using powers of Nijenhuis recursion
operators~$\mathrm{N}\in\End_\BBR\varkappa(\pi)$ with vanishing Nijenhuis
torsion~$\lshad\mathrm{N},\mathrm{N}\rshad^{\text{fn}}=0$.

\section{Compatibility of Frobenius operators}\label{SecCompat}
We introduce two types of compatibility for Frobenius operators
$A_1$,\ $\ldots$,\ $A_N\colon\Omega^1(\xi_\pi)\to\gothg(\pi)$, where
now $\Omega^1(\xi_\pi)=\gf/\bigcap_{i=1}^N\ker A_i$.
The linear compatibility means that
linear combinations of operators remain Frobenius and hence the `individual'
Koszul brackets are correlated. The strong compatibility of~$N$
Frobenius operators means that the sum of their images is an
involutive distribution in the Lie algebra~$\gothg(\pi)$
of evolutionary vector fields.    
We endow the spaces of both linear and strong compatible Frobenius
operators with a bi\/-\/linear bracket that satisfies the Jacobi
identity, and then we relate the Lie\/-\/type algebras of
operators to an affine geometry with bi\/-\/differential Christoffel
symbols. 
Let us consider these notions in more detail.

\subsection{The linear compatibility}\label{SecWeak}
First recall that any linear combination of two compatible Hamiltonian
operators is Hamiltonian by definition and hence Frobenius.
Example~\ref{ExSohr} shows that a Hamiltonian operator can be
decomposed as a sum of operators that satisfy~\eqref{EqDefFrob}, e.g.,
$A_2^{\text{KdV}}=-\tfrac{1}{2}\cdot D_x^3+2\cdot uD_x+u_x$
is a linear combination of $\rme_1=D_x^3$, $\rme_2=uD_x$, and~$\rme_3=u_x$.
The decomposition may not be unique due to the existence of several
linear dependent Frobenius operators that appear in the splitting;
indeed, the operator $A_2^{\text{KdV}}$ can be also obtained
using $A_4=uD_x-u_x$, see~\eqref{KdVWeightOps}.

\begin{define}\label{DefWeak}
Frobenius operators $A_1$,\ $\ldots$,\ $A_N$ are \emph{linear
compatible} if their arbitrary linear combinations
$A_{\vec{\lambda}}=\sum_{i=1}^N\lambda_i A_i$ are Frobenius for any
$\vec{\lambda}\in\BBR
^N$.
The operators are linear compatible \emph{at a point}
$\vec{\lambda}_0\in\BBR
^N$
if $A_{\vec{\lambda}_0}$ is Frobenius for a fixed linear combination.
\end{define}

\enlargethispage{1.3\baselineskip}\begin{example}
There are two classes of pairwise linear compatible Frobenius scalar
operators~\eqref{pHHydro} of first order. The first type of pairs is
$\gamma_2(u)=\text{const}\cdot\gamma_1(u)$ with any $\alpha_1$,\
$\alpha_2\in\BBR$ and arbitrary functions $\beta_1(u)$, $\beta_2(u)$.
The second class is given by letting $\alpha_1=\alpha_2\in\BBR$, while the
functions $\beta_1$,\ $\beta_2$, $\gamma_1$, and~$\gamma_2$ remain arbitrary.
\end{example}

\begin{state}\label{LCOpLCBr}
The Sokolov bracket induced on the domain of a linear combination
$A_{\vec{\lambda}}=\sum_{i=1}^N\lambda_i A_i$ of linear compatible
Frobenius operators is
\[
\ib{\,}{\,}{\sum\limits_{i=1}^N\lambda_i A_i}=
 \sum_{i=1}^N\lambda_i\cdot\ib{\,}{\,}{A_i}.
\]
\end{state}

\begin{proof}
This is readily seen by inspecting the coefficients of $\lambda_i^2$ in
the quadratic polynomials in $\lambda_i$ that appear in both sides of
the equality $\bigl[A_{\vec{\lambda}}(p),A_{\vec{\lambda}}(q)\bigr]=
A_{\vec{\lambda}}\bigl([p,q]_{A_{\vec{\lambda}}}\bigr)$ upon the Koszul
bracket, here $p,q\in\Omega^1(\xi_\pi)$.

Consider the commutator $\bigl[\sum_i\lambda_iA_i(p),\sum_j\lambda_j A_j(q)\bigr]$.
On one hand, it is equal to
\begin{align}
{}&=\sum_{i\neq j}\lambda_i\lambda_j\bigl[A_i(p),A_j(q)\bigr]+
  \sum_i\lambda_i^2 A_i\bigl(\cEv_{A_i(p)}(q)-\cEv_{A_i(q)}(p)+\ib{p}{q}{A_i}\bigr).
\label{SqLambdaApart}\\
\intertext{On the other hand, the linear compatibility of~$A_i$ implies}
{}&=A_{\vec{\lambda}} 
\bigl(\cEv_{A_{\vec{\lambda}} 
(p)}(q)\bigr)  -  A_{\vec{\lambda}}   
\bigl(\cEv_{A_{\vec{\lambda}}   
(q)}(p)\bigr)  +  A_{\vec{\lambda}}   
\bigl(\ib{p}{q}{A_{\vec{\lambda}}   
}\bigr).\notag
\end{align}
The entire commutator is quadratic homogeneous in~$\vec{\lambda}$, whence the
bracket $\ib{\,}{\,}{A_{\vec{\lambda}}}$ is linear in~$\vec{\lambda}$.
From~\eqref{SqLambdaApart} we see that the individual brackets~$\ib{\,}{\,}{A_i}$
are contained in~it. Therefore,
\[
\ib{p}{q}{A_{\vec{\lambda}}}=\sum_\ell\lambda_\ell\cdot\ib{p}{q}{A_\ell}+
  \sum_\ell\lambda_\ell\cdot\gamma_\ell(p,q),
\]
where $\gamma_\ell\colon\Omega^1(\xi_\pi)\times\Omega^1(\xi_\pi)\to
\Omega^1(\xi_\pi)$. We claim that all summands $\gamma_\ell(\cdot,\cdot)$,
which do not depend on $\vec{\lambda}$ at all, vanish. Indeed, assume the converse.
Let there be $\ell\in[1,\ldots,N]$ such that $\gamma_\ell(p,q)\neq0$; without loss
of generality, suppose $\ell=1$. Then set $\vec{\lambda}=(1,0,\ldots,0)$, whence
\begin{multline*}
\Bigl[\sum_i\lambda_iA_i(p),\sum_j\lambda_jA_j(q)\Bigr]=
\Bigl[\bigl(\lambda_1A_1\bigr)(p),\bigl(\lambda_1A_1\bigr)(q)\Bigr]
=\bigl(\lambda_1A_1\bigr)\bigl(\lambda_1\gamma_1(p,q)\bigr)\\
{}+\bigl(\lambda_1A_1\bigr)\Bigl(\cEv_{(\lambda_1A_1)(p)}(q)-
   \cEv_{(\lambda_1A_1)(q)}(p)+ 
    \lambda_1\ib{p}{q}{A_1} 
\Bigr).
\end{multline*}
Consequently, $\gamma_\ell(p,q)\in\ker A_\ell$ for all $p$ and~$q$.
Since each $A_\ell$ is nondegenerate by assumption, we have that
$\gamma_\ell=0$ for all $\ell$, which concludes the proof.
\end{proof}

\begin{cor}[Infinitesimal deformations of Frobenius operators]
Two Frobenius operators are linear compatible iff for any $p$,\
$q\in\Omega^1(\xi_\pi)$ one has
\[
\bigl[B(p),A(q)\bigr]+\bigl[A(q),B(p)\bigr]=
 A\bigl([p,q]_B\bigr)+B\bigl([p,q]_A\bigr),
\]
which is equivalent to the relation
\[
\cEv_{A(p)}(B)(q)+\cEv_{B(p)}(A)(q)-\cEv_{A(q)}(B)(p)-\cEv_{B(q)}(A)(p)=
 A\bigl(\ib{p}{q}{B}\bigr)+B\bigl(\ib{p}{q}{A}\bigr).
\]
\end{cor}

\begin{example}
The three Noether operators $\hat{A}_0$, $A_1$, and $\hat{A}_2$,
see~\eqref{A12dBous} on p.~\pageref{A12dBous},
are linear compatible Frobenius structures for system~\eqref{d-B}.
Any linear combination $\lambda_0\hat{A}_0+\lambda_1A_1+\lambda_2\hat{A}_2$
is Frobenius again, and therefore Sokolov's brackets
for operators~(\ref{A1dBous}--\ref{A2dBous}) are correlated by
\[
\ib{\,}{\,}{\hat{A}_2}=\ib{\,}{\,}{A_1}+\ib{\,}{\,}{A_2},
\]
which we claimed in Example~\ref{ExFrobRec}.
\end{example}

\begin{rem}
The operators remain Frobenius when multiplied by a constant, therefore
pass to the projective frame $\lambda\in\BBR\mathbb{P}^N$ of $N\in\BBN$
Frobenius operators. Then in $\CDiff(\Omega^1(\xi_\pi),\varkappa(\pi))$
there is a basis of Frobenius operators which either are isolated
points or which generate Frobenius cells
with a nontrivial topology of attaching the simplexes together.

An illustration is given by Example~\ref{ExSohr}: For
$\vec{\mathrm{e}}_1=D_x^3$, $\vec{\mathrm{e}}_2=uD_x$, and
$\vec{\mathrm{e}}_3=u_x$, the curve $A_2^{\text{KdV}}=(\lambda:2:1)$
is Hamiltonian and the ray $A_4=f(u)\cdot(0:1:-1)$
is Frobenius.
\end{rem}

\subsection{The strong compatibility}\label{SecStrong}
We impose an additional specification on the structure of the commutators of
evolutionary vector fields whose generating sections belong to images of
several Frobenius operators.

\begin{define}\label{DefStrong}
Frobenius operators $A_1$,\ $\ldots$,\ $A_N\colon\Omega^1(\xi_\pi)\to
\varkappa(\pi)$ are \emph{strong compatible} if the commutators of
evolutionary fields in the images of any two of them belong to the sum of
the images of all the $N$ operators such that, for any
$p,q\in\Omega^1(\xi_\pi)$ and $1\leq i,j\leq N$,
\begin{equation}\label{DefGamma}
[A_i(p),A_j(q)]=
 A_j\bigl(\cEv_{A_i(p)}(q)\bigr)-A_i\bigl(\cEv_{A_j(q)}(p)\bigr)
 +\sum_{k=1}^N A_k\bigl(\Gamma^k_{ij}(p,q)\bigr)
 \in\sum_{\ell=1}^N\text{im}\,A_\ell.
\end{equation}
\end{define}

\begin{example}\label{ExStrongFromLiou}
Consider the Liouville equation $\cU_{xy}=\exp(2\cU)$.
Let $\varsigma_x$~be the same projection onto the $x$-\/characteristics as
in~\eqref{ProjOnX}, and similarly for $\bar{\varsigma}$ and the coordinate~$y$.
Frobenius operators $\square=\bigl(\cU_x+\tfrac{1}{2}D_x\bigr)\circ\varsigma_x$
and $\bar{\square}=\bigl(\cU_y+\tfrac{1}{2}D_y\bigr)\circ\bar{\varsigma}_x$,
c.f.~\eqref{SquareSok}, are strong compatible. The bi\/-\/differential
coefficients~$\Gamma^k_{ij}$ are given in~\eqref{GammaSquareSok}
on p.~\pageref{GammaSquareSok}.
\end{example}

\begin{defNo}[continued]
The common domain $\Omega^1(\xi_\pi)$ of the operators~$A_i$ is
an $\cF(\pi)$-\/submodule. Therefore, in view of the functional arbitrariness
of sections $p$,\ $q\in\Omega^1(\xi_\pi)$, we say that the involutive
distribution of evolutionary vector fields in the images of linear
independent strong compatible operators $A_1$,\ $\ldots$,\ $A_N$ has
\emph{reduced dimension} (the \emph{rank})~$N$.
\end{defNo}

\begin{example}[The Magri schemes]\label{ExMagri}
Completely integrable bi\/-\/Hamiltonian hierarchies determine 
commutative distributions of reduced dimension~$N=2$. All commutators
vanish for the restrictions $A_1$,\ $A_2$ of Hamiltonian operators onto
the hierarchies; here $1\leq i,j,k\leq 2$.

In other words, the strong compatibility of Hamiltonian operators
$A_k\colon\Omega^1(\pi)\to\varkappa(\pi)$ is achieved on linear subspaces
of~$\Omega^1(\pi)$. 
This is valid for the linear span of the gradients $\psi_i=\bE(\cH_i)$
of the Hamiltonians $\cH_i\in\bar{H}^n(\pi)$ which descend from
the Casimirs $\cH_0\in H^{A_1}_0(\pi)$ in the Magri scheme, see
Theorem~\ref{ThMagri}. Indeed, one has
$\text{im}\,A_2\subset\text{im}\,A_1$
whenever both Hamiltonian operators are restricted onto the descendants
of the Casimirs for~$A_1$, and hence the commutators
(however, which vanish\footnote{For example,
the first and second Hamiltonian structures for the KdV equation~\eqref{IKdV}
are not strong
compatible unless restricted onto some subspaces of their
arguments. On the linear subspace of descendants of the Casimir $[w\,dx]$,
we have $\text{im}\,A_2\subset\text{im}\,A_1$ and, since the
image of the Hamiltonian operator~$A_1=D_x$ is closed, we have
$[\text{im}\,A_1$,\ $\text{im}\,A_2]\subset\text{im}\,A_1$.
We emphasize that we do not exploit the commutativity of the flows.}
by the same theorem)
belong to the image of~$A_1$. Thus the iteration~\eqref{MagriResolve}
of the Magri scheme corresponds to involutive distributions
of reduced dimension~two.
\end{example}

Let us formulate the properties of the bi\/-\/differential symbols
$\Gamma_{ij}^k\in\CDiff\bigl(\bigotimes^2\Omega^1(\xi_\pi)$,
$\Omega^1(\xi_\pi)\bigr)$, which are determined by strong compatible
Frobenius operators $A_1$,\ $\ldots$,~$A_N$.
Note that the symbols $\Gamma^k_{ij}$ depend on
a point~$\theta\in J^\infty(\pi)$.

\begin{property}
By construction, for any number~$N$ of strong compatible
Frobenius operators, we have
\begin{equation}\label{DecompositionRequirement}
\Gamma^k_{ii}=\delta_i^k\cdot\ib{\,}{\,}{A_i},\qquad
\text{for each $i$, $1\leq i\leq N$.}
\end{equation}
Hence a Frobenius operator yields the involutive distribution of
reduced dimension one.
\end{property}

\begin{property}
The symbols $\Gamma_{ij}^k$ are not uniquely defined.
Indeed, they are gauged by the conditions
\begin{equation}\label{NonUnique}   
\sum_{k=1}^N A_k\Bigl(\cEv_{A_j(q)}(p)\delta^k_i-\cEv_{A_i(p)}(q)\delta^k_j
 +\Gamma^k_{ij}(p,q)\Bigr)=0,\qquad p,q\in\Omega^1(\xi_\pi);
\end{equation}
again, we assume the semi\/-\/simplicity of all the images in~$\gothg(\pi)$:
$\bigl[A_\ell(\psi),\gothg(\pi)\bigr]=0$ implies $\psi\in\ker A_\ell$.
\end{property}

\begin{property}
If, additionally, two strong compatible Frobenius
operators $A_i$ and $A_j$ are linear compatible,
then their Sokolov's brackets are
\[ 
\ib{p}{q}{A_i}=\Gamma^j_{ij}(p,q)+\Gamma^j_{ji}(p,q)
\quad \text{and}\quad
\ib{p}{q}{A_j}=\Gamma^i_{ij}(p,q)+\Gamma^i_{ji}(p,q)
\]
for any~$p,q\in\Omega^1(\xi_\pi)$.
\end{property}

\begin{proof}
For brevity, denote $A=A_i$, $B=A_j$ and consider the Frobenius linear
combination~$\mu A+\nu B$. By Proposition~\ref{LCOpLCBr}, we have
\begin{multline*}
\bigl(\mu A+\nu B\bigr)\bigl(\ib{p}{q}{\mu A+\nu B}\bigr)={}\\
 {}=\mu^2A\bigl(\ib{p}{q}{A}\bigr)+\mu\nu\cdot A\bigl(\ib{p}{q}{B}\bigr)
 +\mu\nu\cdot B\bigl(\ib{p}{q}{A}\bigr)+\nu^2B\bigl(\ib{p}{q}{A}\bigr).
\end{multline*}
On the other hand,
\begin{multline*}
\Bigl[\bigl(\mu A+\nu B\bigr)(p),\bigl(\mu A+\nu B\bigr)(q)\Bigr]\\
{}=\mu^2\bigl[A(p),A(q)\bigr]+\mu\nu\bigl[A(p),B(q)\bigr]
   -\mu\nu\bigl[A(q),B(p)\bigr]+\nu^2\bigl[B(p),B(q)\bigr].
\end{multline*}
Taking into account~\eqref{DefGamma} and equating the coefficients of~$\mu\nu$,
we obtain
\[
A\bigl(\ib{p}{q}{B}\bigr)+B\bigl(\ib{p}{q}{A}\bigr)=
 A\bigl(\Gamma_{AB}^A(p,q)\bigr)+B\bigl(\Gamma_{AB}^B(p,q)\bigr)-
 A\bigl(\Gamma_{AB}^A(q,p)\bigr)-B\bigl(\Gamma_{AB}^B(q,p)\bigr).
\]
Using the obvious formulas $\Gamma_{AB}^A(q,p)=-\Gamma_{BA}^A(p,q)$ and
$\Gamma_{AB}^B(q,p)=-\Gamma_{BA}^B(p,q)$, see~\eqref{GammaGradedSym} below,
we isolate the arguments of the operators and obtain the assertion.
\end{proof}

\begin{define} 
From now on, we consider Frobenius operators that are both linear and
strong compatible. Such operators will be called \emph{totally compatible}.
By definition, totally compatible operators span linear spaces
\begin{equation}\label{FLSpace}
\EuA=\bigoplus_{k=1}^N A_k\cdot\BBR
\end{equation}
of Frobenius operators.
\end{define}

\begin{example}
The operators $\square$ and $\bar{\square}$, which were introduced in
Example~\ref{ExStrongFromLiou}, are totally compatible.
This construction admits a straightforward generalization~\eqref{Square} for other
Euler\/--\/Lagrange Liouville\/-\/type systems, see Theorem~\ref{IspHO}.
\end{example}

\begin{rem}\label{RemManin}
Within the Hamiltonian approach~\cite{Lstar}, it is very productive to think
that the cosymmetries~$\psi\in\hat{\varkappa}(\pi)$ are \emph{odd}.
Indeed, in this particular situation the homomorphisms
$\psi\in\Hom_{\cF(\pi)}\bigl(\varkappa(\pi),\bar{\Lambda}^n(\pi)\bigr)$
are identified with Cartan $1$-\/forms times the volume form $\Id\volume(M^n)$
for the base of the jet bundle.

We preserve this understanding for domains~$\Omega^1(\xi_\pi)$
of Frobenius operators~\eqref{Frob2} of second kind. The new $\BBZ$-\/grading
must be never mixed\footnote{We owe this remark to Yu.\,I.\,Manin
(private communication).}
with any (e.g., $\BBZ_2$-{}) gradings of any variables on~$J^\infty(\pi)$ and
with the arising gradings of the sections $F\in\gh$ and $\psi\in\hat{\gh}$
(here $\gf=\hat{\gh}$).
Hence if $\pi$~is a super\/-\/bundle with Grassmann\/-\/valued
sections, then Frobenius operators~$A$ are bi\/-\/graded~\cite{JKKersten}.
Their proper $\BBZ$-\/grading is $|A|_{\BBZ}=-1$ because the images
in~$\gothg(\pi)$ have degree zero;
the $\BBZ_2$-\/degree of the operators~$A$ can be arbitrary.
In particular, Hamiltonian operators $P$ produce Cartan's $0$-\/forms by
$\cEv_{P(\cdot)}\colon\Omega^1(\pi)\to\gothg(\pi)$.

In what follows, we assume for simplicity that all coordinates
on~$J^\infty(\pi)$ are permutable, whence $\Omega^1(\xi_\pi)$ is even
w.r.t.\ the $\BBZ_2$-\/grading.
\end{rem}

\begin{property}
For any $i,j,k\in[1,\ldots,N]$ and for arguments $p,q\in\Omega^1(\xi_\pi)$ of
$\BBZ$-\/degree~$1$ for strong compatible Frobenius operators of second kind,
we have
\begin{equation}\label{GammaGradedSym}
\Gamma^k_{ij}(p,q)=-\Gamma_{ji}^k(q,p)
 =(-1)^{|p|_{\BBZ}\cdot|q|_{\BBZ}}\cdot\Gamma^k_{ji}(q,p)
\end{equation}
due to the 
skew\/-\/symmetry of the commutators~\eqref{EvBracket}.
Hence the symbols $\Gamma^k_{ij}$ are symmetric w.r.t.\ the $\BBZ$-\/grading
in the case~\eqref{Frob2}.
\end{property}

Now consider the space of flat connections $\nabla^{A_k}$ defined
by~\eqref{Connection} for each Frobenius operator~$A_k$.
Let us reveal the standard behaviour of the Christoffel symbols~$\Gamma_{ij}^k$.

\begin{property}[Transformations of~$\Gamma_{ij}^k$]
Let $\tilde{w}=\tilde{w}[w]$ be a nondegenerate change of fibre coordinates in the
bundle~$\xi$. Recall that the sections $p,q\in\gf$ are reparametrized by
$p\mapsto\tilde{p}=\Xi(p)$ and $q\mapsto\tilde{q}=\Xi(q)$, where
$\Xi=\ell_{\tilde{w}}^{(w)}$ for Frobenius operators~\eqref{Frob1} of first kind
and $\Xi=\bigl[\bigl(\ell_{\tilde{w}}^{(w)}\bigr)^*\bigr]^{-1}$ for the operators
of second kind. Consequently, Frobenius operators $A_1$,\ $\ldots$,\
$A_N\colon\Omega^1(\xi_\pi)\to\gothg(\pi)$ with a common domain $\Omega^1(\xi_\pi)=
\gf/\bigcap_i\ker A_i$ are transformed by $A_i\mapsto\tilde{A}_i=
A_i\circ\Xi^{-1}\bigr|_{w=w[\tilde{w}]}$. Then the bi\/-\/differential symbols
$\Gamma_{ij}^k\in\CDiff\bigl(\Omega^1(\xi_\pi)\times\Omega^1(\xi_\pi)\to
\Omega^1(\xi_\pi)\bigr)$ obey 
\begin{equation}\label{TransformGamma}
\Gamma_{ij}^k(p,q)\mapsto\Gamma_{\tilde{\imath}\tilde{\jmath}}^{\tilde{k}}
 \bigl(\tilde{p},\tilde{q}\bigr)=
\bigl(\Xi\circ\Gamma_{ij}^{\tilde{k}}\bigr)\bigl(\Xi^{-1}\tilde{p},\Xi^{-1}\tilde{q}\bigr)+
\delta_i^{\tilde{k}}\cdot\cEv_{\tilde{A}_j(\tilde{q})}(\Xi)\bigl(\Xi^{-1}\tilde{p}\bigr)-
\delta_j^{\tilde{k}}\cdot\cEv_{\tilde{A}_i(\tilde{p})}(\Xi)\bigl(\Xi^{-1}\tilde{q}\bigr).
\end{equation}
This is a direct analogue of the standard rules $\tilde{\Gamma}=
\Xi\,\Gamma\,\Xi^{-1}+\Id\Xi\cdot\Xi^{-1}$ for the connection
$1$-\/forms~$\Gamma$ under reparametrizations~$\Xi$.
Formula~\eqref{TransformGamma} represents this rule in the case of 
connections over the infinite jet bundles.
\end{property}

\begin{proof}
Denote $A=A_i$ and $B=A_j$; without loss of generality, assume $i=1$ and~$j=2$.
Let us calculate the commutators of vector fields in
the images of $A$ and~$B$ using two systems of coordinates in the domain.
Then we equate the commutators straighforwardly, because the fibre coordinate in
the images of the operators is not touched at all. So, we have, originally,
\begin{align*}
\bigl[A(p),&B(q)\bigr]=B\bigl(\cEv_{A(p)}(q)\bigr)-A\bigl(\cEv_{B(q)}(p)\bigr)+
 A\bigl(\Gamma_{AB}^A(p,q)\bigr)+B\bigl(\Gamma_{AB}^B(p,q)\bigr)+
 \sum_{k=3}^N A_k\bigl(\Gamma_{AB}^k(p,q)\bigr).\\
\intertext{On the other hand, we substitute $\tilde{p}=\Xi(p)$ and $\tilde{q}=\Xi(q)$
in $\bigl[\tilde{A}(\tilde{p}),\tilde{B}(\tilde{q})\bigr]$, whence, by the Leibnitz
rule, we obtain}
\bigl[\tilde{A}(\tilde{p}),&\tilde{B}(\tilde{q})\bigr]=
\tilde{B}\bigl({\cEv_{\tilde{A}(\tilde{p})}}   
  (\Xi)(q)\bigr)+\bigl({\tilde{B}\circ\Xi}\bigr)  
  \bigl({\cEv_{\tilde{A}(\tilde{p})}}(q)\bigr)-   
\tilde{A}\bigl({\cEv_{\tilde{B}(\tilde{q})}}(\Xi)(p)\bigr)+ 
  \bigl({\tilde{A}\circ\Xi}\bigr)   
  \bigl({\cEv_{\tilde{B}(\tilde{q})}}(p)\bigr)\\  
{}&{}
+\bigl(A\circ\Xi^{-1}\bigr)\bigl(\Gamma_{\tilde{A}\tilde{B}}^{\tilde{A}}(\Xi p,\Xi q)\bigr)
+\bigl(B\circ\Xi^{-1}\bigr)\bigl(\Gamma_{\tilde{A}\tilde{B}}^{\tilde{B}}(\Xi p,\Xi q)\bigr)
+\sum_{\tilde{k}=3}^N\bigl(
                           {A}_{\tilde{k}}\circ\Xi^{-1}\bigr)
  \bigl(\Gamma_{\tilde{A}\tilde{B}}^{\tilde{k}}(\Xi p,\Xi q)\bigr).
\end{align*}
Therefore,
\begin{align*}
\Gamma_{AB}^A(p,q)&=\bigl(\Xi^{-1}\circ\Gamma_{\tilde{A}\tilde{B}}^{\tilde{A}}\bigr)
  (\Xi p,\Xi q) -\bigl(\Xi^{-1}\circ\cEv_{B(q)}(\Xi)\bigr)(p),\\
\Gamma_{AB}^B(p,q)&=\bigl(\Xi^{-1}\circ\Gamma_{\tilde{A}\tilde{B}}^{\tilde{B}}\bigr)
  (\Xi p,\Xi q) +\bigl(\Xi^{-1}\circ\cEv_{A(p)}(\Xi)\bigr)(q),\\
\Gamma_{AB}^k(p,q)&=\bigl(\Xi^{-1}\circ\Gamma_{\tilde{A}\tilde{B}}^{
                                                                    {k}}\bigr)
  (\Xi p,\Xi q)\qquad\text{for $k\geq3$.}
\end{align*}
Acting by~$\Xi$ on these equalities and expressing $p=\Xi^{-1}\tilde{p}$,
$q=\Xi^{-1}\tilde{q}$, we conclude the proof.
\end{proof}

\begin{cor}
The bi\/-\/differential symbols $\Gamma^k_{ij}$ constitute symmetric flat 
connections $\nabla(\vec{\lambda})=\sum_k\lambda_k\nabla^{A_k}$
in the graded triples $\bigl(\Omega^1(\xi_\pi),\gothg(\pi),\EuA\bigr)$
determined by Frobenius operators of second kind.
\end{cor}

\begin{rem}
A straightforward calculation shows that two operators~\eqref{pHHydro}
are strong compatible if and only if they are proportional.
Next, consider operators~\eqref{A12dBous} for the Boussinesq\/-\/type
system~\eqref{d-B}.
If these linear compatible Frobenius operators are not
restricted to a subspace of $\Omega^1(\cE)$, and thus the arguments of
$A_k$ are generic, then the commutators $[\text{im}\,A_i,\text{im}\,A_j]$,
$0\leq i<j\leq2$, are \emph{not} decomposed using~\eqref{DefGamma}
w.r.t.\ the images of any \emph{two} operators $A_i$ and~$A_j$.
We argue that these negative examples
have a very deep motivation, which has been indicated in Example~\ref{ExMagri}.

Indeed, by definition, the
commutator~\eqref{DefGamma} always contains the standard first two
terms in the r.h.s. and takes values in the entire sum of images
$\sum_k\text{im}\,A_k$ for generic $p,q\in\Omega^1(\xi_\pi)$.
Hence the commutation relations, which determine the Lie\/-\/algebraic
type of the involutive distribution in the sum of images of $A_1$,\
$\ldots$,\ $A_N$, depend on a linear subspace $S\subset\Omega^1(\xi_\pi)$
that contains $p$ and~$q$.
We may choose it ourselves in such a way that some of the operators
become restricted on their kernels (up to the condition~\eqref{NonUnique}).

Therefore we expect that the strong compatibility and the decompositions
w.r.t.\ the three linear independent operators~\eqref{A12dBous} are restored
for restrictions of the Frobenius operators onto the spans of
Hamiltonians for the dispersionless $3$-\/component Boussinesq hierarchy.
This will be discussed elsewhere.
\end{rem}

\subsection{Algebras of Frobenius operators}
Linear spaces of recursion operators $R\in\End_\BBR\varkappa(\pi)$ are
equipped with the associative composition~$\circ$, and therefore the recursions
constitute monoids. Their unit is the identity mapping, and there appear
relations between the operators or between their restrictions onto
differential equations. For instance, the structural relations for
for recursion operators of the Krichever\/--\/Novikov equations are described
by hyperelliptic curves, see~\cite{SokDemskojKN}.
Taking compositions of two recursions~$R_i$ and $R_j$, we also obtain
their formal commutators~\eqref{SchoutenBr} by setting
\begin{equation}\label{StandardLieR}
\lshad R_i,R_j\rshad=R_i\circ R_j-R_j\circ R_i.
\end{equation}
Nontrivial examples of relations for the algebra
structures~\eqref{StandardLieR} are known, e.g., one has $\lshad R_1,R_2\rshad
=R_1^2$ for the dispersionless $3$-\/component Boussinesq system~\eqref{d-B},
see~\cite{JK3Bous}. In this way, the Richardson\/--\/Nijenhuis
bracket~\eqref{SchoutenBr} endows the linear spaces of recursions with
a graded Lie algebra structure~\cite{JKKersten,Opava}. If the Nijenhuis torsion
$\lshad\rmN,\rmN\rshad^{\text{fn}}$ vanishes for an
operator~$\rmN\colon\varkappa(\pi)\to\varkappa(\pi)$
and the Fr\"o\-li\-cher\/--\/Nijenhuis bracket $\lshad\,,\,\rshad^{\text{fn}}$,
then it produces trivial infinitesimal deformations~\eqref{BrNijenhuis} of the
standard bracket~$[\,,\,]$ on another linear space, that is, on~$\varkappa(\pi)$.
By their turn, Frobenius recursion operators~\eqref{FrobRec} determine
nontrivial finite deformations~$[\,,\,]_R$ of the Lie algebra structure
of~$\gothg(\pi)$.


We propose a reverse scheme that starts with the standard
structure~\eqref{EvBracket} in the Lie algebra~$\gothg(\pi)$ and then endows
linear spaces of Frobenius operators with a Lie\/-\/type bracket.



First, let Frobenius operators $A_1,\ldots,A_N\colon\gf\to\gothg(\pi)$ be
totally compatible such that each point of the linear space~\eqref{FLSpace} is
a Frobenius operator. Consider the commutation closure
relations~\eqref{DefGamma}, which are specified for vector fields with
generating sections that belong to the images of~$A_i$. These relations
express the decomposition of the commutators in the left\/-\/hand side
w.r.t.\ the images again.
Note that formula~\eqref{DefGamma} is linear w.r.t.\ each operator from~$\EuA$.

We suggest to take the decomposition~\eqref{DefGamma} and define
the commutation rules
\begin{equation}\label{CommuteOperators}
\ob{A_i}{A_j}(p,q)=\bigl[A_i(p),A_j(q)\bigr]
\end{equation}
on~$\EuA$, here $1\leq i,j\leq N$ and $p,q\in\Omega^1(\xi_\pi)=\gf/\bigcap_i\ker A_i$.
The structural constants of the
algebra~$\EuA$ are encoded by the bi\/-\/differential symbols~$\Gamma_{ij}^k$,
whose properties were described in section~\ref{SecStrong}. By construction,
the Jacobi identity holds for the bracket~\eqref{CommuteOperators} which,
under $i\leftrightarrow j$,\ $p\leftrightarrow q$, is skew\/-\/symmetric for
Frobenius operators~\eqref{Frob1} of first kind and which is symmetric (in the
graded sense $|p|_{\BBZ}=|q|_{\BBZ}=1$, see Remark~\ref{RemManin}) for
Frobenius operators~\eqref{Frob2} of second kind.

Consequently, each Frobenius operator $A_i\in\EuA$ spans the
one\/-\/dimensional 
algebra, and the Koszul bracket~$[\,,\,]_{A_i}$ measures its noncommutativity.
The assumption that the operators are linear compatible implies that
each line $\langle\vec{\lambda}\rangle$, where
$
\langle\vec{\lambda}\rangle\stackrel{\text{def}}{{}={}}
\BBR\cdot\sum_{i=1}^N\lambda_iA_i\subset\EuA$, 
is a one\/-\/dimensional subalgebra of~$\EuA$.

By this argument, we obtain the third operation on the space~\eqref{FLSpace}
of totally compatible Frobenius recursion operators
$\varkappa(\pi)\to\varkappa(\pi)$, in addition to
the composition~$\circ$ and the commutation~\eqref{StandardLieR}.
At the same time, if the Frobenius operators are not recursions,
then the compositions of the
operators $A_i\colon\gf\to\varkappa(\pi)$ are not defined,
and the algebra structure~\eqref{CommuteOperators}
on the linear space~$\EuA$ is the only one remaining.

\begin{example}
Proposition~\ref{ThDorfmanTriplette} proves the
existence of one\/-\/dimensional algebras of Hamiltonian Frobenius
operators of second kind.
Theorem~\ref{ThMagri} yields the two\/-\/dimensional 
algebras~$\EuA$, see Example~\ref{ExMagri}.
Restricting the Poisson pencils~$(A_1,A_2)$ onto the hierarchies,
we obtain analogues of the solvable two\/-\/dimensional Lie algebra
with a relation $[a_1,a_2]=a_1$. Moreover, the Magri scheme shows that
these algebras~$\EuA$ are commutative.
\end{example}

Second, let us consider a wider class of operators $A_i\colon\gf\to
\varkappa(\pi)$ that may not be Frobenius. Namely, let all the operators
be transformed by either~\eqref{Frob1} or~\eqref{Frob2}, which makes them well
defined, 
and suppose that the commutation closure 
\begin{equation}\label{CommutClosure}
\Bigl[\sum\nolimits_i\img A_i,\sum\nolimits_j\img A_j\Bigr]\subseteq
\sum\nolimits_k\img A_k
\end{equation}
is valid for the images of the whole 
$N$-\/tuple $(A_1,\ldots,A_N)$. 
This incorporates the previous case of totally compatible
Frobenius operators that satisfy~\eqref{EqDefFrob}.
Now we introduce a Lie\/-\/type structure on the linear subspace
$\EuA\subseteq\CDiff\bigl(\gf\to\varkappa(\pi)\bigr)$.

Let $\ob{A_i}{A_j}$ be the same bracket~\eqref{CommuteOperators} of the
operators $A_1$,\ $\ldots$,~$A_N\in\EuA$ at hand.
In other notation, by~\eqref{CommutClosure} we assume that\footnote{%
Thus we postulate that the bi\/-\/differential dependence on the arguments
in~\eqref{CommutC} is entirely absorbed by the structural
constants~$\bc_{ij}^k$.%
}
\begin{equation}\label{CommutC}
\ob{A_i}{A_j}=\sum_{k=1}^N A_k\circ\bc_{ij}^k,
\end{equation}
where $\bc_{ij}^k\colon\gf\times\gf\to\gf$ are the bi\/-\/differential
structural constants of the new algebra~$\EuA$. Obviously, the Christoffel
symbols~$\Gamma^k_{ij}$, which were introduced in~\eqref{DefGamma},
are encoded by $\bc_{ij}^k$ up to the gauge~\eqref{NonUnique}.
The structural constants are symmetric or skew\/-\/symmetric
w.r.t.\ the lower indexes simultaneouly with~$\Gamma_{ij}^k$.
Admitting a slight abuse of language, we continue calling these 
algebras~$\EuA$ with relations~\eqref{CommutC} by algebras of Frobenius
operators, although the image of an operator~$A_\ell$ may be not closed under the
commutation for some~$\ell\in[1,\ldots,N]$.

For instance, we have~$\bc_{ij}^k\equiv0\mod\ker A_k$ for the algebras~$\EuA$
of rank two that are generated by restrictions of Poisson pencils~$(A_1,A_2)$
onto the Magri schemes, see Example~\ref{ExMagri}.
Other examples of such algebras~$\EuA$ 
will be given in the next section, 
where we assign Frobenius operators~\eqref{Square} of second
kind to Liouville\/-\/type integrable systems (in particular, to the 2D~Toda
chains associated with complex semi\/-\/simple Lie algebras, 
see~\eqref{IEToda}). 

Thus we arrive at the structural theory problem for the operator 
algebras~$\EuA$ with 
generators~$A_i\colon\gf\to\varkappa(\pi)$ 
and relations~\eqref{CommutC}.

\begin{conjecture}
There are horizontal $\cF(\xi)$\/-modules~$\gf\hookrightarrow 
\Gamma\bigl(\pi_\infty^*(\xi)\bigr)$ and linear differential
operators $A_1$,\ $\ldots$,\ $A_N\colon\gf\to\gothg(\pi)$
such that, by a choice of appropriate subspaces $S\subset\gf$ in their
domain, one recovers a bi\/-\/differential extension of
the classical structural theory of (e.g., nilpotent,
semi\/-\/simple, or solvable) Lie algebras of reduced dimension~$N$
for the linear spaces~$\EuA$, which are generated by the
restrictions of the operators $A_1,\ldots,A_N$ onto~$S$, 
and for the algebra structures~\eqref{CommutC} on them.
\end{conjecture}

\begin{rem}
Recursion operators for differential equations $\cE=\{F=0\}$
can be understood as B\"ack\-lund autotransformations between symmetries,
which are solutions of the linearized systems $\ell_F(\vph)\doteq0$,
see~\cite{ClassSym,JKKersten}. Hence the structures~\eqref{CommutC}
of the algebras~$\EuA$ of
Frobenius recursion operators for~$\cE$ are inherited by these classes of
B\"ack\-lund autotransformations. This determines the brackets between
non\/-\/abelian coverings over the equations~$\cE$.
\end{rem}

\section{Factorizations of symmetries of Liouville\/-\/type systems}\label{SecLiou}
In this section we describe an infinite class of
Frobenius operators $\square$ and calculate
the brackets $\ib{\,}{\,}{\square}$ induced by them. These operators
appear in the description of sym\-met\-ries 
of the hyperbolic Liouville\/-\/type Euler\/--\/Lagrange nonlinear
systems~\cite{Demskoi2004, Shabat, Startsev2006, SokolovUMN}.

To start with, we extend the collection of known Frobenius operators
with the one that factors point symmetries of the
non\/-\/evolutionary $(2+1)$-dimensional $A_\infty$-\/Toda equation.

\begin{example}[\cite{JMathSci2004}] 
   \label{ExHeav}
Generators of the point symmetry algebra for the
`heavenly' Toda equation $u_{xy}=\exp(-u_{zz})$ have the form
$\vph^x=\hat{\square}^x\bigl(\phi(x)\bigr)$ or
$\vph^y=\hat{\square}^y\bigl(\bar\phi(y)\bigr)$, where
$\phi,\bar\phi\in C^\infty(\BBR)$ and
\begin{equation}\label{pHOHeav}
\hat{\square}^x=u_x+\tfrac{1}{2}z^2\,D_x,\quad 
\hat{\square}^y=u_y+\tfrac{1}{2}z^2\,D_y=
  (x\leftrightarrow y)\bigl(\hat{\square}^x\bigr).
\end{equation}
Clearly, the commutator of any two point symmetries of $x$-{} or
$y$-\/type is a 
point symmetry again such that the action of the operators
$\hat{\square}$ on the
spaces of the free functional parameters~$\phi$,\ $\bar{\phi}$ is given
by the Wronskian, $\ib{\phi_1}{\phi_2}{\hat{\square}^x}=
\phi_1\cdot{(\phi_2)}_x-{(\phi_1)}_x\cdot\phi_2$.
\end{example}

Operators~\eqref{pHOHeav} factor the right\/-\/hand sides of the symmetry flows
$u_t=\hat{\square}(\phi)$ 
on the heavenly equation. 
This means that a bigger differential equation is the representing object for
a transformation group generated by the flows.
This approach to constructing Frobenius operators is very productive.
First, let us recall a procedure that assigns hyperbolic Euler\/--\/Lagrange
systems to hierarchies which are Hamiltonian w.r.t.\
$\hat{A}_1=\text{const}\cdot D_x$, see~\cite{TMPhGallipoli,JMathSci2004}
and also Remark~\ref{RemFalqui} on p.~\pageref{RemFalqui}.
The method is based on the canonical coordinate\/-\/momenta
formalism~\cite{Dirac} for~PDE and on a representation of the KdV\/-\/type
hierarchies as commutative Lie subalgebras of Noether symmetries for the
hyperbolic systems.

\begin{example}\label{ExAmbientd3Bous}
The dispersionless $3$-\/component Boussinesq\/-\/type system,
see p.~\pageref{d-B},
\begin{equation}\tag{\ref{d-B}}
u_t=ww_x+v_x,\quad v_t=-uw_x-3u_xw,\quad w_t=u_x,
\end{equation}
is not written here in the form of a conserved current only due to an
unfortunate choice of local coordinates for it and for
structures~\eqref{A12dBous} in~\cite{JK3Bous}. Indeed, let us return to the
notation of~\cite{Nutku} and let the new dependent variables
$\gm=(\uu,\uv,\uw)$ be
\begin{equation}\label{Momenta}
\uu=v+w^2,\qquad \uv=w,\qquad \uw=u,
\end{equation}
which are densities of the Casimirs w.r.t.\ the first Hamiltonian
structure~\eqref{dBA0} for~\eqref{d-B}.
Thence we obtain the system
\begin{equation}
\label{CF}
\begin{pmatrix}\uu_t\\ \uv_t\\ \uw_t\end{pmatrix}=
\begin{pmatrix}0&1&0\\ 1&0&0\\ 0&0&1\end{pmatrix} D_x
\begin{pmatrix}\delta/\delta\uu\\ \delta/\delta\uv\\ \delta/\delta\uw
 \end{pmatrix}\bigl(\cH_1\bigr),
\end{equation}
where $\cH_1=\bigl[(\uu\uw-\tfrac{1}{2}\uv^2\uw)\,\Id x\bigr]$.
Likewise, the dispersionless Benney equation~\cite[Eq.~(19)]{Nutku} acquires
the same form~\eqref{CF} in these coordinates.

Now we introduce the conjugate variables $\bU=(U,V,W)$ such that
\[
\Bigl(\ell_\gm^{(\bU)}\Bigr)^*=-\left(\begin{smallmatrix}0&1&0\\ 1&0&0\\
  0&0&1\end{smallmatrix}\right)\cdot D_x;
\]
in other words, the adjoint linearization $\bigl(\ell_\gm^{(\bU)}\bigr)^*$
of the canonical momenta~$\gm$ w.r.t.\ the canonical coordinates~$\bU$
is proportional to the first Hamiltonian structure for~\eqref{CF}. Thus we set
\begin{equation}\label{Coordinates}
\uu=V_x,\qquad \uv=U_x,\qquad \uw=W_x.
\end{equation}
The potential variables satisfy the system
\begin{equation}\tag{\ref{CF}${}'$}\label{CFp}
U_t=W_x,\qquad V_t=-U_xW_x,\qquad W_t=-\tfrac{1}{2}U_x^2+V_x.
\end{equation}
Using the notation~\eqref{Momenta} and~\eqref{Coordinates} together, we
cast~\eqref{CF} and~\eqref{CFp} to the canonical form
\begin{equation}\label{CFBoth}
U_t=\frac{\delta\cH_1}{\delta\uu},\quad
V_t=\frac{\delta\cH_1}{\delta\uv},\quad
W_t=\frac{\delta\cH_1}{\delta\uw};\qquad
\uu_t=-\frac{\delta\cH_1}{\delta U},\quad
\uv_t=-\frac{\delta\cH_1}{\delta V},\quad
\uw_t=-\frac{\delta\cH_1}{\delta W},
\end{equation}
where $\cH_1=\bigl[(V_xW_x-\tfrac{1}{2}U_x^2W_x)\,\Id x\bigr]$.

Next, let us find the hyperbolic Euler\/--\/Lagrange equation
\[
\cEEL=\Bigl\{\bE_\bU(\cL)=0\ \mid\ \cL=\bigl[L\,\Id x\Id y\bigr],
L=-\tfrac{1}{2}\langle\gm,\bU_y\rangle-\mathrm{H}_\IL(\bU)\Bigr\}
\]
such that the bi\/-\/Hamiltonian hierarchy of commuting flows
for~\eqref{CFBoth} is composed by symmetries of~$\cEEL$. This is done
straightforwardly. We note that system~\eqref{CFBoth} is scaling\/-\/invariant
w.r.t.\ the homogeneity weights $|\uu|=2$,\ $|\uv|=1$,\ $|\uw|=3/2$;
$|U|=0$,\ $|V|=1$,\ $|W|=1/2$; and $|D_x|=1$, $|D_t|=3/2$.
The symmetries\footnote{For $|s|=2$, the symmetry of~\eqref{CFp} is
$U_s=V_x$, $V_s=\tfrac{1}{3}U_x^3-\tfrac{1}{2}W_x^2$, $W_s=-U_xW_x$;
the flow with $|\tau|=3$ is given by $U_\tau=2U_xV_x+W_x^2$,
$V_\tau=\tfrac{1}{2}U_x^4+V_x^2-2U_xW_x^2$, $W_\tau=2V_xW_x-2U_x^2W_x$.
There is no symmetry of~\eqref{CFp} with time weight~$5/2$.}
of~\eqref{CFBoth} with time weights $3/2=|D_t|$, $2$,
and~$3$ fix the Hamiltonian $\mathrm{H}_\IL(\bU)$ for~$\cEEL$ uniquely.
The only system that is ambient w.r.t.\ the whole hierarchy of~\eqref{CFBoth}
is the wave equation
\[
\cE_\varnothing=\bigl\{U_{xy}=0,\ V_{xy}=0,\ W_{xy}=0\bigr\}.
\]

The wave equation~$\cE_\varnothing$ is exactly solvable, because the conditions
\begin{equation}\label{WaveIntegrals}
U_x,V_x,W_x\in\ker D_y\bigr|_{\cE_\varnothing},\quad
U_y,V_y,W_y\in\ker D_x\bigr|_{\cE_\varnothing}
\end{equation}
can be integrated immediately, and we
obtain~$\bU=\text{\textbf{\textit{f}}}(x)+\text{\textbf{\textit{g}}}(y)$.
At the same time, by the argument illustrated in this example,
we have naturally arrived to the definition
of a class of nonlinear hyperbolic systems such that
conditions~\eqref{WaveIntegrals} become nontrivial.
\end{example}

\begin{define}[\cite{SokolovUMN}]\label{DefLiouType}
A \emph{Liouville\/-\/type system} $\cEL$ is a system
$\{u_{xy}=F(u,u_x,u_y;x,y)\}$ of hyperbolic equations which possesses
the \emph{integrals} $w_1$,\ $\ldots$,\ $w_r$; $\bar{w}_1$,\ $\ldots$,\
$\bar{w}_{\bar{r}}\in C^\infty(\cEL)$ such that the relations
${D_y\bigr|}_{\cEL}(w_i)\doteq0$ and ${D_x\bigr|}_{\cEL}(\bar{w}_j)\doteq0$ hold by virtue ($\doteq$) of~$\cEL$, and such that
all conservation laws for $\cEL$ are of the form
$\bigl[f(x,[w])\,\Id x\bigr]\oplus\bigl[g(y,[\bar{w}])\,\Id y\bigr]$.
\end{define}     

\begin{example}
The $m$-\/component 2D~Toda chains~\eqref{IEToda} associated with semi\/-\/simple
complex Lie algebras~\cite{Leznov} constitute an important class
of Liouville\/-\/type systems, here $u=(u^1,\ldots,u^m)$.
By~\cite{Shabat}, they possess the complete sets of $2m$
integrals $w_1,\ldots,w_m$; $\bar{w}_1,\ldots,\bar{w}_{m}$ and are
integrable iff $K$ is the Cartan matrix. This class is covered by the
ansatz in Proposition~\ref{NoetherSymTh} below.
\end{example}

\begin{rem}\label{Vessiot}
Consider Liouville\/-\/type systems that possess complete sets
of $2m$ integrals: $r=m$ and $\bar{r}=m$.
By slightly narrowing the class of such equations, let us consider the
systems~$\cEL$ whose general solutions are parameterized by arbitrary
functions $f^1(x)$, $\ldots$, $f^m(x)$ and $g^1(y)$, $\dots$, $g^m(y)$.
These equations are represented as the diagrams~\cite{KarabanovaKiselev}
\[
\bigoplus\nolimits_{i=1}^m J^\infty(\pi^x)\oplus J^\infty(\pi^y)\xrightarrow{\text{sol}}\cEL
\xrightarrow{\text{int}}\bigoplus\nolimits_{i=1}^m J^\infty(\pi^x)\oplus J^\infty(\pi^y),
\]
where $\pi^x$ and $\pi^y$ are the trivial fibre bundles
$\BBR\times\BBR\to\BBR$ such that $f$, $w$ and $g$, $\bar{w}$
determine their sections, respectively;
the first arrow is given by the formulas for exact solutions and the
second arrow is determined by the integrals.
\end{rem}

\begin{rem}\label{RemLiouVsSym}
Consider now a class of hyperbolic quasilinear systems~$\cE_\infty$ whose
symmetries $\vph=\square(\phi)$ are determined by Frobenius linear differential
operators $\square$, c.f.\ Remark~\ref{RemIntegralManifolds} on
p.~\pageref{RemIntegralManifolds}. Definition~\ref{DefLiouType} of the
Liouville\/-\/type systems, which postulates the existence of the integrals,
is not equivalent to the definition of~$\cE_\infty$ with the functional freedom
in symmetries. However, the two notions are very close.
In Corollary~\ref{SymStructure}, see below, we prove that symmetries $\vph$ of
Euler\/--\/Lagrange Liouville\/-\/type systems~$\cEL$ are factored by differential
operators $\square$, $\bar{\square}$ that originate from
the integrals~$w$,\ $\bar{w}$ for~$\cEL$: we have\footnote{It is believed
that Liouville\/-\/type systems do not possess any symmetries other
than~\eqref{SymForLiou}, unlike, e.g., the KdV equation~\eqref{IKdV} that admits
the scaling and the Galilean symmetries in addition to its commutative
hierarchy~$\gA$.}
\begin{equation}\label{SymForLiou}
\vph=\square\bigl(\phi(x,[w])\bigr),\qquad
\bar{\vph}=\bar{\square}\bigl(\bar{\phi}(y,[\bar{w}])\bigr),
\end{equation}
where components of the sections $\phi$ and~$\bar{\phi}$ are arbitrary
smooth functions. We show that the operators $\square$, $\bar{\square}$ are
Frobenius, which yields decomposable operator
algebras~$\EuA=\square\cdot\BBR\oplus\bar{\square}\cdot\BBR$ of rank~two.
\end{rem}

Let us remark on the history of the problem. 
Scalar Liouville\/-\/type equations were studied
in~\cite{SokolovUMN} and their symmetries have been further analyzed
in~\cite{Demskoi2004}. Operators~$\square$ that factor Noether
symmetries of Euler\/--\/Lagrange Liouville\/-\/type systems $u_{xy}=F(u;x,y)$
were constructed in~\cite{TMPhGallipoli}.
The general case $u_{xy}=F(u$,\ $u_x$,\ $u_y$;\ $x,y)$ of the
Euler\/--\/La\-g\-ran\-ge
Liouville\/-\/type systems (see~\eqref{SquareForEL})
was considered 
in~\cite{Startsev2006};
however, no method for reconstructing the brackets
$\ib{\,}{\,}{\square}$ is described there.

The problem of construction of operators~$\square$ that assign
(possibly, not all) symmetries
of non\/--\/Euler\/--\/Lagrange Liouville\/-\/type systems to their integrals $w$,
$\bar{w}$ is much less transparent. A considerable progress has been achieved here
in a recent paper~\cite{SokStar}, see section~\ref{SecNonEL}, where it is shown
that the existence of differential operators~$\square$ is based on the existence
of~$\bar{w}$, and respectively for $\bar{\square}$ and~$w$.
We analyse mainly the former case of Euler\/--\/Lagrange systems.

\begin{example}\label{ExLiouE}
Consider the 
parametric extension of the scalar Liouville equation~\eqref{ELiou},
\begin{equation}\label{LiouE}
u_{xy}=\exp(2u)\cdot\sqrt{1+4\veps^2u_x^2}.
\end{equation}
This equation is ambient w.r.t.\ the hierarchy of Gardner's
deformation of the potential modified KdV equation,
see~\cite{KarabanovaKiselev}.
The contraction $\cU=\cU(\veps,[u(\veps)])$
from~\eqref{LiouE} to the
non\/-\/extended equation $\cU_{xy}=\exp(2\cU)$ is
$\cU=u+\tfrac{1}{2}\arcsinh(2\veps u_x)$; it
determines the third order integral
for~\eqref{LiouE} using the integral~\eqref{KdVSubst} at $\veps=0$,
see Example~\ref{ExLiou}. However, the regularized minimal integral
of second order for~\eqref{LiouE} is
\begin{equation}\label{MinIntLiouE}
w=\bigl(1-\sqrt{1+4\veps^2u_x^2}\bigr)\bigr/{2\veps^2}
   +u_{xx}\bigr/\sqrt{1+4\veps^2u_x^2};
\end{equation}
such that all other $x$-\/integrals for~\eqref{LiouE} are differential
functions of~\eqref{MinIntLiouE}.
The second integral for~\eqref{LiouE} is
$\bar{w}=u_{yy}-u_y^2-\veps^2\cdot\exp(4u)\in\ker D_x\bigr|_{\cE}$.
The operators $\bar{\square}=u_y+\tfrac{1}{2}D_y$ and
\begin{equation}\label{SquareE}
\square=
\tfrac{1}{2}(1+4 \veps^2 u_x^2-2 \veps^2 u_{xx})\cdot D_x+
u_x+4 \veps^2 u_x^3-2 \veps^2 u_{xxx}+
   \frac{12 \veps^4 u_x u_{xx}^2}{1+4 \veps^2 u_x^2}
\end{equation}
assign symmetries~\eqref{SymForLiou} of~\eqref{LiouE} to its integrals.
We emphasize that operators in the
family~\eqref{SquareE} assign \emph{higher} symmetries
$\vph=\square\bigl(\phi(x)\bigr)$ of~\eqref{LiouE} to functions on the
base of the jet bundle whenever $\veps\neq0$, while the operator
$\bar{\square}$ always determines point 
symmetries~$\bar{\vph}=\bar{\square}\bigl(\bar{\phi}(y)\bigr)$.

Both operators $\square$ and $\bar{\square}$ 
satisfy~\eqref{EqDefFrob}.
The bracket $\ib{p}{q}{\bar{\square}}=p_yq-pq_y$ for $\bar{\square}$
is familiar; the bracket induced in the inverse image of $\square$ is
calculated in appendix~\ref{AppLiouEBracket}.
The surprisingly high
differential orders of $\ib{\,}{\,}{\square}$ with respect to its
arguments and coefficients is motivated by the presence of higher
order derivatives of~$u$ in~\eqref{SquareE}.
\end{example}

\begin{rem} 
Deformations of the algebras of Frobenius operators appear by virtue of the Gardner
deformations for differential equations, see Example~\ref{ExLiouE}
and~\cite{KarabanovaKiselev}.
Let $\cE(\mu)$ and $\cE(\nu)$ be the extensions of an equation $\cE(0)$
and let $\gm_\veps\colon\cE(\veps)\to\cE(0)$ be the Miura contraction
at $\veps\in I\subset\BBR$.
Symmetries $\vph_\mu\in\sym\cE(\mu)$ and $\vph_\nu\in\sym\cE(\nu)$
induce the symmetries of~$\cE(0)$ through the contraction, although the
induced flows can be formal sums of infinitely many generators of
higher orders. However, consider the commutator of the (formal) symmetries.
For all $\veps(\mu,\nu)$ such that the commutator is lifted to a
true symmetry of~$\cE(\veps)$, we define the product
$\diamond\colon(\mu,\nu)\mapsto\mu\diamond\nu=\veps(\mu,\nu)$. Thus we
obtain the multiplication $M\diamond N=\{\veps\}\subset\BBR$ on the sets
$M=\{\mu\}$, $N=\{\nu\}\subset\BBR$. For the Liouville\/-\/type
Gardner's extension~\eqref{LiouE}, we further obtain the product
$\diamond\colon\bigl(\square(\mu),\square(\nu)\bigr)\mapsto
\square(\mu\diamond\nu)$ of the Frobenius operators~\eqref{SquareE}.
\end{rem}

\subsection{Frobenius operators for Lagrangian systems}
For Euler\/--\/Lagrange Liouville\/-\/type systems
$\cEEL=\{F\equiv\bE(\cL)=0\}$, the existence of
factorizations~\eqref{SymForLiou} for at least a part of symmetries
is rigorous and can be readily seen as follows.
For integrals~$w$ such that $D_y(w)=\nabla(F)$ and for any $I(x,[w])$,
the generating section $\psi_I=\Bigl[\nabla^*\circ\bigl(\ell_w^{(u)}\bigr)^*
\circ\bigl(\ell_I^{(w)}\bigr)^*\Bigr](1)$ for a conservation law
$[I\,\Id x]$ solves the equations $\ell_{\bE(\cL)}^*(\psi_I)\doteq0$
on~$\cEEL$. The Helmholtz condition~\eqref{Helmholz} for $\ell_{\bE(\cL)}$ implies
that the vector
\begin{equation}\label{SquareForEL}
\vph[\phi]=\Bigl[\nabla^*\circ\bigl(\ell_w^{(u)}\bigr)^*\Bigr]\bigl(
\phi(x,[w])\bigr)\in\ker\ell_{\bE(\cL)}\bigr|_{\cEEL}
\end{equation}
is a symmetry
of~$\cEEL$ for any $\phi=\bigl(\ell_I^{(w)}\bigr)^*(1)=\bE_w(I\,\Id x)$.
A standard reasoning (see Lemma~\ref{LBaseEuler} or
Corollary~\ref{SymStructure} below) shows that formula~\eqref{SquareForEL}
yields symmetries of the 
sys\-tem~$\cEEL$ even if sections~$\phi$
do not belong to the image of the variational derivative~$\bE_w$.

In this section, we construct a class of \emph{Frobenius}
operators~$\square$ associated with 
symmetries of Euler\/--\/Lagrange Liouville\/-\/type systems.
We express the operators~$\square$ and the
brackets~$\ib{\,}{\,}{\square}$ in terms of minimal integrals~$w$.
This is done by the following argument.

\begin{state}[\cite{TMPhGallipoli}]\label{NoetherSymTh}
Let $\kappa$ be a nondegenerate symmetric constant real $(m\times
m)$-\/matrix. Suppose that $\cL=\bigl[L\,\Id x\Id y\bigr]$ with the density
$L=-\tfrac{1}{2}\sum_{i,j}\kappa_{ij}u^i_xu^j_y-H_{\IL}(u;x,y)$ is the
Lagrangian of a Liouville\/-\/type equation $\cEL=\{\bE(\cL)=0\}$.
Let $\gm=\dd L/\dd u_y$ be Dirac's momenta~\cite{Dirac}
and $w(\gm)=(w^1,\ldots,w^r)$ be the minimal set of integrals for~$\cEL$
that belong to the kernel of ${D_y\bigr|}_{\cEL}$.
Then the adjoint linearization
\begin{equation}\label{Square}
\square=\bigl(\ell_w^{(\gm)}\bigr)^*
\end{equation}
of the integrals w.r.t.\ the momenta factors
all Noether symmetries~$\vph_{\cL}$ of~$\cEL$, which are given by
\begin{equation}\label{NoetherSym}
\vph_\cL = \square\bigl({\delta\cH}/{\delta w}\bigr)
\end{equation}
for any $\cH=\bigl[H(x,[w])\,\Id x\bigr]$.
\end{state}

\begin{proof}[Outline of the proof]
The assertion follows from
\begin{itemize}
\item 
the structure
$\psi=-D_x^{-1}\bigl(\bE_u(\cH)\bigr)$
of the generating sections of conservation laws for hyperbolic
systems~$u_{xy}=\kappa^{-1}\bigl(\delta H_{\IL}/\delta u\bigr)$
resolved w.r.t.\ the second\/-\/order derivatives,
\item 
the correlation
\[\frac{\delta}{\delta u}=\bigl(\ell_\gm^{(u)}\bigr)^*\circ
\bigl(\ell_w^{(\gm)}\bigr)^*\circ\frac{\delta}{\delta w}\]
between the variational derivatives w.r.t.\ $u$ and~$w$, and
\item 
the correlation $\psi=\kappa\vph_\cL$ between generating sections of
conservation laws and Noether symmetries~$\vph_\cL$
of systems~$u_{xy}=\kappa^{-1}\bigl(\delta H_{\IL}/\delta u\bigr)$,
see Theorem~\ref{ThNoether}.
\end{itemize}
\end{proof}

\begin{cor}\label{SymStructure}
Under the assumptions and notation of Proposition~\textup{\ref{NoetherSymTh},}
the section 
\[
\vph=\square\bigl(\vec{\phi}(x,[w])\bigr)\in\varkappa(\pi)
\]
is a 
symmetry of the Liouville\/-\/type equation~$\cEL$ for any
$\vec{\phi}={}^t(\phi^1,\ldots,\phi^r)\in\gf$.
\end{cor}

\begin{proof}\label{ProofBase}
The proof is standard and analogous to the one for
Lemma~\ref{LBaseEuler} 
with the only alteration in the jet space at hand.
Consider the jet bundle $J^\infty(\xi)$ over the fibre bundle
\begin{equation}\label{KdVBundle}
\xi\colon\BBR^r\times\BBR\to\BBR
\end{equation}
with the base $\BBR\ni x$
and the fibres $\BBR^r$ with coordinates $w^1,\ldots,w^r$.
By Proposition~\ref{NoetherSymTh}, the statement of the theorem is valid
for any $\smash{\vec{\phi}}\in\gf=\hat{\varkappa}(\xi)$
in the image of the variational derivative $\bE_w$, where $w=w[\gm]$.
Obviously, the image contains all
$r$\/-\/tuples $\smash{\vec{\phi}}$ whose components $\phi^i(x)\in
C^\infty(\BBR)$ are functions on the base of the new jet bundle.
The substitution $w=w\bigl[\gm[u]\bigr]\colon\gf\to\varkappa(\pi)$
converts sections~$\vec{\phi}\in\gf$
to elements of~$\Gamma\bigl(\pi_\infty^*(\xi)\bigr)$.
Now recall that $\square$ is an operator in total
derivatives~\eqref{CartanConnection}, whose action on $r$-\/tuples of
functions on the jet space $J^\infty(\pi)$
is defined by~\eqref{DefDxByJs} through their restriction onto
the jets of sections $w=\smash{\vec{\phi}}(x)$, whence the assertion follows.
\end{proof}

\begin{theor}\label{IspHO}
The operators~\eqref{Square} are Frobenius of second kind if the integrals~$w$ are
minimal. 
\end{theor}

This assertion is continued in Proposition~\ref{FormulaForSquare}
on p.~\pageref{FormulaForSquare}.

\begin{proof}[Proof of Theorem~\textup{\ref{IspHO}}]
The commutator of two Noether symmetries $\vph_{\cL}',\vph_{\cL}''$ is
a Noether symmetry~$\vph_\cL$,
and hence the conservation law corresponds to it. The geometry of the
Euler\/--\/Lagrange Liouville\/-\/type equations
$\cEL\simeq\{\kappa^{-1}\bE_u(\cL)=0\}$ is such that the
conservation law is represented
by an integral, $D_y(H)\doteq0$ on~$\cEL$. By assumption, the
integrals~$w$ that specify the symmetries
$\vph_{\cL}'=\square\bigl(\vec{\phi}'[w]\bigr)$ and
$\vph_{\cL}''=\square\bigl(\vec{\phi}''[w]\bigr)$ are minimal, meaning that
\emph{any} integral is a differential function of them, hence
$H=H(x,[w])$. Let the gauge of the minimal integrals be fixed.
Then the factorization~\eqref{NoetherSym} for the new
symmetry~$\vph_\cL$ follows from Proposition~\ref{NoetherSymTh}.
Under differential reparametrizations $w=w[\tilde{w}]$ of the integrals,
the sections $\phi=\delta\cH/\delta w$ are transformed by
$\phi=\bigl(\ell_{\tilde{w}}^{(w)}\bigr)^*(\tilde{\phi})$,
thence $\square$~is a well defined Frobenius operatos of second kind~\eqref{Frob2}.
\end{proof}


\begin{rem} 
Frobenius operators~$\square$ generate symmetries of Euler\/--\/Lagrange
Liouville\/-\/type systems using arbitrary $r$-\/tuples $\vec{\phi}$ of integrals.
Therefore let us consider columns of these $(m\times r)$-\/matrix operators
separately:
\begin{equation}\label{ExpandSquare}
\square=\Bigl(\bigl(\square_1\bigr),\cdots,\bigl(\square_r\bigr)\Bigr).
\end{equation}
Generally, the image of a $k$-th column $\square_k$ is not closed under the
commutation. This is the case when
the $k$-th components $\phi^k$ of sections $\vec{\phi}\in\gf$ are
coupled not only in the $k$-th component of the bilinear
bracket~$\ib{\,}{\,}{\square}$. Likewise, the commutators of symmetries
in the images of any two columns $\square_i$,\ $\square_j$ are decomposed
with respect to the images of other operators~$\square_1$,\ $\ldots$,\
$\square_r$ as~well.

However, the decomposition~\eqref{ExpandSquare} does not produce operator
algebras~$\EuA$ of rank~$r$. Indeed, under a reparametrization $\tilde{w}=
\tilde{w}[w]$ in the domain of a Frobenius operator~$\square$ of second kind,
which yields
$\bigl[\square\circ\bigl(\ell_{\tilde{w}}^{(w)}\bigr)^*\bigr](\tilde{\phi})=
\tilde{\square}(\tilde{\phi})$, the columns $\square^k$ are not transformed
individually. Therefore they are not an $r$-\/tuple of objects, but they do
constitute a single well\/-\/defined operator~\eqref{ExpandSquare}.
By construction, it generates the algebra~$\EuA$ of rank~$1$.
\end{rem}

\begin{example}\label{ExA2}
Consider the Euler\/--\/Lagrange 2D~Toda system associated with the simple Lie
algebra~$\mathfrak{sl}_3(\BBC)$, see~\cite{Leznov,Shabat},
\begin{equation}\label{A2Toda}
\cEToda=\Bigl\{ U_{xy}=\exp(2U-V),\ V_{xy}=\exp(-U+2V),\qquad
K=\left(\begin{smallmatrix}\phantom{+}2&-1\\
 -1&\phantom{+}2\end{smallmatrix}\right)\Bigr\}.
\end{equation}
The minimal integrals for~\eqref{A2Toda} are~\cite{LeznovSmirnovShabat,Shabat}
\begin{subequations}\label{IntA2}
\begin{align}
w^1&=U_{xx}+V_{xx}-U_x^2+U_xV_x-V_x^2,\label{IntA2First}\\
w^2&=U_{xxx}-2U_xU_{xx}+U_xV_{xx}+U_x^2V_x-U_xV_x^2.
\end{align}
\end{subequations}
All symmetries (
up to $x\leftrightarrow y$) of~\eqref{A2Toda} are of the form
$\vph=\square\bigl(\vec{\phi}\bigl(x,[w^1],[w^2]\bigr)\bigr)$, where
$\vec{\phi}={}^t(\phi^1,\phi^2)\in\hat{\varkappa}(\xi)$
and the $(2\times2)$-\/matrix operator in total derivatives is\footnote{%
\label{CorrectForBous}
For convenience, we have multiplied the second column of 
operator~\eqref{SquareA2} by $-3$, which of course does not preclude it from assigning symmetries of~\eqref{A2Toda} to \emph{arbitrary} arguments~$\phi^2$.
However, this correction, which is introduced by hand, is the only reason 
why the operator $(A_{ij})$ in Example~\ref{ExA2TodaPseudoAk} on
p.~\pageref{ExA2TodaPseudoAk} is not Hamiltonian;
see~\cite{TMPhGallipoli,Protaras} for the bi\/-\/Hamiltonian pair of the
Boussinesq hierarchy associated with the semi\/-\/simple Lie 
algebra~$\mathfrak{sl}_3(\BBC)$.}
\begin{equation}\label{SquareA2}
\square=\begin{pmatrix}
U_x+D_x & 2D_x^2+3U_xD_x +U_x^2+2U_xV_x-2V_x^2-U_{xx}+2V_{xx} \\
V_x+D_x & D_x^2 -2U_{xx}+V_{xx}+2U_x^2-2U_xV_x-V_x^2 \end{pmatrix}.
\end{equation}
By Theorem~\ref{IspHO}, the image of operator~\eqref{SquareA2} is closed
w.r.t.\ the commutation. The image of the first column, of first order, is
itself closed under commutation (see, e.g., \cite{TMPhGallipoli}),
but the image of the second column of
$\square=\bigl(\square_1,\square_2\bigr)$ is not.
Indeed, the bracket~$\ib{\,}{\,}{\square}$ in the domain of the entire
operator~$\square$ is equal to
\begin{align}
\ib{\vec{p}}{\vec{q}}{\square}^1&=\boxed{p^1_xq^1-p^1q^1_x}
  +3p^2q^1_{xx}-3p^1_{xx}q^2\notag\\
{}&{}\qquad{}+\underline{6p^2q^2_{xxx}-6p^2_{xxx}q^2
  +6w^1\cdot\bigl(p^2q^2_x-p^2_xq^2\bigr) },\label{A2SokBr}\\
\ib{\vec{p}}{\vec{q}}{\square}^2&=p^2_xq^1-p^1q^2_x+2p^1_xq^2-2p^2q^1_x
 +\underline{3p^2_{xx}q^2-3p^2q^2_{xx}}.\notag
\end{align}
Here we box the individual bracket $\ib{\,}{\,}{\square_1}$ for the
$(2\times1)$-\/matrix operator~$\square_1$, and we underline the couplings
of components in the domain of~$\square_2$; under commutation, they hit both images
of~$\square_1$ and~$\square_2$. The remaining terms specify the
bi\/-\/differential structural constants, which we denote by $\Gamma_{ij}^k$
with an obvious meaning. We have
\[
\Gamma^1_{12}=-3D_x^1\otimes\bun,\ \Gamma^1_{21}=3\cdot\bun\otimes D_x^2,
\ \Gamma^2_{12}=-\bun\otimes D_x+2D_x\otimes\bun,
\ \Gamma^2_{21}=D_x\otimes\bun-2\cdot\bun\otimes D_x.
\]
In Proposition~\ref{FormulaForSquare} we explain why the coefficient~$6w^1$
in~$\Gamma^1_{22}$ is expressed by a function of integrals for
the Liouville\/-\/type system~\eqref{A2Toda}.

Examples of Frobenius operators associated with the Dynkin
diagrams via the 2D~Toda chains~\eqref{IEToda}
are discussed in~\cite{Protaras}. 
\end{example}

The integrals $w[\gm]$ of Euler\/--\/Lagrange Liouville\/-\/type
systems~$\cEL$ determine the Miura substitutions from commutative
mKdV\/-\/type Hamiltonian hierarchies~$\gB$ of Noether symmetries
on~$\cEL$ to completely integrable KdV\/-\/type hierarchies~$\gA$ of
higher symmetries of the multi\/-\/component wave
equations~$\cE_\varnothing=\{s_{xy}=0\}$, see below. The potential
modified KdV equation~\eqref{IpmKdV}, which is transformed to the KdV
equation~\eqref{IKdV} by~\eqref{KdVSubst}, gives a natural example.
This is discussed in detail in~\cite{TMPhGallipoli}.

\enlargethispage{0.7\baselineskip}
The hierarchies share the Hamiltonians
$\cH_i[\gm]=\cH\bigl[w[\gm]\bigr]$ through the Miura substitution.
The Hamiltonian structures for the Magri schemes of $\gA$ and~$\gB$ are
correlated by the diagram
\begin{equation}\label{Diag}
\begin{diagram}
\cH_0 &&&& \cH_1 &&&& \cH_2 && {} \\
\dto_{\bE_s} & \rdto^{\bE_w} &&&
  \dto_{\bE_s} & \rdto^{\bE_w} &&&
  \dto_{\bE_s} & \rdto^{\bE_w} & {} \\
\Phi_0 & \pile{\lto^{\hat A_1} \\ \rto_{A_1}}
 & \phi_0 & \rto^{\hat A_k} &
 \Phi_1 & \pile{\lto^{\hat A_1} \\ \rto_{A_1}}
 & \phi_1 & \rto^{\hat A_k} &
 \Phi_2 & \pile{\lto^{\hat A_1} \\ \rto_{A_1}}  & \phi_2
\lefteqn{{}\quad\text{(hierarchy~$\gA$)}}\\
&& \dto^{\square} & & \uto_{\ell_w^\gm} &&
 \dto^{\square} && \uto_{\ell_w^\gm} &&
 \dto_{\square}  \\
&& \vph_1 & \pile{\rto^{\hat B_1} \\ \lto_{B_1}}
 & \psi_1 & \rto_{B_{k'}}  &
 \vph_2 & \pile{\rto^{\hat B_1} \\ \lto_{B_1}}
 & \psi_2 & \rto_{B_{k'}} & {\ldots}
\lefteqn{{}\quad\text{(hierarchy~$\gB$).}}\\
&&&\luto_{\bE_\gm} & \uto_{\bE_{u}} &&&
\luto_{\bE_\gm} & \uto_{\bE_{u}} \\
&&&& \cH_0 &&&&\cH_1 && {}
\end{diagram}
\end{equation}
Frobenius operators~$\square$ map cosymmetries~$\phi_i$
for the hierarchy~$\gA$ to symmetries~$\vph_{i+1}$ that belong to the
modified hierarchy~$\gB$. The first Hamiltonian structure
$\hat{B}_1=(\ell_\gm^{(u)})^*$ for~$\gB$ originates from the
differential constraint $\gm=\dd L/\dd u_y$ upon the coordinates~$u$
and the momenta~$\gm$ for~$\cEL$.

\begin{lemma}
Introduce the linear differential operator
\begin{equation}\label{Quattro}
\hat{A}_k=\square^*\circ\hat{B}_1\circ\square, 
\end{equation}
which is completely determined by the Euler\/--\/Lagrange
Liouville\/-\/type system~$\cE_\IL$.
The operator~\eqref{Quattro} is Hamiltonian and determines
a higher (that is, $k=k(\square,\gm)\geq2$) 
Poisson structure for the KdV\/-\/type hierarchy~$\gA$.
\end{lemma}

\begin{proof}
By construction, the bracket~\eqref{PoissonEquiv} given by~$\hat{A}_k$
satisfies the equality
\[
\bigl\{\cH_1[w],\cH_2[w]\bigr\}_{\hat{A}_k}=
   \bigl\{\cH_1[w[\gm]],\cH_2[w[\gm]]\bigr\}_{\hat{B}_1}.
\]
Therefore the left\/-\/hand side of the above equality is
bi\/-\/linear, skew\/-\/symmetric, and satisfies the Jacobi identity.
Fourth, it measures the velocity of the integrals~$w$ along a Noether
symmetry of~$\cEL$. Since evolutionary derivations are permutable with
the total derivative~$D_y$, the velocity $\{\cH_1,\cH_2\}_{\hat{A}_k}$
lies in $\ker D_y\bigr|_{\cEL}$ and hence its density is a differential function
of the minimal integrals~$w$.
\end{proof}

\begin{state}\label{FormulaForSquare}
The bracket $\ib{\,}{\,}{\square}$ in the inverse image of
operator~\eqref{Square}
is equal to the bracket $\smash{\ib{\,}{\,}{\hat{A}_k}}$, which is
induced by the Hamiltonian operator $\smash{\hat{A}_k}$ and which
is calculated by formula~\eqref{EqDogma},
\begin{equation}\label{CalculateSokBrViaHam}
\ib{\phi'}{\phi''}{\square}=\ib{\phi'}{\phi''}{\hat{A}_k}=
\Bigl(\ell_{\phi',\hat{A}_k}^{(w)}\Bigr)^*(\phi''),\qquad
\phi',\phi''\in\cosym\gA\subset\sym\cE_\varnothing.
\end{equation}
The coefficients of the Hamiltonian operator~$\hat{A}_k$ and of the
bilinear terms in the bracket $\ib{\,}{\,}{\square}$ are
differential functions of the integrals~$w$.
\end{state}

The multi\/-\/component wave equation $\cE_\varnothing=\{s_{xy}=0\}$,
whose symmetries contain the hierarchy~$\gA$ and such that 
$\hat{A}_1=\bigl(\ell_w^{(s)}\bigr)^*$ potentiates the image
of the Miura substitution, is not \emph{a priori} unique. Again, the
constraint between the coordinates~$s$ and the momenta~$w$ for~$\gA$
determines the first Hamiltonian operator $\hat{A}_1$ for~$\gA$,
but the constraint
appears apparently from nowhere\footnote{This difficulty of the theory
was pointed out to us by B.\,A.\,Dubrovin (private communication).};
the shift of the field or the frozen point argument~\cite{ArnoldKhesin}
are customary procedures here. Our paradoxal conclusion is
that the first structure $\hat{A}_1=A_1^{-1}$ for~$\gA$ is chosen such
that~$A_1$ factors the higher Hamiltonian structure for~$\gB$.
We thus have
\[
B_{k'}=\square\circ A_1\circ\square^*,\quad
k'=k'\bigl(\square,\bigl(\ell_w^{(s)}\bigr)^*\bigr)\geq2,
\]
which specifies the required nonlocalities.

This means that Frobenius operators and the factorizations they provide
are helpful in the bi\/-\/Hamiltonianity tests for integrable systems.
It is likely that one can reveal a similar origin of the nonlocal first
Hamiltonian structures for the Drinfeld\/--\/Sokolov
hierarchies~\cite{DSViniti84} associated with the Kac\/--\/Moody
algebras, whose Cartan matrices are degenerate.

Reciprocally, formula~\eqref{CalculateSokBrViaHam} calculates the
bi\/-\/differential 
brackets~$\ib{\,}{\,}{\square}$ 
induced by 
Frobenius operators~\eqref{Square}.
Setting to zero all but one components of sections $\phi'\cdot\bun_i$,
$\phi''\cdot\bun_j\in\gf=\hat\varkappa(\xi)$ in domains of
operators~$\square$ (respectively, except $i$-th and $j$-th components,
$1\leq i,j\leq r$), we obtain the 
coefficients in the bracket~$\ib{\,}{\,}{\square}$.
Proposition~\ref{FormulaForSquare} shows that the functional class for these
coefficients 
is very narrow.

\begin{example}[The modified Kaup\/--\/Boussinesq equation]\label{ExIra}
Consider an Euler\/--\/Lagrange extension
of the scalar Liouville equation~\cite{KarabanovaKiselev},
\begin{equation}\label{EqIra}
A_{xy}=-\tfrac{1}{8}A\exp\bigl(-\tfrac{1}{4}B\bigr),\qquad
B_{xy}=\tfrac{1}{2}\exp\bigl(-\tfrac{1}{4}B\bigr).
\end{equation}
Denote the 
momenta by
\begin{equation}\label{m2KBConstraint}
a=\tfrac{1}{2}B_x\quad\text{and}\quad b=\tfrac{1}{2}A_x.
\end{equation}
The minimal integrals of system~\eqref{EqIra} are
\[ 
w_1=-\tfrac{1}{4}a^2-a_x,\qquad w_2=ab+2b_x,
\] 
such that $\bar{D}_y(w_{i})\doteq0$ on~\eqref{EqIra}, $i=1,2$.
Hence the operator
\begin{equation}\label{SquareKBous}
\square=\Bigl(\ell_{w_1,w_2}^{(a,b)}\Bigr)^*=
\begin{pmatrix} -\tfrac{1}{4}B_x+D_x & \tfrac{1}{2}A_x\\
  0 & \tfrac{1}{2}B_x-2\,D_x \end{pmatrix}
\end{equation}
factors (Noether) symmetries of~\eqref{EqIra}.
The bracket $\ib{\,}{\,}{\square}$ induced in the inverse image of the
Frobenius operator~$\square$ is
\begin{equation}\label{SokKBous}  
\ib{\vec{\psi}}{\vec{\chi}}{\square} = \tfrac{1}{2}\cdot\begin{pmatrix}
 \psi^2_x\,\chi^1-\psi^1\,\chi^2_x +\psi^1_x\,\chi^2-\psi^2\,\chi^1_x
 \mathstrut \\
 \psi^2_x\,\chi^2-\psi^2\,\chi^2_x\mathstrut \end{pmatrix},
\end{equation}
where $\vec{\psi}={}^t(\psi^1,\psi^2)$ and $\vec{\chi}={}^t(\chi^1,\chi^2)$.
Split the operator $\square=\bigl(\square_1,\square_2\bigr)$ to two columns.
The 
bracket~\eqref{SokKBous} determines the structural constants~$\Gamma^k_{ij}$
for the commutation of images of $\square^1$,\ $\square^2$.
For example, setting the components $\psi^2,\chi^2$ and $\psi^1,\chi^1$
in~\eqref{SokKBous} to zero, respectively, we isolate the brackets
\[
\ib{\,}{\,}{\square_1}=0,\qquad
\ib{\,}{\,}{\square_2}=\tfrac{1}{2}\bigl(D_x\otimes\bun-\bun\otimes D_x\bigr)
\]
for~$\square_1$ and~$\square_2$.
Thus we have obtained an extension of the Wronskian\/-\/based bracket
for the second Hamiltonian structure of~KdV, see
Remark~\ref{RemManyOps}.
At the same time, bracket~\eqref{SokKBous} contains all other structural
constants for the rank one algebra generated
by~$\square=\bigl(\square_1,\square_2\bigr)$, e.g.,
$
\Gamma^1_{12}=\tfrac{1}{2}\bigl(D_x\otimes\bun-\bun\otimes D_x\bigr)$, and
$\Gamma^2_{12}=0$,
whence $\Gamma^1_{21}$ and~$\Gamma^2_{21}$ are deduced by the skew\/-\/symmetry.

Consider a symmetry of~\eqref{EqIra},
\begin{equation}\label{pm2KB}
A_t=\tfrac{1}{2}A_xA_{xx}+\tfrac{1}{2}\left(\tfrac{1}{4}A_x^2-1\right)B_x,\qquad
B_t=-2A_{xxx}+\tfrac{1}{8}A_xB_x^2-\tfrac{1}{2}A_xB_{xx}.
\end{equation}
Using the constraint~\eqref{m2KBConstraint}
between the coordinates~$A$,\ $B$ and the momenta
$a$,\ $b$, we cast~\eqref{pm2KB} to the canonical form
\begin{equation}\label{EqExCanon}
A_t=\frac{\delta\cH}{\delta a},\quad a_t=-\frac{\delta\cH}{\delta
A};\qquad
B_t=\frac{\delta\cH}{\delta b},\quad b_t=-\frac{\delta\cH}{\delta
B},
\end{equation}
where
$\cH=\bigl[\bigr(\tfrac{1}{32}A_x^2B_x^2+\tfrac{1}{4}A_xA_{xx}B_x
 +\tfrac{1}{2}A_{xx}^2-\tfrac{1}{8}B_x^2\bigr)\,\Id x\bigr]$.

Let us 
choose an equivalent 
pair of integrals
$
u=w_2$, 
$v=w_1+\tfrac{1}{4}w_2^2  
$. 
It is remarkable that the evolution of~$u$ and~$v$ along~\eqref{pm2KB},
which equals (see p.~\pageref{KB})
\begin{equation}\tag{\ref{KB}}
u_t=uu_x+v_x,\qquad v_t=(uv)_x+ u_{xxx},
\end{equation}
is the Kaup\/--\/Boussinesq system,
and \eqref{pm2KB}~is actually the potential twice\/-\/modified
Kaup\/--\/Boussinesq equation.
The right hand side of the integrable system~\eqref{pm2KB} belongs to
the image of the adjoint linearization
$\tilde{\square}=\bigl(\ell_{(u,v)}^{(a,b)}\bigr)^*$.
The Frobenius operator $\tilde{\square}$ factors
the \emph{third} Hamiltonian structure
$\hat{A}_3^{\text{KB}} = \tilde{\square}^* \circ
\bigl(\ell_{(a,b)}^{(A,B)}\bigr)^* \circ \tilde{\square}$
for~\eqref{KB}; we have $k=3$ and
\[
\hat{A}_3^{\text{KB}}=
\begin{pmatrix}
u\,D_x+\tfrac{1}{2}u_x & D_x^3+(\tfrac{1}{4}u^2+v)\,D_x+\tfrac{1}{4}(u^2+2v)_x \\
D_x^3+(\tfrac{1}{4}u^2+v)\,D_x+\tfrac{1}{2}v_x &
   \tfrac{1}{2}(2u\,D_x^3+3u_x\,D_x^2+(3u_{xx}+2uv)D_x+u_{xxx}+(uv)_x)
\end{pmatrix}.
\]
By Proposition~\ref{FormulaForSquare},
the bracket $\ib{\,}{\,}{\tilde{\square}}$ is equal to
$\ib{\,}{\,}{\hat{A}_3^{\text{KB}}}$,
which is given by formula~\eqref{EqDogma}. We obtain
\begin{equation}\label{AlternativeNotation}
\ib{\vec{\psi}}{\vec{\chi}}{\tilde{\square}} =
\ib{\vec{\psi}}{\vec{\chi}}{\hat{A}_3^{\text{KB}}} =
\begin{pmatrix}
\vec\psi\cdot\nabla_1(\vec\chi)-\nabla_1(\vec\psi)\cdot\vec\chi \\
\vec\psi\cdot\nabla_2(\vec\chi)-\nabla_2(\vec\psi)\cdot\vec\chi
\end{pmatrix},
\end{equation}
where
$\nabla_1= - \tfrac{1}{2}\left(\begin{smallmatrix} D_x & 0 \\
   u\,D_x & D_x^3+v\,D_x \end{smallmatrix}\right)$
and 
$\nabla_2= - \tfrac{1}{2}\left(\begin{smallmatrix} 0 & D_x \\
   D_x & u\,D_x \end{smallmatrix}\right)$.
Here we use an alternative notation~\eqref{AlternativeNotation}
for the components of the bracket $\ib{\,}{\,}{\tilde{\square}}$
that acts by the differential operators $\nabla_1$ and~$\nabla_2$
on one factor in each coupling.

Finally, the operator $\hat{A}_1=\left(\begin{smallmatrix}0 & D_x\\
D_x & 0\end{smallmatrix}\right)$ is the first Hamiltonian structure
for~\eqref{KB}; its inverse $A_1=\hat{A}_1^{-1}$ factors the second
Hamiltonian structure $B_2=\tilde{\square}\circ A_1\circ\tilde{\square}$
for~\eqref{pm2KB}.
\end{example}

We recall that the operators~\eqref{Square} determine
recursions for the symmetry algebras~$\sym\cEEL$ of Liouville\/-\/type
systems.

\begin{state}[see~\cite{JMathSci2004,TodaLawsActa}]
Let $\cEEL$~be an Euler\/--\/Lagrange Liouville\/-\/type system that
meets the assumptions of Proposition~\ref{NoetherSymTh},
$w[\gm]$~be the integrals, and $\square$~be Frobenius operators~\eqref{Square}.
Let the $r$-\/tuples $\vec{\phi}={}^t(\phi^1,\ldots,\phi^r)\in\gf=
\hat{\varkappa}(\xi)\bigr|_{w}$ 
be the same as in Corollary~\ref{SymStructure}, see~\eqref{KdVBundle} on
p.~\pageref{KdVBundle}. Then the $m$-\/tuples
\[
\mathbf{\omega}_{\vec{\phi}}=\square\Bigl(\Id_\cC\vec{\phi}
 \bigl(x,\bigl[w[\gm]\bigr]\bigr)\Bigr)
\]
of Cartan's $1$-\/forms are generating sections of local recursion operators
$R_{\vec{\phi}}\colon\vph\mapsto\cEv_\vph\inner\mathbf{\omega}_{\vec{\phi}}$
for $\vph\in\sym\cEEL$.
\end{state}

\begin{conjecture}
Under the above assumptions and notation,
the recursion operators~$R_{\vec{\phi}}$ constitute infinite\/-\/dimensional
Frobenius\/--\/Lie algebras by
\[
\bigl[R_{\vec{\phi}'}(\vph'),R_{\vec{\phi}''}(\vph'')\bigr]=
 \int\limits_{\vec{\phi}\in\gf}
R_{\vec{\phi}}\Bigl(
\cEv_{R_{\vec{\phi}'}(\vph')}(\vph'')\cdot\delta(\vec{\phi}-\vec{\phi}'')
-\cEv_{R_{\vec{\phi}''}(\vph'')}(\vph')\cdot\delta(\vec{\phi}-\vec{\phi}')
+\Gamma^{\vec{\phi}}_{\vec{\phi}',\vec{\phi}''}\bigl(\vph',\vph''\bigr)
\Bigr)\,\Id\vec{\phi},
\]
where $\vec{\phi}',\vec{\phi}''\in\gf=\hat{\varkappa}(\xi)\bigr|_{w}$
and $\vph',\vph''\in\sym\cEEL$ are any symmetries of the
Euler\/--\/Lagrange Liouville\/-\/type systems~$\cEEL$.
\end{conjecture}

\subsection{Non\/-\/Lagrangian Liouville\/-\/type systems}%
\label{SecNonEL}
Let $\cEL=\{F=0\}$ be a Liouville\/-\/type system; now it may not be
Euler\/--\/Lagrange. Let $w\in\ker D_y\bigr|_{\cEL}$ be a section composed by
minimal integrals for~$\cEL$, thence $D_y(w)=\nabla(F)$, see~\eqref{Divergence}
on p.~\pageref{Divergence}. By construction of Liouville\/-\/type systems,
hyperbolic equations in~$\cEL$ are independent from each other. Therefore
$\cEL$ is both normal and $\ell$-\/normal~\cite{ClassSym,Opava}, meaning that
$\Delta(F)=0$ or $\Delta\circ\ell_F\doteq0$ on~$\cEL$ implies $\Delta=0$,
respectively. By this argument, a section $\vph\in\varkappa(\pi)$ is a
symmetry of a Liouville\/-\/type system~$\cEL$ if and only if the evolutionary
vector field~$\cEv_\vph$ preserves the integrals,
\[
D_y\bigl(\cEv_\vph(w)\bigr)=\cEv_\vph(\nabla)(F)+\nabla\bigl(\cEv_\vph(F)\bigr)
\doteq\nabla\bigl(\ell_F(\vph)\bigr)\text{ on~$\cEL$.}
\]
Consider the operator equation
\[
D_y\circ\ell_w\doteq\nabla\circ\ell_F\text{ on~$\cEL$.}
\]
If, hypothetically, a differential
operator $\square\colon\gf\to\varkappa(\pi)$
such that $\ell_w\circ\square\colon\ker D_y\bigr|_{\cEL}\to\ker D_y\bigr|_{\cEL}$
were constructed, which thus resembles the right inverse of~$\ell_w$, then it
would assign symmetries of a Liouville\/-\/type system~$\cEL$ to arbitrary
$r$-\/tuples of the integrals, see~\eqref{SymForLiou}.

In a recent publication~\cite{SokStar}, the following fundamental
result has been obtained: 
\begin{quote}
\emph{Suppose $\cEL=\{F=0\}$ is an $m$\/-\/component
Liouville\/-\/type system that admits
complete sets $w\in\ker D_y\bigr|_{\cEL}$ and $\bar{w}\in\ker D_x\bigr|_{\cEL}$
of~$m$ integrals which are independent 
on a dense open subset of~$\cEL$.
Let an integer $k\gg1$ be sufficiently large, obeying an estimate in
Lemma~\textup{4} of \textit{loc.\ cit}. Consider the operator equation
for $\square\colon\gf\to\varkappa(\pi)$,
\begin{equation}\label{OpEqSS}
\ell_w\circ\square=\bun_{m\times m}\cdot D_x^k\mod\CDiff_{<k}\Bigl(\ker
  D_y\bigr|_{\cEL}\to\ker D_y\bigr|_{\cEL}\Bigr),
\end{equation}
whose right\/-\/hand side is not fixed \textit{a priori}. Consider the
approximation
\[\tilde{\square}
\approx\sum_{i=0}^{k-\ord_x(w)-1}
   \alpha_i\cdot D_x^{k-\ord_x(w)-i}+O(1)
\]
of a formal inverse to~$\ell_w^{(u)}$, and modify its columns following
the constructive procedure of Lemma~\textup{4} in~\textit{loc.\ cit}.
This yields the linear differential operator $\square$
in total derivatives that solves equation~\eqref{OpEqSS} exactly.}
\end{quote}
The profound result quoted above establishes the truncation from below for
the se\-qu\-en\-ce of coefficients 
of~$\square$ that satisfy
some recurrence relations; most remarkably, this is a consequence of the
presence of a complete set $\bar{w}\in\ker D_x\bigr|_{\cEL}$
of $y$-\/integrals for~$\cEL$.

Whenever a local operator in total derivatives~$\tilde{\square}=
\bigl(\tilde{\square}^1,\ldots,\tilde{\square}^r\bigr)$ is known, the differential
order of its $i$-th column~$\tilde{\square}^i$ can be decreased if the
corresponding integral $\tilde{w}^i$ is not minimal. Indeed,
the reduction $\tilde{w}^i=D_x(w^i)$ and the relation $\ell_{\tilde{w}^i}=
D_x\circ\ell_{w^i}$ means that the order~$k$ of the right\/-\/hand side of the
equation $D_x\circ\ell_{w^i}\circ\tilde{\square}^i=D_x^k\mod\CDiff\bigl(\ker
D_y\bigr|_{\cEL}\to\ker D_y\bigr|_{\cEL}\bigr)$ has been increased artificially.

Therefore the most general form of the determining equation~\eqref{OpEqSS} is
\begin{equation}\tag{\ref{OpEqSS}${}'$}\label{OpEq}
\ell_w\circ\square\in\CDiff\Bigl(\ker
  D_y\bigr|_{\cEL}\to\ker D_y\bigr|_{\cEL}\Bigr),
\end{equation}
with an \emph{a priori} unfixed operator $\sum_{i=0}^{k\gg1}\Omega_i\bigl(x,
[w]\bigr)\cdot D_x^{k-i}$ in the right\/-\/hand side, here
$\Omega_i\in\ker D_y\bigr|_{\cEL}$ are $(r\times r)$-\/matrices.
Likewise, the operator~$\square$ itself is defined up to right multiplication
by $\CDiff$-\/operators from the same ring.
As soon as the minimal integrals~$w$ are differentiated a suitable number
of times such that their differential orders coincide in the end, and
under assumption that the determinant
$\det\Omega_k
$ does not vanish almost
everywhere on~$\cEL$, the topmost coefficient in the r.h.s.\ of~\eqref{OpEq}
can be set equal to unit, as in~\eqref{OpEqSS}. Indeed, this is achieved by
a solution $\tilde{\square}\mathrel{{:}{=}}\square\circ\Omega_k^{-1}$
of~\eqref{OpEq}.
The coefficients $(\Omega_i)_{\mu\nu}$ of the product
$\ell_{\tilde{w}}\circ\tilde{\square}$ are generally not equal to the
Kronecker symbols $\delta_{\mu\nu}\cdot\delta^k_i$. Thus the
operator $\square\circ D_x^{-k}$ is not the right inverse of~$\ell_w$.
We conclude that equation~\eqref{OpEq} describes the problem of
``right\/-\/inverting'' operators in a ring with nontrivial zero divisors.
A constructive algorithm for solving it has been given
in~\cite{SokStar}. However, in most cases the image of the resulting
operator~$\tilde{\square}$ does not span the entire $x$\/-\/component
of the Lie algebra $\sym\cEL$, and such operators are generally not
Frobenius.

\begin{example}
Consider the Liouville equation $\cE_{\text{Liou}}=\{u_{xy}=\exp(2u)\}$,
see~\eqref{ELiou}, its minimal integral~$w=u_{xx}-u_x^2$, and the operator
$\square=u_x+\tfrac{1}{2}D_x$ that assigns symmetries to integrals.
A straightforward calculation yields that
\[
\ell_{2w}\circ\square=D_x^3+4wD_x+2w_x
=\bun\cdot D_x^3\mod\CDiff\bigl(\ker D_y\bigr|_{\cELiou}\to\ker D_y\bigr|_{\cELiou}\bigr).
\]
\end{example}

\begin{example}\label{ExA2TodaPseudoAk}
Similarly, consider the 2D~Toda system~\eqref{A2Toda}, its minimal
integrals~\eqref{IntA2}, and the operator~$\square=\bigl(\square^1,
\square^2\bigr)$, which is given in~\eqref{SquareA2}. Then we have
\begin{align*}
\ell_{\frac{1}{2}w^1}\circ\square^1&= D_x^3+w^1D_x+\tfrac{1}{2}w^1_x,\\
\ell_{\frac{1}{3}w^1}\circ\square^2&= D_x^4+w^1D_x^2
  +\bigl[2w^1_x-3w^2\bigr]\cdot D_x +\bigl[w^1_{xx}-2w^2_x\bigr],\\
\ell_{w^2}\circ\square^1&= D_x^4+w^1D_x^2+3w^2D_x+w^2_x,\\
\ell_{\frac{1}{2}w^2}\circ\square^2&= D_x^5+2w^1D_x^3+3w^1_xD_x^2
  +\Bigl[3\bigl(w^1_{xx}-w^2_x\bigr)-(w^1)^2\Bigr]\cdot D_x
  +\Bigl[w^1_{xxx}-\tfrac{3}{2}w^2_{xx}+w^1w^1_x\Bigr].
\end{align*}
After the correction $\tilde{w}^1=w^1_x$, $\tilde{\square}^1=\square^1\circ
D_x$ of differential order for the first integral~\eqref{IntA2First},
we obtain the matrix $\Omega_5=\left(\begin{smallmatrix}2&3\\
1&2\end{smallmatrix}\right)$ with unit determinant at the top of the operator
in the right\/-\/hand side of~\eqref{OpEq}. This yields a solution
$\tilde{\square}\circ\Omega_5^{-1}$ of equation~\eqref{OpEqSS}.

We refer to footnote~\ref{CorrectForBous} on p.~\pageref{CorrectForBous} and 
to~\cite{Protaras} for further comments on this example, which is related to the Boussinesq hierarchy. Actually, the operator $\ell_w\circ\square$ is
\emph{Hamiltonian} whenever the second column, $\square^2$ in~$\square$, 
is taken in its original scaling (here it has been divided by~$-\tfrac{1}{3}$).
\end{example}

\begin{lemma}\label{LCoeffInKernel}
If a solution~$\square$ of~\eqref{OpEq} for a Liouville\/-\/type system~$\cEL$
is Frobenius, then all coefficients
of the bracket~$\ib{\,}{\,}{\square}$ belong to~$\smash{\ker D_y\bigr|_{\cEL}}$.
\end{lemma}

\begin{proof}
By assumption, we have that $\bigl(D_y\circ\ell_w^{(u)}\circ\square\bigr)
\bigl([\phi',\phi'']_\square\bigr)\doteq0$ for
$\phi',\phi''(x,[w])\in\gf$. This equals
\[
0\doteq\bigl(D_y\circ\underline{\ell_w^{(u)}\circ\square}\bigr)
\Bigl(\cEv_{\square(\phi')}(\phi'')-\cEv_{\square(\phi'')}(\phi')+
\ib{\phi'}{\phi''}{\square}\Bigr)
\doteq\bigl(\ell_w^{(u)}\circ\square\bigr)\bigl(D_y
\bigl(\ib{\phi'}{\phi''}{\square}\bigr)\bigr),
\] 
because the underlined composition satisfies~\eqref{OpEq}. Clearly,
$D_y(\phi')$ and $D_y(\phi'')$ vanish on~$\cEL$ for arbitrary
$\phi',\phi''\in\gf$. For the same reason, not only the whole bracket
$\ib{\phi'}{\phi''}{\square}$, but each particular coefficient
standing at the bilinear terms lies in $\ker D_y\bigr|_{\cEL}$.
\end{proof}

\newpage
\section*{Conclusion}
We introduced the coordinate\/-\/independent definition of
Frobenius linear differential
operators $A\colon\Omega^1(\xi_\pi)\to\gothg(\pi)$
in total derivatives, whose images constitute Lie subalgebras
in the algebra $\gothg(\pi)$ of
evolutionary vector fields. We demonstrated that,
in presence of Miura\/-\/type
substitutions~$w\colon J^\infty(\pi)\to\Gamma(\xi)$,
there are two main classes of such operators.
They generalize recursion and Hamiltonian operators, whenever sections of
fibre bundles in their domains~$\Omega^1(\xi_\pi)$ obey vector and covector
transformation laws under a change of coordinates, respectively.
We defined the notions of linear and strong compatibile Frobenius operators
and gave examples of compatible Frobenius structures of each kind,
indicating both the matrix operators~$A$ and the types of transformations
$\psi\mapsto\tilde{\psi}$ for pre\/-\/images~$\psi$ of vector
fields~$\cEv_{A(\psi)}$.

We solved a long\/-\/standing problem in geometry of integrable systems:
We gave a complete description of higher symmetry algebras for hyperbolic
Euler\/--\/Lagrange equations~$\cEL$ of Liouville\/-\/type (in particular, for
all 2D~Toda chains associated with semi\/-\/simple complex Lie algebras of rank~$r$).
To this end, we obtained an explicit formula~\eqref{Square} for
Frobenius differential operators~$\square$ of second kind
that assign (Noether) symmetries
$\vph=\square\bigl(\vec{\phi}\bigl)$ to arbitrary $r$-\/tuples
$\vec{\phi}\bigl(x,[w]\bigr)$ (respectively, to $\vec{\phi}=\delta\cH/\delta w$
for any~$\cH\bigl(x,[w]\bigr)$).
Second, we calculated the commutation relations in $\sym\cEL$,
and we proved that the arising coefficients of the brackets
$\ib{\,}{\,}{\square}$ are differential functions of the substitutions~$w$.

We suggested a new algebraic operation on the linear spaces
$\EuA=\bigoplus_{i=1}^r\BBR\cdot A_i$ of operators
$A_i\colon\Omega^1(\xi_\pi)\to\gothg(\pi)$ whenever the spaces
$\sum_i\img A_i$ are closed under commutation. In particular, we considered
the spaces of both linear and strong compatible Hamiltonian operators for
the Magri schemes. Independently, we associated another
class of such spaces to Liouville\/-\/type systems.
Using the Lie algebra structure of vector fields~$\cEv_\vph$ with
$\vph\in\img A_\ell$, we defined a Lie\/-\/type bracket
$\ob{A_i}{A_j}=\sum_{k=1}^r A_k\circ\bc_{ij}^k$ of operators~$A_i,A_j\in\EuA$.
This operation on linear spaces of Frobenius operators extends the
associative composition 
and the Schouten bracket 
of recursions~$\gothg(\pi)\to\gothg(\pi)$.

We have analysed the algebraic and geometric properties of Frobenius
structures.
We described flat connections in the triples $\bigl(\Omega^1(\xi_\pi),
\gothg(\pi),A\bigr)$ such that bi\/-\/differential Christoffel symbols
are encoded by the structural constants~$\bc_{ij}^k$ of the operator
algebras~$\EuA$.
We derived a condition when Frobenius recursions,
which are solutions of the classical Yang\/--\/Baxter
equation~\eqref{IFrobRec} for the Lie algebra~$\gothg(\pi)$,
generate chains of nontrivial deformations for the standard 
structure 
on~$\gothg(\pi)$. By our Counterexample~\ref{StuckAt}, the concept of
Lie algebroids over the infinite jet spaces does not repeat the case of
finite\/-\/dimensional base manifolds, when the anchors are determined by
the Lie algebra homomorphisms~$A$.
Therefore we assigned formal differential complexes over Lie algebras
$\Omega^1(\xi_\pi)$ to Frobenius operators~$A$, such that
the representations of the differentials in the
complexes through homological vector fields
restores one of equivalent constructions of Lie algebroids.

Our approach not only incorporates the standard Hamiltonian formalism for PDE,
but also generates new nonlinear models.
A rigorous parallel between the bi\/-\/differential and classical
affine geometry allows to interpret known equations of string theory as
the master equations in the setting over the jet spaces.
The suggested algebraic technique can be especially efficient
whenever off\/-\/shell considerations are in order, without any restriction
onto differential equations. With all this,
we intend to contribute to a profound relation between 
integrable systems 
and Lie algebras, which is well acknowledged in mathematical
and theoretical physics~\cite{KontsZaboronsky,Dorfman,DSViniti84,
VdLeurKac,MiShSok,Reyman}.  



\subsection*{Discussion}
Open problems concerning with Frobenius structures can be
assembled in two categories: questions about geometry of involutive
distributions generated by evolutionary fields in the images of Frobenius
operators, and application of the formalism to integrable systems.

\subsubsection*{Geometric aspects}
Answering a motivating question of M.~Kontsevich, who asked us whether
Frobenius operators
are the anchors in Lie algebroids over infinite jet spaces, we were stuck
at Counterexample~\ref{StuckAt} for $\hat{A}_1^{\text{KdV}}$. We proposed
a way out, which is based on the formal complex~\eqref{FrobComplex}
and the representation of its differential via the
homological vector fields (in the spirit of the Koszul\/--\/Tate
resolution for Lagrangian systems). At the same time, the difficulty means that
a correct \emph{variational} analogue of the Leibnitz
rule~\eqref{LeibnitzInAlg} has not been recovered yet. Variational
analogues on~$J^\infty(\pi)$ are known from~\cite{Lstar,Opava} for vector fields, forms, commutators and Schouten brackets, for symplectic structures,
etc.\ (see Table~\ref{BiTableEnv} on p.~\pageref{BiTableEnv}) but,
to the best of our knowledge, the bare
Leibnitz rule itself has never been under debate. Hence a profound analysis
of Koszul\/--\/Schouten structures on~$\Omega^k(M^n)$, which was performed
in~\cite{YKSMagri}, remains without an analogue on
\[
\Omega^k(\xi_\pi)\stackrel{\text{def}}{{}={}}
\bigotimes\nolimits^k\Omega^1(\xi_\pi),
\]
leaving our \emph{ad hoc} construction of Lie algebroids
over~$J^\infty(\pi)$ so implicit.

We have already indicated that standard classification problems arise for
the algebras~$\EuA$ of Frobenius operators with the brackets
$\ob{A_i}{A_j}=\sum_k A_k\circ\bc_{ij}^k$.
For Hamiltonian operators~$P$, we know two ways, \eqref{EqDogma}
and~\eqref{TwoGrads}, of calculating Dorfman's brackets $[\,,\,]_P$
on~$\Omega^1(\pi)$.  Is there a similar construction that yields
the bi\/-\/differential Christoffel
symbols~$\Gamma^k_{ij}\colon\Omega^1(\pi)\times\Omega^1(\pi)\to\Omega^1(\pi)$
for totally compatible Hamiltonian structures? More globally, is there any
homological obstruction for the existence of compatible Hamiltonian or recursion operators that constitute non\/-\/commutative 
algebras? 
We expect to use here some techniques which will be more refine than
the ones used in
the proof of Lemma~\ref{L3Vect}. We notice that the assertion
of~\cite{JKVerb3Vect} was a reformulation of a much stronger
result in~\cite{Gessler} that states non\/-\/existence of nontrivial
variational $k$-\/vectors and $k$-\/forms at all~$k\geq3$ under given assumptions.
However, this is not our case of the Christoffel symbols $\Gamma^k_{ij}$,
which are not skew\/-\/adjoint in each argument and which are not tensors,
obeying the transformation rules for connection $1$-\/forms.
Let us indicate that, in these terms, the approach of geodesic motion is reproduced
for the problems of optimization and control in the theory of partial differential
equations.

The use of coordinate\/-\/independent constructions of geometry of
integrable systems~\cite{ClassSym,Dubr,Lstar,Opava}, see
section~\ref{SecHam}, allowed us to introduce the well\/-\/defined
notion of Frobenius operators. This would have been impossible within a
`simple' approach of local coordinates. We emphasize that now the
definition itself manifests the two main classes of the operators,
whose domains are the analogues of tangent and cotangent bundles to
smooth manifolds. The general case of Frobenius operators and their
Lie\/-\/type algebras will be the object of another paper.

\subsubsection*{Field\/-\/theoretic viewpoint}
The definition of Frobenius operators does not depend on
the number~$n$ of independent variables. In this paper, we passed from
examples of Frobenius operators for $1$D~evolutionary hierarchies
(the KdV\/-\/type 
equations) to factorizations of symmetry generators
for $2$D~Toda chains, which interpolate between KP and~mKP
(see~\cite{VdLeurKac}). We expect that Frobenius operators can serve a key
to construction of integrable systems in multidimension. By this argument,
we arrive at the problem of finding Frobenius structures for the flows of
the KP~hierarchy.

Another recent 
construction may have applications in this theory, as soon as examples of
Frobenius operators are known (e.g., for $(1+n)$-\/dimensional hydrodynamic
chains). 
Namely, consider Frobenius complexes~\eqref{FrobComplex}
with the differentials~$\mathbf{d}$, and
study their strong homotopy deformations, which involve
Schlessinger\/--\/Stasheff's algebras with Lie\/-\/type brackets of many
arguments (see~\cite{KontsZaboronsky,BarKonts,ForKac} and references therein).
We know that in our situation, which is based on evolutionary fields on the jet spaces,
the Wronskian determinants play a central role.
Hence their homology\/-\/preserving generalizations for functions of
many variables, which were defined in~\cite{ForKac}, will be useful.

Returning one step back from the homotopy Lie algebras to homological vector
fields, we recall their immanent presence in the Batalin\/--\/Vilkovisky
formalism, see~\cite{KontsZaboronsky}. A description of its ingredients
over the infinite jet spaces was our second motivation.

In this jet setting, Frobenius recursion operators are solutions of the
classical Yang\/--\/Baxter equation~\eqref{IFrobRec} for the Lie
algebra~$\gothg(\pi)$ of evolutionary vector fields. It is still not clear
how far one can pattern upon the standard approach to the Yang\/--\/Baxter
equations, what systems $\cE\subset J^\infty(\pi)$ can be obtained, are there
any reductions for them, and what is the physical significance of the models
whose state functions are transformed by derivations that belong
to~$\gothg(\pi)$.

Finally, we note once again that we have always remained in the non\/-\/graded
setting and that the differential operators were local.
We are advised 
to apply the formalism of Frobenius structures towards the study of
noncommutative differential equations, see Remark~\ref{RemManin}
on p.~\pageref{RemManin}, and, first, reveal these structures 
for integrable evolution equations on associative algebras~\cite{MikhAssoc}.
In addition, the use of difference Frobenius operators can be a fruitful
intermediate idea for discretization of integrable systems with free
functional parameters in their symmetries (e.g., Liouville\/-\/type difference
systems). The restriction~\eqref{IDefFrob} produces narrow classes
of difference operators such that the discrete systems are integral objects for
the symmetry algebras. 

\smallskip
We have described a generator of operator algebras~$\EuA$
with the commutation relations~\eqref{ICommuteOperators}.
In section~\ref{SecLiou}, we managed to
calculate the bi\/-\/differential structural constants $\bc_{ij}^k$
by using the fact that the integrals~$w$ for the
Liouville\/-\/type systems~$\cEL$ induce Miura's transformations
$w=w[u]\colon\gB\to\gA$ between bi\/-\/Hamiltonian hierarchies $\gA$ and
$\gB\subseteq\sym\cEL$. By~\cite{SokolovUMN}, \emph{many} substitutions are obtained in this way. Similarly, the class of
structures~\eqref{ICommuteOperators} given by Dynkin diagrams via the 2D~Toda
chains may be \emph{standard} in the following sense: Is it true that for all
spaces~$\EuA$ of Frobenius operators $A_i\colon\Omega^1(\xi_\pi)\to\gothg(\pi)$
of second kind there are Hamiltonian operators
$P\colon\hat{\varkappa}(\xi)\bigr|_{w}\to
\varkappa(\xi)\bigr|_{w}$ such that $\ib{\,}{\,}{A_i}=\ib{\,}{\,}{P}$\,?

\begin{Finalrem}[M.~Kontsevich, private communication]
Algebras with bi\/-\/differential structural constants appear naturally
in the BRST-\/formalism. Hence 
the algebras~$\EuA$ of Frobenius operators may have applications in the string theory.
\end{Finalrem} 


\newpage
\appendix
\section{Analogy between Hamiltonian ODE and PDE}\label{BiTable}
In Table~\ref{BiTableEnv} we track the geometric correspondence
between Hamiltonian ODEs and PDEs; we adapt the analogy to our needs,
and therefore it is forced to remain
incomplete. The distinction between the coordinates and momenta
in the PDE framework, which is implemented in section~\ref{SecLiou}
to (symmetries of) Euler\/--\/Lagrange
systems, is addressed in~\cite{Dirac,TMPhGallipoli}. The concept of
$\Delta$-\/coverings over PDE, which is convenient in
practical calculations that arise here, is developed
in~\cite{Lstar}, see Remark~\ref{RemLstar} on p.~\pageref{RemLstar}.
\begin{table}[hb]\caption{Hamiltonian ODE and PDE.}\label{BiTableEnv}
\begin{tabular}{cll}
$\bullet$ & \parbox[t]{59mm}{No internal structure of a time point
$t\in\BBR$;} & \parbox[t]{85mm}{\strut Smooth base
manifold $M^n$ and a fibre bundle $\pi\colon E^{n+m}\to M^n$.}\\
$\bullet$ & \parbox[t]{59mm}{Symplectic manifold
$M^{2n}\ni(p,q)$;} & \parbox[t]{85mm}{\strut Infinite jet space
$J^\infty(\pi)\xrightarrow{\pi_\infty}M^n$.}\\
$\bullet$ & \parbox[t]{59mm}{Components $X^i$ of vector fields
   $X\in\Gamma(TM^{2n})$ on $M^{2n}$;}
& \parbox[t]{85mm}{\strut Sections $\vph=(\vph^1[u]$,\ $\ldots$,\
$\vph^m[u])\in\varkappa(\pi)=\Gamma(\pi_\infty^*(\pi))$
of the induced fibre bundle.}\\
$\bullet$ & \parbox[t]{59mm}{Vector fields
   $X\in\Gamma(TM^{2n})$;} & \parbox[t]{85mm}{\strut Evolutionary
vector fields $\cEv_\vph\in\ID^v(J^\infty(\pi))$.}\\
$\bullet$ & \parbox[t]{59mm}{Integral trajectories
   $\bigl(p(t)$, $q(t)\bigr)\subset M^{2n}$ of the field $X$;} &
\parbox[t]{85mm}{\strut Solutions of the
autonomous equation $u_t=\vph$.}\\
$\bullet$ & \parbox[t]{59mm}{Lie algebra $\bigl(TM^{2n}$,
   $[\,,\,]\bigr)$ of vector fields on $M^{2n}$;} &
\parbox[t]{85mm}{\strut Lie algebra $\gothg(\pi)$ of evolutionary
   derivations.}\\
$\bullet$ & \parbox[t]{59mm}{The de Rham differential $\Id$;} &
\parbox[t]{85mm}{\strut The de Rham differential $\Id=\Id_h+\Id_\cC$
 split to horizontal and vertical parts w.r.t.~$\pi$.}\\
$\bullet$ & \parbox[t]{59mm}{The de Rham complex;} & \parbox[t]{85mm}{The
variational bi\/-\/complex and the $\cC$-\/spectral sequence.}\\
$\bullet$ & \parbox[t]{59mm}{The space of Hamiltonians
$\cH\in C^\infty(M^{2n})$;} & \parbox[t]{85mm}{\strut The highest
horizontal cohomology $\bar{H}^n(\pi)\ni\cH$.}\\
$\bullet$ & \parbox[t]{59mm}{The cotangent bundle
 $T^*M^{2n}$ and its sections $\psi\colon
 TM^{2n}\to C^\infty(M^{2n})$;} & \parbox[t]{85mm}{\strut The dual module
$\hgf=\Hom_{\cF(\pi)}\bigl(\gf(\pi)$,$\bar{\Lambda}^n(\pi)\bigr)$ for a
horizontal module of sections of a bundle over $J^\infty(\pi)$.}\\
$\bullet$ & \parbox[t]{59mm}{The differential $\Id\cH$ of a
Hamiltonian $\cH$;} & \parbox[t]{85mm}{\strut Euler's operator $\bE$
as a restriction of $\Id_\cC$ to $\bar{H}^n(\pi)$, the `gradient'
$\bE(\cH)\in\hat{\varkappa}$.}\\
$\bullet$ & \parbox[t]{59mm}{Symplectic $2$-\/form
   $\omega\in\Omega^2(M^{2n})$,
   Poisson bi\/-\/vector $\cP\in\Gamma\bigl(\bigwedge^2TM^{2n}\bigr)$;}
& \parbox[t]{85mm}{\strut Hamiltonian
   operator $A\in\CDiff(\hgf,\varkappa)$ in total
   derivatives; $\gf\simeq\varkappa(\pi)$ for evolution equations.}\\
$\bullet$ & \parbox[t]{59mm}{Hamiltonian vector field $X_\cH$ such that
   $X_\cH\inner\omega=\Id\cH$;} & \parbox[t]{85mm}{\strut Sections
   $\vph=A(\psi)$, $\psi\in\hat{\varkappa}(\pi)$.}\\
$\bullet$ & \parbox[t]{59mm}{\strut The Poisson bracket
   $\{\cH_1$,\ $\cH_2\}=\omega\bigl(X_{\cH_1}$,\ $X_{\cH_2}\bigr)=
     X_{\cH_1}\inner\Id\cH_2$;} & \parbox[t]{85mm}{\strut The Poisson
  bracket $\{\cH_1$,\ $\cH_2\}_A=
  \bigl\langle\bE(\cH_1)$,\ $A\bigl(\bE(\cH_2)\bigr)\bigr\rangle
     =\cEv_{A(\bE(\cH_2))}(\cH_1)$.}
\end{tabular}
\end{table}

\section{Reconstruction
of the brackets~$\ib{\cdot}{\cdot}{A}$}\label{AppLiouEBracket}
In this appendix we describe an inductive procedure that assigns the
bracket $\ib{\cdot}{\cdot}{A}$ to a nondegenerate Frobenius
operator~$A$, see~\eqref{EqOplusB}.
The bracket $\ib{\,}{\,}{A}$ may be not contained in our knowledge that $A$ is
Frobenius if, e.g., the operator determines the factorization of
symmetries of a Liouville\/-\/type system and has minimal differential
order. 
This is precisely the case of operator~\eqref{SquareE}, which is used
in Example~\ref{ExBrByA} as an illustration.

\begin{rem}\label{RemLstar}
The algorithm we suggest is based on the use of the
$\Delta$-\/coverings~\cite{Lstar} over the jet spaces~$J^\infty(\pi)$.
In our case, it amounts to `forgetting'
the $\cF(\pi)$-\/(sub)module structure
of~$\Omega^1(\xi_\pi)$ and treating it
as a jet (super-)\/bundle over $J^\infty(\pi)$,
see~\cite{Lstar} for details. Hence, instead of calculating
$D_x(\psi[u])$ for $\psi\in\Omega^1(\xi_\pi)$, one introduces the variable
$\psi_x$ and so on,\footnote{Note that an additional relation $\psi_y=0$
can be introduced for the new variables~$\psi$
that imitate the integrals. These integrals are the sections
in domains of Frobenius operators
(see~\eqref{SymForLiou} and~\eqref{KdVBundle}) for Liouville\/-\/type systems.}
setting the derivatives $\cEv_\vph(\psi)$ in the
Koszul bracket to zero, see~\eqref{EqOplusB}. Consequently, only the
derivatives of coefficients of the operator~$A$ contribute to
the left\/-\/hand side of~\eqref{EqOplusBBoth} while its right\/-\/hand
side becomes $A(\ib{\psi_1}{\psi_2}{A})$. This is the structure of
equations (\ref{EqBase}--\ref{EqStep}) below: they do not contain any
evolutionary summands involving domains of~$A$ in the right\/-\/hand sides.

Also, the nature of the assumption~\eqref{NoKernelsIntersect} becomes clear.
Indeed, this construction tells us that there are no notrivial kernels
for restrictions of the nondegenerate operators onto
the new jet bundles with sections~$\psi_1$,~$\psi_2$.
Note that the same operators~$A$ may have kernels which are spanned by
certain sections $\psi[u]\in\Gamma(\pi_\infty^*(\xi))$ of the induced
fibre bundles, see Example~\ref{ExNondegenerateWithKernel} on
p.~\pageref{ExNondegenerateWithKernel}.
\end{rem}

For brevity, let us technically assume that $A=\|\sum_k A^{ij}_k\,D_x^k\|$ is a
matrix operator in~$D_x$, where $A^{ij}_k\in\cF(\pi)$.
We use the notation $\bun_i$ for the basic sections
$\psi={}^t(0_1,\ldots,0_{i-1},1_i,0_{i+1},\ldots,0_r)$
of~$\Omega^1(\xi_\pi)$, whence
\begin{equation}\label{LookFromOutside}
\psi=\sum_{i=1}^r\psi^i\cdot\bun_i,\qquad \psi^i\in\cF(\pi).
\end{equation}
Suppose further that the operator~$A$ is nondegenerate,
see p.~\pageref{NoKernelsIntersect},
\begin{equation}\tag{\ref{NoKernelsIntersect}}
\bigcap_k\ker A_k=\{0\}\quad\text{for }A=\sum_k A_k\cdot D_x^k.
\end{equation}
Hence we encounter no difficulties when resolving inhomogeneous equations
$A\bigl(\{\!\{\psi_1$, $\psi_2\}\!\}_{A}\bigr)=\vph$ w.r.t.\ the brackets
$\ib{\,}{\,}{A}\in\CDiff(\bigwedge^2\Omega^1(\xi_\pi),\Omega^1(\xi_\pi))$,
which are bilinear in $\psi_1,\psi_2\in\Omega^1(\xi_\pi)$.

Let the (yet unknown) bracket be
\[
\ib{\psi_1}{\psi_2}{A}=\sum_{i,j,k=1}^r
  c_{ijk}^{\alpha\beta}\cdot(\psi_1^i)_\alpha
(\psi_2^j)_\beta\cdot\bun_k,    
\]
where $c_{ijk}^{\alpha\beta}\in\cF(\pi)$ and the condition
$c_{ijk}^{\alpha\beta}=-c_{jik}^{\beta\alpha}$ follows from the
skew\/-\/symmetry of the bracket~$\ib{\,}{\,}{A}$.
The coefficients $c_{ijk}^{00}$ can be nontrivial if the
dimension~$r$ of the fibres of~$\xi$
is $r>1$, see Example~\ref{ExdBousHuman} below.

The base of the algorithm is given by the Jacobi bracket of
the sections~$\bun_i$,~$\bun_j$: 
\begin{equation}\label{EqBase}
\bigl[A(\bun_i),A(\bun_j)\bigr] = A(c_{ijk}^{00}\cdot\bun_k).
\end{equation}
The choice $1\leq i\leq j\leq r$ yields $r(r-1)/2$ compatible systems of
$m$ equations. The components of sections in domains of~$A$
in the right\/-\/hand side of~\eqref{EqBase} are enumerated by~$k$.
Since $A$~is nondegenerate, the equations are solvable.
Actually, these systems are overdetermined
whenever the differential order of~$A$ is positive and hence the
left\/-\/hand sides of~\eqref{EqBase} and~\eqref{EqStep}, see below,
contain higher order derivatives of~$u$ that are not present among the
arguments of~$c^{\alpha\beta}_{ijk}[u]$.

The inductive step is made by using the sections $x^\alpha\cdot\bun_i$
and $x^\beta\cdot\bun_j$. We obviously have
\begin{equation}\label{EqStep}
\bigl[A(x^\alpha\cdot\bun_i),A(x^\beta\cdot\bun_j)\bigr] -
  A\Bigl(\sum_{0\leq\alpha'+\beta'<\alpha+\beta}
   (x^\alpha)^{(\alpha')}(x^\beta)^{(\beta')}c_{ijk}^{\alpha'\beta'}
   \cdot\bun_k\Bigr)
 = A\bigl(\alpha!\beta!c_{ijk}^{\alpha\beta}\cdot\bun_k\bigr),
\end{equation}
whence the coefficients $c^{\alpha\beta}_{ijk}$ on the diagonal
$\alpha+\beta=\text{const}$ are obtained one by one. 
Having passed through the diagonal $0\leq\alpha+\beta=\text{const}$, with
$\alpha\geq\beta$ or $\alpha\leq\beta$ in view of the relation
$c^{\beta\alpha}_{jik}=-c^{\alpha\beta}_{ijk}$, we check the condition
\begin{equation}\label{TerminalCondition}
\bigl[A(\psi),A(\chi)\bigr] =
   A\Bigl(\sum_{\alpha+\beta\leq\text{const}}
   c_{ijk}^{\alpha\beta}\cdot\psi_\alpha^i\chi_\beta^j\Bigr)
\end{equation}
that terminates the algorithm when holds. The differential order of
the bracket $\ib{\,}{\,}{A}$ with respect to its arguments is estimated
by calculating the Lie bracket $[A(\psi),A(\chi)]$
and taking into account the Leibnitz rule
in the right\/-\/hand side of~\eqref{EqOplusBBoth}.
We remark that the representation of jet coordinates $u_\sigma$ using
powers $x^\sigma$ of base variables is standard in geometry of differential
equations~\cite{ClassSym}.

\begin{example}\label{ExdBousHuman}
The operator $A_1$ for the dispersionless $3$-\/component Boussinesq
system~\eqref{d-B} has order zero and its matrix~\eqref{A1dBous} is
nondegenerate almost everywhere,
hence the condition~\eqref{NoKernelsIntersect} is valid.
We reconstruct the bracket $\ib{\,}{\,}{A_1}$ for this operator
performing two steps of the above algorithm.

The first step involves six combinations of~$\psi_1=\bun_i$
and~$\psi_2=\bun_j$ with $1\leq i\leq j\leq3$.
The second step repeats the first, but now $\psi_2$ is multiplied by~$x$,
and we have nine combinations $\psi_1=\bun_i$ and~$\psi_2=x\cdot\bun_j$,
$1\leq i,j\leq3$.
Then the terminal check~\eqref{TerminalCondition} is fulfilled.
This proves that the components of the bracket $\ib{\,}{\,}{A_1}$ are
\begin{subequations}\label{SokBrABoth}
\begin{align}
\ib{p}{q}{A_1}^u&= p^w_xq^u-p^uq^w_x
 +3w(p^uq^v_x -p^v_xq^u) +3w(p^vq^u_x-p^u_xq^v)\notag\\
{}&\quad{} +p^u_xq^w-p^wq^u_x
  +2w_x(p^uq^v-p^vq^u)+u(p^vq^v_x-p^v_xq^v),\notag\\
\ib{p}{q}{A_1}^v&= p^u_xq^u-p^uq^u_x +4w(p^vq^v_x-p^v_xq^v)
 +p^v_xq^w-p^wq^v_x +p^w_xq^v-p^vq^w_x,\notag\\
\ib{p}{q}{A_1}^w&= u(p^uq^v_x-p^v_xq^u) +3w^2(p^vq^v_x-p^v_xq^v)
 +2u_x(p^vq^u-p^uq^v)\notag\\
{}&\quad{}+w(p^u_xq^u-p^uq^u_x)
 +u(p^vq^u_x-p^u_xq^v) +p^w_xq^w-p^wq^w_x.\\
\intertext{The first order operator~\eqref{A2dBous} is also
nondegenerate, and we obtain}
\ib{p}{q}{A_2}^u&= 3w(p^uq^v_x-p^v_xq^u)
 +2w(p^vq^u_x-p^u_xq^v) +2w_x(p^uq^v-p^vq^u)\notag\\
{}&\quad{} +p^w_xq^u-p^uq^w_x +2u(p^vq^v_x-p^v_xq^v),\notag\\
\ib{p}{q}{A_2}^v&= 4w(p^vq^v_x-p^v_xq^v) +p^u_xq^u-p^uq^u_x
 +2(p^w_xq^v-p^vq^w_x),\notag\\
\ib{p}{q}{A_2}^w&= 8v(p^vq^v_x-p^v_xq^v) +8u(p^vq^u_x-p^u_xq^v)
 +6w^2(p^vq^v_x-p^v_xq^v)\notag \\
{}&\quad{}+w(p^u_xq^u-p^uq^u_x)
 +2u_x(p^vq^u-p^uq^v) +u(p^uq^v_x-p^v_xq^u).
\end{align}
\end{subequations}
Sokolov's bracket for Hamiltonian operator~\eqref{A2HamdBous} is equal to
the sum of brackets~\eqref{SokBrABoth}, because the operators are linear
compatible. The result agrees with~\eqref{EqDogma}.
\end{example}

\begin{example}\label{ExBrByA}
Following the above algorithm and using the package~\cite{Jets}, we
obtain the bracket on the domain of the
Frobenius operator~\eqref{SquareE}: for any
$p,q\in\Omega^1(\cE^\veps_\IL)$, we have
\begin{align*} 
{}&\ib{p}{q}{\square} =
  \veps^2\cdot\bigl(p_{xx} q_x-p_x q_{xx}\bigr)
 -2\veps^2\cdot\bigl(p_{xxx} q-p q_{xxx}\bigr)\\
&\quad{}-12 \veps^4\cdot\bigl(8 \veps^2 u_x^3 u_{xx}-4 \veps^2 u_x^2 u_{xxx}+4 \veps^2 u_x u_{xx}^2
   +2 u_x u_{xx}-u_{xxx}\bigr)\\
&\qquad{}\times\bigl[1+8 \veps^2 u_x^2+16 \veps^4 u_x^4-2 \veps^2 u_{xx}
 -8 \veps^4 u_x^2 u_{xx}\bigr]^{-1}
 \cdot\bigl(p_{xx} q-p q_{xx}\bigr)\\
%
{}&\quad{}+\bigl(\underline{1}+288 \veps^4 u_x^4-288 \veps^4 u_x^2 u_{xx}+28 \veps^2 u_x^2-16 \veps^2 u_{xx}-288 \veps^6 u_x u_{xx} u_{xxx}\\
&\qquad{}-96 \veps^6 u_{xx}^3+3072 \veps^{10} u_x^{10}+24 \veps^6 u_{xxx}^2+24 \veps^4 u_{4x}+1408 \veps^6 u_x^6
  +3328 \veps^8 u_x^8\\
&\qquad{}-768 \veps^{10} u_{4x} u_{xx} u_x^4 -384 \veps^8 u_{4x} u_x^2 u_{xx}
  -2304 \veps^8 u_x^3 u_{xx} u_{xxx}+384 \veps^8 u_{xx}^2 u_x u_{xxx}\\
&\qquad{}-4608 \veps^{10} u_x^5 u_{xx} u_{xxx}+16 \veps^4 u_{xx}^2-5632 \veps^8 u_x^6 u_{xx}
  -1920 \veps^6 u_{xx} u_x^4 +3328 \veps^8 u_x^4 u_{xx}^2\\
&\qquad{}+512 \veps^6 u_{xx}^2 u_x^2+384 \veps^{10} u_x^4 u_{xxx}^2
  -960 \veps^{10} u_{xx}^4 u_x^2-48 \veps^4 u_x u_{xxx}-3072 \veps^{10} u_x^7 u_{xxx}\\
&\qquad{}+3072 \veps^{10} u_{xx}^3 u_x^4
  -2304 \veps^8 u_x^5 u_{xxx}-576 \veps^6 u_x^3 u_{xxx}+288 \veps^6 u_{4x} u_x^2
  +384 \veps^8 u_x^2 u_{xx}^3\\
&\qquad{}+6144 \veps^{10} u_{xx}^2 u_x^6-6144 \veps^{10} u_{xx} u_x^8+1152 \veps^8 u_{4x} u_x^4
  +1536 \veps^{10} u_{4x} u_x^6
  +192 \veps^8 u_{xxx}^2 u_x^2\\
&\qquad{}+240 \veps^8 u_{xx}^4+1536 \veps^{10} u_{xx}^2 u_x^3 u_{xxx}
  -48 \veps^6 u_{4x} u_{xx}\bigr)\\
&\qquad{}\times\bigl[\underline{1}+96 \veps^4 u_x^4+256 \veps^6 u_x^6
  +256 \veps^8 u_x^8+4 \veps^4 u_{xx}^2-48 \veps^4 u_x^2 u_{xx}
  +32 \veps^6 u_{xx}^2 u_x^2\\
&\qquad\quad{}-4 \veps^2 u_{xx}-256 \veps^8 u_x^6 u_{xx}+64 \veps^8 u_x^4 u_{xx}^2
  -192 \veps^6 u_{xx} u_x^4+16 \veps^2 u_x^2\bigr]^{-1}\cdot\bigl(p_x
q-p q_x\bigr).
\end{align*}
The two underlined units correspond to the bracket $p_xq-pq_x$ on the
domain of the operator $\square=\cU_x+\tfrac{1}{2}D_x$ that
factors symmetries of the Liouville equation $\cU_{xy}=\exp(2\cU)$
at~$\veps=0$. In agreement with Lemma~\ref{LCoeffInKernel},
the non\/-\/constant coefficients of bilinear terms
$p_{xx}q-pq_{xx}$ and $p_xq-pq_x$ belong to~$\smash{\ker D_y\bigr|_{\cEL^\veps}}$.
\end{example}

\begin{rem}
The classification problem for Frobenius operators $A$ and the task of
reconstruction of the associated brackets $\ib{\,}{\,}{A}$ can be
performed using any software capable for calculation of the
commutators, e.g., \cite{SsTools,Jets} designed for symmetry analysis
of evolutionary (super-)\/PDE.
The implementation of technique of the
$\Delta$-\/coverings \cite{Lstar} is extremely productive here;
see~\cite{SsTools} for numerous examples.
In this paper, we considered not only
the $\ell$-{} and $\ell^*$-\/coverings, which correspond to Frobenius
recursions~\eqref{FrobRec} for symmetries of PDE
and to Noether operators 
for determined evolutionary systems, respectively.
\end{rem}


\section{Bi\/-\/differential representations of the
brackets~\protect{$\ib{\,}{\,}{A}$}}\label{AppBiDiffA12}\label{AppBiDiff}
The components of Sokolov's brackets
$\ib{\,}{\,}{A}\in\CDiff\bigl(\bigwedge^2\Omega^1(\xi_\pi),
  \Omega^1(\xi_\pi)\bigr)$ are matrix bi\/-\/differential operators in
total derivatives w.r.t.\ the components of sections that belong
to~$\Omega^1(\xi_\pi)$. We illustrate this for the
brackets~\eqref{SokBrABoth}, which were calculated in
Example~\ref{ExdBousHuman} on p.~\pageref{ExdBousHuman}
for Frobenius operators~\eqref{A1dBous} and~\eqref{A2dBous}.

The notation means that the differential operators standing in the
first and second tensor factors act, respectively, on the
first and second arguments of the
coupling $\langle\psi_1\mid\ib{\,}{\,}{A}\mid\psi_2\rangle$.

The  $u$-, $v$-, and $w$-components of the 
bracket $\{\!\{\ ,\ \}\!\}_{A_1}$
on the domain $\hat{\varkappa}(\pi)$
of~$A_1$ are the bi\/-\/differential operators
{\small\begin{gather*}
\begin{pmatrix}
0&3w\cdot(\bun\otimes D_x-D_x\otimes\bun)+2w_x\cdot\bun\otimes\bun&
   D_x\otimes\bun-\bun\otimes D_x\\
3w\cdot(\bun\otimes D_x -D_x\otimes\bun)-2w_x\cdot\bun\otimes\bun&
   u\cdot(\bun\otimes D_x-D_x\otimes\bun)&0\\
D_x\otimes\bun-\bun\otimes D_x&0&0
\end{pmatrix},\\
\begin{pmatrix}
D_x\otimes\bun-\bun\otimes D_x&0&0\\
0&4w\cdot(\bun\otimes D_x-D_x\otimes\bun)&
   D_x\otimes\bun-\bun\otimes D_x\\
0&D_x\otimes\bun-\bun\otimes D_x&0
\end{pmatrix},\\
\begin{pmatrix}
w\cdot(D_x\otimes\bun-\bun\otimes D_x)&
  u\cdot(\bun\otimes D_x-D_x\otimes\bun)-2u_x\cdot\bun\otimes\bun&0\\
u\cdot(\bun\otimes D_x-D_x\otimes\bun)+2u_x\cdot\bun\otimes\bun&
  3w^2\cdot(\bun\otimes D_x-D_x\otimes\bun)&0\\
0&0&D_x\otimes\bun-\bun\otimes D_x
\end{pmatrix}.
\end{gather*}}
The components of the bracket $\{\!\{\,,\,\}\!\}_{A_2}$
associated with the operator~$A_2$ are
{\small\begin{gather*}
\begin{pmatrix}
0&3w\cdot\bun\otimes D_x-2w\,D_x\otimes \bun+2w_x\cdot\bun\otimes \bun
   &-\bun\otimes D_x\\
2w\cdot\bun\otimes D_x-3w\,D_x\otimes \bun-2w_x\cdot\bun\otimes \bun&
   2u\cdot(\bun\otimes D_x-D_x\otimes\bun)&0\\
D_x\otimes \bun&0&0
\end{pmatrix},\\
\begin{pmatrix}
D_x\otimes \bun-\bun\otimes D_x&0&0\\
0&4w\cdot(\bun\otimes D_x-D_x\otimes \bun)&-2\cdot\bun\otimes D_x\\
0&2D_x\otimes \bun&0
\end{pmatrix},\\
\begin{pmatrix}
w\cdot(D_x\otimes \bun-\bun\otimes D_x)&
  u\cdot\bun\otimes D_x-8u\,D_x\otimes \bun-2u_x\cdot\bun\otimes \bun&0\\
2u_x\cdot\bun\otimes \bun+8u\cdot\bun\otimes D_x-u\,D_x\otimes \bun&
 (8v+6w^2)\cdot(\bun\otimes D_x-D_x\otimes \bun)&0\\
0&0&0 \end{pmatrix}.
\end{gather*}}
Both bi\/-\/differential matrix representations are
skew\/-\/symmetric in their arguments.

\section{Calculation of $\ib{\,}{\,}{R}$ for Frobenius
recursions~$R$}\label{AppRecJets}
The following program for \textsc{Jets} environment~\cite{Jets}
under \textsc{Maple} calculates Sokolov's bracket $\ib{\,}{\,}{R_0}$
for Frobenius recursion~$R_0=\hat{A}_0\circ A_1^{-1}$, see~\eqref{FRec},
which is factored by the first Hamiltonian operator~\eqref{dBA0}
and the inverse of Noether operator~\eqref{A1dBous} for the
dispersionless $3$-\/component Boussinesq\/-\/type system~\eqref{d-B}.
\begin{verbatim}
> read `Jets.s`;
> coordinates([x],[u,v,w,p1,p2,p3,q1,q2,q3],5):
\end{verbatim}
The dependent coordinates $\mathtt{p}^i$, $\mathtt{q}^j$ denote
the components of the fibre coordinates $\xi_1,\xi_2$ in the
new jet space, which is the $\ell$-\/covering over~$J^\infty(\pi)$,
see Remark~\ref{RemLstar} on p.~\pageref{RemLstar} and~\cite{Lstar}.
The notation is in parallel with formula~\eqref{ChainRuleFormula}.
\begin{verbatim}
> A:=Matrix(3,3): ia:=Matrix(3,3): M:=Matrix(3,3): N:=Matrix(3,3):
> A:=<<w*w_x+v_x,-3*w*u_x-u*w_x,u_x>|
  <-3*w*u_x-u*w_x,-3*w^2*w_x-4*w*v_x-u*u_x,v_x>|<u_x,v_x,w_x>>;
\end{verbatim}
We have assigned $\mathtt{A}=A_1$, see~\eqref{A1dBous}.
\begin{verbatim}
> with(LinearAlgebra):
> ia:=MatrixInverse(A):
\end{verbatim}
This yields $\mathtt{ia}=A_1^{-1}\colon\sym\cE\to\cosym\cE$, which we also denote
by~$\omega$. Now put $\mathtt{m}=A_1^{-1}(\xi_1)$ and $\mathtt{n}=A_1^{-1}(\xi_2)$.
\begin{verbatim}
> m1 := simplify(ia[1,1]*p1+ia[1,2]*p2+ia[1,3]*p3):
> m2 := simplify(ia[2,1]*p1+ia[2,2]*p2+ia[2,3]*p3):
> m3 := simplify(ia[3,1]*p1+ia[3,2]*p2+ia[3,3]*p3):
>
> n1 := simplify(ia[1,1]*q1+ia[1,2]*q2+ia[1,3]*q3):
> n2 := simplify(ia[2,1]*q1+ia[2,2]*q2+ia[2,3]*q3):
> n3 := simplify(ia[3,1]*q1+ia[3,2]*q2+ia[3,3]*q3):
\end{verbatim}
We have computed $\psi_i=\omega(\xi_i)$; recall that $\psi\in\cosym\cE$.

Next, we apply the Hamiltonian operator~$\hat{A}_0$ and obtain
symmetries of~\eqref{d-B},
$\mathtt{k}=\hat{A}_0(\mathtt{m})=\hat{A}_0(A_1^{-1}(\xi_1))$
and $\mathtt{l}=\hat{A}_0(\mathtt{n})=\hat{A}_0(A_1^{-1}(\xi_2))$.
\begin{verbatim}
> k1 := simplify(evalTD(TD(m1,x))):
> k2 := simplify(evalTD(-4*w*TD(m2,x)-2*w_x*m2+TD(m3,x))):
> k3 := simplify(evalTD(TD(m2,x))):
>
> l1:=simplify(evalTD(TD(n1,x))):
> l2:=simplify(evalTD(-4*w*TD(n2,x)-2*w_x*n2+TD(n3,x))):
> l3:=simplify(evalTD(TD(n2,x))):
\end{verbatim}

Now we act by evolutionary derivations on the coefficients of
the operator~$\omega=A_1^{-1}$, see~\eqref{ChainRuleFormula}.
We set $\mathtt{M}=\cEv_{R_0(\xi_1)}(\omega)$; note that the
matrix~$\mathtt{M}$ is symmetric.
\begin{verbatim}
> M[1,1]:=simplify(evalTD(k1*pd(ia[1,1],u)+TD(k1,x)*pd(ia[1,1],u_x)+
  k2*pd(ia[1,1],v)+TD(k2,x)*pd(ia[1,1],v_x)+k3*pd(ia[1,1],w)+
  TD(k3,x)*pd(ia[1,1],w_x))):
> M[1,2]:=simplify(evalTD(k1*pd(ia[1,2],u)+TD(k1,x)*pd(ia[1,2],u_x)+
  k2*pd(ia[1,2],v)+TD(k2,x)*pd(ia[1,2],v_x)+k3*pd(ia[1,2],w)+
  TD(k3,x)*pd(ia[1,2],w_x))):
> M[1,3]:=simplify(evalTD(k1*pd(ia[1,3],u)+TD(k1,x)*pd(ia[1,3],u_x)+
  k2*pd(ia[1,3],v)+TD(k2,x)*pd(ia[1,3],v_x)+k3*pd(ia[1,3],w)+
  TD(k3,x)*pd(ia[1,3],w_x))):
> M[2,2]:=simplify(evalTD(k1*pd(ia[2,2],u)+TD(k1,x)*pd(ia[2,2],u_x)+
  k2*pd(ia[2,2],v)+TD(k2,x)*pd(ia[2,2],v_x)+k3*pd(ia[2,2],w)+
  TD(k3,x)*pd(ia[2,2],w_x))):
> M[2,3]:=simplify(evalTD(k1*pd(ia[2,3],u)+TD(k1,x)*pd(ia[2,3],u_x)+
  k2*pd(ia[2,3],v)+TD(k2,x)*pd(ia[2,3],v_x)+k3*pd(ia[2,3],w)+
  TD(k3,x)*pd(ia[2,3],w_x))):
> M[3,3]:=simplify(evalTD(k1*pd(ia[3,3],u)+TD(k1,x)*pd(ia[3,3],u_x)+
  k2*pd(ia[3,3],v)+TD(k2,x)*pd(ia[3,3],v_x)+k3*pd(ia[3,3],w)+
  TD(k3,x)*pd(ia[3,3],w_x))):
\end{verbatim}
In the same way, we define the symmetric matrix
$\mathtt{N}=\cEv_{R_0(\xi_2)}(\omega)$.
\begin{verbatim}
> N[1,1]:=simplify(evalTD(l1*pd(ia[1,1],u)+TD(l1,x)*pd(ia[1,1],u_x)+
  l2*pd(ia[1,1],v)+TD(l2,x)*pd(ia[1,1],v_x)+l3*pd(ia[1,1],w)+
  TD(l3,x)*pd(ia[1,1],w_x))):
> N[1,2]:=simplify(evalTD(l1*pd(ia[1,2],u)+TD(l1,x)*pd(ia[1,2],u_x)+
  l2*pd(ia[1,2],v)+TD(l2,x)*pd(ia[1,2],v_x)+l3*pd(ia[1,2],w)+
  TD(l3,x)*pd(ia[1,2],w_x))):
> N[1,3]:=simplify(evalTD(l1*pd(ia[1,3],u)+TD(l1,x)*pd(ia[1,3],u_x)+
  l2*pd(ia[1,3],v)+TD(l2,x)*pd(ia[1,3],v_x)+l3*pd(ia[1,3],w)+
  TD(l3,x)*pd(ia[1,3],w_x))):
> N[2,2]:=simplify(evalTD(l1*pd(ia[2,2],u)+TD(l1,x)*pd(ia[2,2],u_x)+
  l2*pd(ia[2,2],v)+TD(l2,x)*pd(ia[2,2],v_x)+l3*pd(ia[2,2],w)+
  TD(l3,x)*pd(ia[2,2],w_x))):
> N[2,3]:=simplify(evalTD(l1*pd(ia[2,3],u)+TD(l1,x)*pd(ia[2,3],u_x)+
  l2*pd(ia[2,3],v)+TD(l2,x)*pd(ia[2,3],v_x)+l3*pd(ia[2,3],w)+
  TD(l3,x)*pd(ia[2,3],w_x))):
> N[3,3]:=simplify(evalTD(l1*pd(ia[3,3],u)+TD(l1,x)*pd(ia[3,3],u_x)+
  l2*pd(ia[3,3],v)+TD(l2,x)*pd(ia[3,3],v_x)+l3*pd(ia[3,3],w)+
  TD(l3,x)*pd(ia[3,3],w_x))):
\end{verbatim}
We act by the operators $\mathtt{M},\mathtt{N}$ on $\xi_2$ and $\xi_1$,
respectively, and calculate the difference
$\mathtt{e}=\cEv_{R_0(\xi_1)}(\omega)(\xi_2)
  -\cEv_{R_0(\xi_2)}(\omega)(\xi_1)$.
\begin{verbatim}
> e1:=simplify(M[1,1]*q1+M[1,2]*q2+M[1,3]*q3-N[1,1]*p1-N[1,2]*p2-N[1,3]*p3):
> e2:=simplify(M[1,2]*q1+M[2,2]*q2+M[2,3]*q3-N[1,2]*p1-N[2,2]*p2-N[2,3]*p3):
> e3:=simplify(M[1,3]*q1+M[2,3]*q2+M[3,3]*q3-N[1,3]*p1-N[2,3]*p2-N[3,3]*p3):
\end{verbatim}

Next, we recall that Sokolov's bracket for~$\hat{A}_0$
is equal to~\eqref{A0Bracket}, and substitute $\psi_i=\omega(\xi_i)$ in it.
Thus we put
$\mathtt{s}=\mathtt{e}+\{\!\{\mathtt{m},\mathtt{n}\}\!\}_{\hat{A}_0}$.
\begin{verbatim}
> s1:=simplify(e1):
> s2:=simplify(e2):
> s3:=simplify(e3+evalTD(2*(m2*TD(n2,x)-TD(m2,x)*n2))):
\end{verbatim}
We can now check that formula~\eqref{A0Bracket} is correct.
\begin{verbatim}
> J:=Jacobi([k1,k2,k3,0,0,0,0,0,0],[l1,l2,l3,0,0,0,0,0,0]):
> D1:=simplify(evalTD(TD(s1,x))):
> D2:=simplify(evalTD(-4*w*TD(s2,x)-2*w_x*s2+TD(s3,x))):
> D3:=simplify(evalTD(TD(s2,x))):
> simplify(evalTD(J[1]-D1));
                                  0
> simplify(evalTD(J[2]-D2));
                                  0
> simplify(evalTD(J[3]-D3));
                                  0
\end{verbatim}

Finally, we act onto the right\/-\/hand side of~\eqref{ChainRuleFormula}
by the operator~$\omega^{-1}=A_1$, and thus we obtain the components
$\mathtt{Z}=A_1(\mathtt{s})$ of Sokolov's bracket~$\ib{\,}{\,}{R_0}$.
\begin{verbatim}
> Z1:=simplify((w*w_x+v_x)*s1+(-3*w*u_x-u*w_x)*s2+u_x*s3);
> Z2:=simplify((-3*w*u_x-u*w_x)*s1+(-3*w*w*w_x-4*w*v_x-u*u_x)*s2+v_x*s3);
> Z3:=simplify(u_x*s1+v_x*s2+w_x*s3);
\end{verbatim}
The result is somewhat a surprise: the output contains more than 15,000
lines. Be that as it may, the operator $R_0=\hat{A}_0\circ\omega$ is
the first known example of a well\/-\/defined Frobenius recursion for
an integrable system.

\newpage
\subsection*{Acknowledgements}
The authors thank Y.~Kosmann\/--\/Schwarzbach, V.\,V.\,Sokolov,
B.\,A.\,Dub\-ro\-vin, M.
~Kon\-tse\-vich, I.\,S.\,Kra\-sil'\-sh\-chik,
and A.~M.~Verbovetsky for helpful discussions and constructive criticisms.
The authors are grateful to the referees for important comments and
suggestions, and also thank E.\,N.\,Mo\-re\-va for assistance in programming.
This work has been partially supported by the European Union through
the FP6 Marie Curie RTN \emph{ENIGMA} (Contract
no.\,MRTN-CT-2004-5652), the European Science Foundation Program
{MISGAM}, and by NWO grant~B61--609, TUBITAK, and NWO VENI grant 639.031.623.
A part of this research was done while A.\,K.\ was visiting
at~$\smash{\text{IH\'ES}}$ and~SISSA.

\enlargethispage{1.1\baselineskip}

\end{document}